\documentclass[letterpaper,11pt, nopreprintline,author,numbers,sort&compress]{elsarticle}

\usepackage{graphicx}
\usepackage{subcaption}
\usepackage{color}
\usepackage{tikz}
\usepackage{newlfont}
\usepackage{multirow}
\usepackage{array}

\usepackage{color}
\usepackage{hyperref}

\usepackage[font=footnotesize]{caption} 
\DeclareCaptionStyle{period-newline}%
[labelsep=period,justification=centering]%
{labelsep=period-newline}
\usepackage{longtable} 
\usepackage{ifthen}
\usepackage{alltt}
\usepackage{float}
\usepackage{enumerate}
\usepackage[text={6.5in,8.5in},centering]{geometry}
\newcolumntype{C}[1]{>{\centering\let\newline\\\arraybackslash\hspace{0pt}}m{#1}}
\newcolumntype{L}[1]{>{\raggedright\let\newline\\\arraybackslash\hspace{0pt}}m{#1}}
\newcolumntype{R}[1]{>{\raggedleft\let\newline\\\arraybackslash\hspace{0pt}}m{#1}}
\newcommand{\ba}{\begin{array} }
\newcommand{\ea}{\end{array} }
\newcommand{\bae}{\begin{eqnarray}}
\newcommand{\eae}{\end{eqnarray}}
\newcommand{\bea}{\begin{eqnarray*}}
\newcommand{\eea}{\end{eqnarray*}}
\newcommand{\be}{\begin{equation}}
\newcommand{\ee}{\end{equation}}

\newcommand{\red}[1]{\textcolor{black}{#1}}

\newcommand{\pr}{{\bf Proof}~~}

\def\to{{\rightarrow}}

\usepackage{enumitem}
\usepackage{lipsum}

\newlength\mylen
\newlist{mycases}{enumerate}{1}
\setlist[mycases,1]{label=\textbf{Case~\arabic*.}, 
  labelwidth=\dimexpr-\mylen-\labelsep\relax,leftmargin=0pt,align=right}

\def\v{\vspace{1ex}}

\def\o{{\omega}}

\def\to{{\rightarrow}}

\usepackage{amsfonts}
\usepackage{amssymb}
\usepackage{amsthm}
\usepackage{ mathrsfs }
 \usepackage{ae} 
\usepackage[T1]{fontenc}
\usepackage[ansinew]{inputenc}
\usepackage{amsmath}
\usepackage{eucal}

\usepackage[english]{babel}

\usepackage{color}
\usepackage{hyperref}
\hypersetup{
     colorlinks   = true,
     citecolor    = blue
}
\usepackage{lscape}
\usepackage{epstopdf}	
\newtheorem{theorem}{\hskip\parindent\bf Theorem}[section]

\newtheorem{proposition}{\bf Proposition}[section]

\usepackage{booktabs}
\usepackage{bm}
\usepackage[cal=cm]{mathalfa} 

\makeatletter
\ifcase \@ptsize \relax
  \newcommand{\miniscule}{\@setfontsize\miniscule{2}{3}}
\or
  \newcommand{\miniscule}{\@setfontsize\miniscule{2}{3}}
\or
  \newcommand{\miniscule}{\@setfontsize\miniscule{2}{3}}
\fi
\makeatother
\usepackage{lipsum}

\usepackage{float}

\usepackage{titlesec}
\titleformat*{\section}{\large\bfseries}
\titleformat*{\subsection}{\large\bfseries}

\usepackage{relsize}

\usepackage{pbox} 

\usepackage{pifont}

\date{}

\begin{document}

\begin{frontmatter}

\title{How to model honeybee population dynamics: stage structure and seasonality}
\author[1]{Jun Chen} \ead{jchen152@asu.edu}
\author[1]{Komi Messan}\ead{kmessan@asu.edu}
\author[2]{Marisabel Rodriguez Messan} \ead{marisabel@asu.edu}
\author[3]{Gloria DeGrandi-Hoffman } \ead{gloria.hoffman@ars.usda.gov}
\author[4,5]{Dingyong Bai}\ead{baidy@gzhu.edu.cn}
\author[6]{Yun Kang}\ead{yun.kang@asu.edu}

\address[1]{Simon A. Levin Mathematical and Computational Modeling Sciences Center, Arizona State University, \\Tempe, AZ 85281, USA.}
\address[2]{Department of Ecology and Evolutionary Biology, Brown University, \\Providence RI 02912}
\address[3]{Carl Hayden Bee Research Center, United States Department of Agriculture-Agricultural Research Service,\\ Tucson, AZ  85719 , USA.}
\address[4]{School of Mathematics and Information Science, Guangzhou University,\\ Guangzhou 510006, PR China.}
\address[5]{Center for Applied Mathematics, Guangzhou University,\\ Guangzhou 510006, PR China}
\address[6]{Sciences and Mathematics Faculty, College of Integrative Sciences and Arts, Arizona State University, \\Mesa, AZ 85212, USA.}

\date{}



\begin{abstract}
Western honeybees (\textit{Apis Mellifera}) serve extremely important roles in our ecosystem and economics as  they are responsible for pollinating \$ 215 billion dollars annually over the world.  Unfortunately,  honeybee population and their colonies have been declined dramatically. The purpose of this article is to explore how we should model honeybee population with age structure and validate the model using empirical data so that we can identify different factors that lead to the survival and healthy of the honeybee colony.  Our theoretical study combined with simulations and data validation suggests that the proper age structure incorporated in the model  and seasonality are important for modeling honeybee population.  Specifically, our work implies that the model assuming that (1) the adult bees are survived from the {egg population} rather than the brood population; and (2) seasonality in the queen egg laying rate, give the better fit than other honeybee models. The related theoretical and numerical analysis of the most fit model indicate that (a) the survival of honeybee colonies requires a large queen egg-laying rate and smaller values of the other life history parameter values in addition to proper initial condition; (b) both brood and adult bee populations are increasing with respect to the increase in the {egg-laying rate} and the decreasing in other parameter values; and (c) seasonality may promote/suppress the survival of the honeybee colony. \\
 \end{abstract}


\end{frontmatter}

%
%
%
%
%
%
%
%
%
%
%
	

\section{Introduction}

Western honeybee (\textit{Apis Mellifera}) is an eusocial insect that has an advanced level of social organization. In honeybee colony, the queen produces the offspring and non-reproductive individuals cooperate in caring for the young ones, that forms complex colonies \cite{winston1991biology}.
Honeybees play indispensable and important roles in human life, economy, and agriculture.  For example, honeybees not only produce valuable products, such as honey, royal jelly, bee wax and propolis in the market, but also are responsible for pollinating crops such as blueberries, cherries, and almonds, that is worth \$215 billion annually worldwide \cite{smith2013pathogens}. If there is no honeybee, it likely leads to changes in human diets and a disproportionate expansion of agricultural land in order to fill this shortfall in crop production by volume \cite{potts2016safeguarding}.
Unfortunately, honeybee  population has been decreasing globally  \cite{smith2013pathogens}. In the United States, the total number of honeybee colonies has been reduced approximate 40\% to 50\%, while in the rest of the world the total number of colonies is reduced by 5\% to 10\% \cite{iniciativasCol}.  The important and critical causes for honeybee colony mortalities include diseases, land-use change, pesticides, pathogens and parasites, and poor beekeeping management \cite{smith2013pathogens, degrandi2013effects,perry2015rapid,oldroyd2007s,degrandi2019economics}.\\
The purpose of this article is to explore how we could better model colony population dynamics to help us understand the honeybee colony  mortalities.\\

 
Honeybee colony itself  is a complex adaptive system with its own resilience to disturbances, whose survival depends on its individual quality, its adaptive capacity and its threshold of resilience to pressures \cite{fredrick2017review}. On average, a colony has about 10,000 to 60,000 bees, that consists of a queen (fertile female) who produces all offspring, a few hundred drones (males) and thousands of workers (sterile females).  Generally, a queen may lay approximate 1000-2000 eggs per day in the peak period \cite{coffey2007parasites}. Due to aging or disability of the queen bee, beekeepers will replace the queen every 1-2 years \cite{coffey2007parasites}. Each honeybee goes through four stages of development: egg, larva, pupa and adult \cite{coffey2007parasites}.  For worker bees, they need 21 days to eclosion to adult bees \cite{winston1991biology,harris1980population,degrandi1989beepop}, and drones need 24 days to mature \cite{degrandi1989beepop}. Population size at each stage and the related maturation time have huge influences on the colony development and its population dynamics \cite{fredrick2017review}. Needless to say, age is linked to division of labor in honeybees \cite{robinson1992colony}. Young workers In the colony, young workers prefer to perform nursing tasks, while older workers prefer foraging activities. However, colonies can accelerate, delay, or even reverse their recruitment behavior as the internal or external environment changes \cite{huang1996regulation}.  \\

Not only the age structure will affect the honeybees colony, but the change of season, temperature, weather, etc. also will influence the honeybees \cite{johnson2002reallocation,coffey2007parasites,degrandi1989beepop,seeley1985survival}. Through experiments and observations, honeybee population present periodic fluctuations due to different reasons. {For instance, we  observe great foraging activity  during spring, summer and fall but the highest activity during the summer \cite{coffey2007parasites}. During spring and summer, pollen and nectar from diverse floras are in great abundance, giving rise to an increase honeybee population. Therefore, given that temperature is one of the main factors in honeybee food availability and thus brood production, honeybee population size is smaller during the winter \cite{seeley1985survival, degrandi1989beepop}.} Thus, the peak of the population is achieved in late June until middle of summer as it starts to decline \cite{seasonality}. The temperature in the colony also will influence honeybee, middle-age honeybees will respond to the heat stress in order to perform \cite{johnson2002reallocation}.  Thus, it is very important to include age structure and the seasonality in studying of honeybee population dynamics and the factors that affect healthy of honeybee colonies. 
Research has shown that the major problems threatening the survival of honeybee colonies could link to: 1) environmental stressors, such as habitat destruction (urbanization, deforestation, forest fires); 2) parasites and pathogens, such varroosis and virus; 3) genetic variation and vitality, like limited importation \cite{perry2015rapid,oldroyd2007s,smith2013pathogens}. In order to quantify the problems and consider the difficulty of directly observing the dynamics of bee populations, mathematical models can be a powerful tool to help us understand how the bee population change and predict the fate of the colony. \\

Mathematical models indeed have been developed to study bee populations dynamics and the related stressors, particularly the effects of pathogens, parasites and nutrient stress factors \cite{russell2013dynamic,kribs2014modeling,khoury2011quantitative,khoury2013modelling,perry2015rapid,betti2014effects,camazine1990mathematical,kang2015ecological,eberl2010importance,aronstein2012varroa,smith2013pathogens}.  DeGrandi-Hoffman \cite{degrandi1989beepop} proposed a first simulation model for honeybee colony dynamics that includes many important factors such weather, egg-laying rate, the age of queen, foraging and brood life cycles. 
There are some previous work focusing on how the death rate of foragers impacts colony viability \cite{khoury2011quantitative,khoury2013modelling}. Khoury  \cite{khoury2011quantitative} published a compartmental model based on these circumstances. The model includes three states, brood, hive and foragers, and incorporates the recruitment process to study the forager death rate. There is a work \cite{russell2013dynamic} that investigated seasonal food availability and transition of hive to foragers. The most recent recent works \cite{messan2018effects,ratti2015mathematical} consider the seasonality in the queen egg-laying rate. Messan et al \cite{messan2018effects} applied seasonality effects into the pollen collection rate that has annual periodicity by the first order harmonic. Ratti et al \cite{ratti2015mathematical} also agrees that seasonality affects dynamics of honeybee and its parasitic virus. This article \cite{ratti2015mathematical} incorporated  seasonality in varroa treatment control as the treatment is applied with four seasons: spring, summer, fall, and winter \cite{ratti2015mathematical}.\\
 
Motivated by the previous work on honeybee population models with age structure \cite{khoury2011quantitative,khoury2013modelling,kang2015ecological} and seasonality \cite{russell2013dynamic,messan2018effects,ratti2015mathematical}, we propose and study honeybee population models with different delay terms to include age structure. We use data to validate our models and explore which model would be more appreciated and the importance of incorporating seasonality in the honeybee population model.
More specifically, the objective of our paper is to develop a proper honeybee population dynamical model with age structure to understand important factors for colony survival, and to explore how seasonality  may affect the colony dynamics and its survival.\\

The remaining of the article is shown as follows: In Section 2, we derive two honeybee colony dynamics models that incorporate varied delay terms. In Section 3, we perform rigorous mathematical analysis for those two models and compare their dynamics. In Section 4, we validate our two models with real honeybee data. Our study shows the importance of seasonality and suggests that one of those two proposed models would be more appropriated for studying honeybee population dynamics. In Section 5, we conclude our study. In the last section, we provide detailed proofs of our theoretical results.\\ 



	\section{Model Derivations}

In this section, we focus on modeling of honeybee colony dynamics with age structure. For convenience, we divide the population of  honeybee colony into brood and adult bees.  Let $B(t),\, H(t)$ be the population of brood and adult bees in a given hive at time $t$, respectively. We assume that:

 \begin{itemize}

	\item[\textbf{A1:}] The daily egg laying rate of honeybee queen is $r$ with the survival rate of $\frac{H^2}{K+H^2+\alpha B} $ where the parameter $K$ is the population of adult bee needed for half of the maximum brood survival rate and $\alpha$ represents the regulation effects from brood population $B$. The term $\frac{H^2}{K+H^2+\alpha B} $ reflects (1) the cooperative brood care from adult bees that perform nursing and collecting food for brood; and (2) the queen and workers that regulate the actual egg laying/survival  rate based on the current available brood population $B$,  which has been supported by the literature work \cite{messan2018effects, schmickl2007hopomo, kang2016disease,eischen1984some}.\\

	\item[\textbf{A2:}] We assume that both brood and adult bees have constant mortality, $d_b$ and $d_h$ respectively.  The maturation time from brood $B$ to adult bee $H$ is denoted by $\tau$ ($\tau=16$ for queen, $\tau=21$ for workers, and $\tau=24$ for drones \cite{coffey2007parasites,khoury2013modelling}), thus the maturation rate is termed as follows:
	$$\underbrace{e^{-d_b\tau}}_{\text{survival rate of brood during time $\tau$}}\underbrace{\frac{r H(t-\tau)^2}{K+H(t-\tau)^2+\alpha B(t-\tau)}}_{\text{new brood at $t-\tau$}}$$

   \end{itemize}

 The two assumptions above lead to the following  non-linear delayed differential equations of honeybee population dynamics (Model \eqref{BH}):\\

\begin{equation}\label{BH}
\begin{array}{lcl}
\frac{dB}{dt}&=&\frac{rH(t)^2}{K+H(t)^2+\alpha B}-d_b B- e^{- d_b \tau} \frac{rH(t-\tau) ^2}{K+H(t-\tau)^2+\alpha B(t-\tau)}   \\\\
\frac{dH}{dt}&=& e^{- d_b \tau} \frac{rH(t-\tau) ^2}{K+H(t-\tau)^2+\alpha B(t-\tau)}-d_h H\\
\end{array}
\end{equation}
where we assume that  the initial condition for $H(t)$ is a nonnegative continuous function when  $t\in [-\tau,0]$ and $B(0)=\int^0_{-\tau}\frac{rH^2(s)e^{d_bs}}{K+H^2(s)+\alpha B(s)}ds$. The biological meaning of each parameter of the proposed model \eqref{BH} is listed in Table \ref{tb1}. In the case $\alpha=0$, the model \eqref{BH} reduces to the following Model \eqref{BH3}

\begin{eqnarray}\label{BH3}
\begin{array}{lcl}
\frac{dB}{dt}&=&\frac{rH(t)^2}{K+H(t)^2}-d_b B- e^{- d_b \tau}\frac{rH(t-\tau)^2}{K+H(t-\tau)^2} \\
\frac{dH}{dt}&=& e^{- d_b \tau} \frac{rH(t-\tau)^2}{K+H(t-\tau)^2}-d_h H(t)
\end{array}
\end{eqnarray}

\noindent\textbf{Notes:} Our proposed model \eqref{BH3} (when $\alpha=0$ in the model \eqref{BH}) is a single specie model with brood $B$ and adult  $H$ stage where these two stages seem to be decoupled. Thus, we could study the dynamics of  Model \eqref{BH3} by exploring the dynamics of $H$ first, then the dynamics of $B$ is totally determined by $H$. We would see the analytical results in the next section.\\

\renewcommand{\arraystretch}{1.5}
\begin{table}[ht]
\begin{center}
\tabcolsep=0.1cm
	\scalebox{0.6}{
		\begin{tabular*}{1.8\textwidth}{@{\extracolsep{\fill} } l l l l l }
        	\hline
     Parameters of Models With and Without Seasonality \\
        \hline
			\hline
            Parameter & Description & Estimate/Units & Reference\\ \hline
            $r$ & Daily egg-laying rate of Queen  & [500 10,000] bees/day & Estimated \\
           $\alpha$ & the regulation effects of brood & [0 20] & Estimated\\
			$d_b$ & Death rate of the brood &[0, 0.3] $day ^{-1}$ & Estimated\\
			$d_h$ & Death rate of the adult bees & [0, 0.3] $day ^{-1}$ & Estimated \\
            $\gamma$ &The length of seasonality & [170, 365] days & Estimated\\
			$\sqrt{K}$ & Colony size at which brood survival rate is half maximum & for $K$ \begin{tabular}[c]{@{}l@{}}[50,000, 1,300,000](model \ref{BH2})\\  $[1*(10^7), 1*(10^8)]$(model \ref{BH}) bees/day\end{tabular} &Estimated \\
		
              $\tau$ & Time spent in brood  & 21 days &Khoury 2013\\
              	  $\psi$ & the time of the maximum laying rate & 12 days  & Harris 1980 \\
			\hline
\end{tabular*}}
\caption{Biological meanings and references of parameters in the proposed model \eqref{BH}.}
		\label{tb1}
\end{center}
\end{table}	

 In literature (e.g, see  \cite{tang2002density}), researchers have been using the compartmental models through ODEs to model population dynamics with age and/or stage structure. Motivated by this,  we have the following delay model

\begin{eqnarray}\label{BH2}
\begin{array}{lcl}
\frac{dB}{dt}&=&\frac{rH(t)^2}{K+H(t)^2}-d_b B- e^{- d_b \tau} B(t-\tau)   \\
\frac{dH}{dt}&=& e^{- d_b \tau} B(t-\tau)-d_h H(t)
\end{array}
\end{eqnarray}where the term $e^{- d_b \tau} B(t-\tau)$ describes the maturation entry rate coming from the juvenile stage with a survival rate $e^{- d_b \tau} $ during the juvenile period $\tau$.

More specifically, the model \eqref{BH2} above assumes that the adult $H(t)$ matures from the survived brood population at time $t-\tau$. In the following two sessions, we will compare population dynamics of Model \eqref{BH2} and the proposed model \eqref{BH} to address the importances of deriving proper delay population models due to the outcomes of dynamics and  model validations. \\

\section{Mathematical Analysis}


{The state space of the proposed model \eqref{BH} is $\mathbb X=C([-\tau,0], \mathbb R_+) \times C([-\tau,0], \mathbb R_+).$  We first show that the proposed model \eqref{BH} is positive invariant and bounded in $\mathbb X$ as the following theorem:}\\

\noindent \begin{theorem}\label{th1:pb}
\noindent Assume that the initial condition $H(t)$ is a nonnegative continuous function defined in $t\in [-\tau,0]$ with $B(0)=\int^0_{-\tau}\frac{rH^2(s)e^{d_bs}}{K+H^2(s)+\alpha B(s)}ds$, then the proposed model  \eqref{BH} is positive invariant and bounded in $\mathbb X$. \end{theorem}

\noindent\textbf{Notes:} The detailed proof of Theorem \ref{th1:pb} is in the last section. Theorem \ref{th1:pb} implies that our proposed model is biologically well defined. The model \eqref{BH} always has the extinction equilibrium $E_e=(0,0)$ which would be locally or globally stable as stated in the next theorem:\\

\begin{theorem}\label{th2:Ee}
[\textit{Stability of Extinction Equilibrium}] The extinction equilibrium $E_e$ of Model \eqref{BH} is always locally asymptotically stable. If the inequality $d_h > \frac{re^{-d_b\tau}}{2\sqrt{K}}$ holds, the extinction equilibrium $E_{e}$  is globally stable.
\end{theorem}

\noindent\textbf{Notes:} The detailed proof of Theorem \ref{th2:Ee} is in the last section. 
Theorem \ref{th2:Ee} indicates that the large maturation time $\tau$ or the mortality at different stages $d_h, d_b$ can lead to the collapsing of the colony.\\

Now we focus on the condition of the colony survival. Let $(B,H)$ be an interior equilibrium of Model \eqref{BH}. Then it satisfies that the following equations:

\begin{eqnarray}
0&=& \frac{rH(t)^2}{K+H(t)^2+\alpha B(t)}-d_b B- e^{- d_b\tau}\frac{rH(t)^2}{K+H(t)^2+\alpha B(t)} \Rightarrow d_b B=\frac{[1-e^{- d_b\tau}]rH^2}{K+H^2+\alpha B}\\
0&=& e^{- d_b \tau} \frac{rH(t)^2}{K+H(t)^2+\alpha B(t)}-d_h H(t) \Rightarrow d_h H=\frac{e^{- d_b \tau} rH^2}{K+H^2+\alpha B}\label{ht1}
\end{eqnarray}
which gives
\begin{eqnarray}\label{eq-BH}
B& =&\frac{d_h[e^{d_b\tau}-1]}{d_b}
\end{eqnarray}
 and
\begin{eqnarray}\label{eq-H}
0& =&\frac{e^{-d_b\tau}r H d_b}{d_b(K+H^2)+\alpha d_h (e^{d_b\tau}-1)H}-d_h.
\end{eqnarray}

Solving the equation \eqref{eq-H} gives
$$H^*_1=\frac{e^{-d_b \tau } \left(d_b r-\alpha  d_h^2 e^{d_b \tau }\left[e^{d_b \tau }-1\right]-\sqrt{\left(d_b r-\alpha  d_h^2 e^{d_b \tau } \left[e^{d_b \tau }-1\right]\right)^2-4 d_b^2 d_h^2 K e^{2 d_b \tau }}\right)}{2 d_b d_h}$$ and

$$H^*_2=\frac{e^{-d_b \tau } \left(d_b r-\alpha  d_h^2 e^{d_b \tau }\left[e^{d_b \tau }-1\right]+\sqrt{\left(d_b r-\alpha  d_h^2 e^{d_b \tau } \left[e^{d_b \tau }-1\right]\right)^2-4 d_b^2 d_h^2 K e^{2 d_b \tau }}\right)}{2 d_b d_h}$$ with $H^*_1\leq H^*_2$.  Now we have the following proposition:\\

\noindent \begin{proposition}\label{p:eq1}
[\textit{Existence of Interior Equilibria}] If $\frac{d_b r-d_h^2e^{d_b \tau }[e^{d_b \tau }-1]\alpha}{2d_h d_b e^{d_b\tau}\sqrt{K}}>1$, then Model \eqref{BH} has  two interior equilibria $E_i, i=1,2$:
$$ E_i=(B^*_i,H^*_i)=\left(\frac{d_h[e^{d_b\tau}-1]}{d_b} H^*_i, H_i^*\right)$$ {where $H_1^*$ is an increasing function of $\alpha$, $K$, $d_b$ and $d_h$, and $H_2^*$ is a decreasing function of $\alpha$, $K$, $d_b$ and $d_h$; whereas, $H_1^*$ is a decreasing function of $r$, and $H_2^*$ is an increasing function of $r$.} In the case that $\frac{d_b r-d_h^2e^{d_b \tau }[e^{d_b \tau }-1]\alpha}{2d_h d_b e^{d_b\tau}\sqrt{K}}=1$, then Model \eqref{BH} has an unique interior equilibrium
$$ E_i=(B^*,H^*)=\left(\frac{d_h[e^{d_b\tau}-1]}{d_b} H^*, H^*\right)\mbox{ with }H^*=H^*_1=H^*_2=\frac{e^{-d_b \tau } \left(d_b r-\alpha  d_h^2 e^{d_b \tau }\left[e^{d_b \tau }-1\right]\right)}{2 d_b d_h}.$$

\end{proposition}

\noindent\textbf{Notes:} Proposition \ref{p:eq1} implies that one of the necessary conditions for the honeybee colony survival is $\frac{d_b r-d_h^2e^{d_b \tau }[e^{d_b \tau }-1]\alpha}{2d_h d_b e^{d_b\tau}\sqrt{K}}>1$ which requires large values of the queen egg laying rate $r$, and the smaller values of the maturation time $\tau$ and the brood regulation effect $\alpha$. In addition, Proposition \ref{p:eq1} indicates that at the interior equilibrium, the ratio of brood $B$ to adult population $H$ is determined by their mortality and maturation time through the equation $\frac{d_h\left[e^{d_b\tau}-1\right]}{d_b}$. 
Based on simulations and analytical results, the interior equilibrium $H^*_2$ is always locally stable if it exists while $H_1^*$ is locally unstable. If $\frac{d_b r-d_h^2e^{d_b \tau }[e^{d_b \tau }-1]\alpha}{2d_h d_b e^{d_b\tau}\sqrt{K}}>1$, then by simple calculations, we have  $\frac{dH^*_1}{d\alpha}>0$ and $\frac{dH^*_2}{d\alpha}<0$. This implies that the brood regulation coefficient $\alpha$ has negative effects on brood and adult population sizes. In the case that $\alpha =0$, then if $\frac{d_b r}{2d_h d_b e^{d_b\tau}\sqrt{K}}>1$ holds, the interior equilibria  $H^*_i, i=1,2$  have the following expressions:

\bae\label{alpha0}
H_1^*&=&\frac{e^{-d_b \tau } \left(d_b r-\sqrt{\left(d_b r\right)^2-4 d_b^2 d_h^2 K e^{2 d_b \tau }}\right)}{2 d_b d_h}\\
H_2^*&=&\frac{e^{-d_b \tau } \left(d_b r+\sqrt{\left(d_b r\right)^2-4 d_b^2 d_h^2 K e^{2 d_b \tau }}\right)}{2 d_b d_h}
\eae

In order to study the stability of the interior equilibrium $E_i=(B^*,H^*)$, we start with the characteristic equation of the interior equilibrium $E_i={B^*_i,H^*_i}$ as follows by letting $A =  \frac{  H r}{\left(\alpha  B+H^2+K\right)^2}$

\begin{equation}\label{th1:ch}
\begin{aligned}
C(\lambda)
                     &= det \left(  \left[ \begin{array}{lcr}
               -\frac{\alpha  H^2 r}{\left(\alpha  B+H^2+K\right)^2}-d_b & \frac{2 H r (\alpha  B+K)}{\left(\alpha  B+H^2+K\right)^2}\\
               0 &  -d_h
              \end{array} \right]
	    +  \left[ \begin{array}{lcr}
		   -\frac{\alpha  H^2 r e^{-d_b \tau }}{\left(\alpha  B+H^2+K\right)^2} & -\frac{2 H r e^{-d_b \tau } (\alpha  B+K)}{\left(\alpha  B+H^2+K\right)^2} \\
		    \frac{\alpha  H^2 r e^{-d_b \tau }}{\left(\alpha  B+H^2+K\right)^2}& \frac{2 H r e^{-d_b \tau } (\alpha  B+K)}{\left(\alpha  B+H^2+K\right)^2}
		  \end{array} \right]
		  \ast e^{-\lambda\tau} - \lambda \mathscr{I} \right)\\
 &= det \left(  \left[ \begin{array}{lcr}
              -\alpha AH-d_b & 2 A (\alpha  B+K)\\
               0 &  -d_h
              \end{array} \right]
	    +  \left[ \begin{array}{lcr}
		    -\alpha AHe^{-d_b \tau } &  -2 A (\alpha  B+K)e^{-d_b \tau } \\
		    \alpha AHe^{-d_b \tau }&  2 A (\alpha  B+K)e^{-d_b \tau }
		  \end{array} \right]
		  \ast e^{-\lambda\tau} - \lambda \mathscr{I} \right)\\
	&=det ( \left[ \begin{array}{lcr}
                -\alpha AH(1+e^{-(\lambda+d_b)\tau})-d_b  -\lambda & 2 A (\alpha  B+K)(1-e^{-(\lambda+d_b)\tau})\\   \alpha AHe^{-d_b \tau } & -d_h+ 2 A (\alpha  B+K)e^{-(\lambda+d_b)\tau}-\lambda
              \end{array}\right]\\
              &=(-\alpha AH(1+e^{-(\lambda+d_b)\tau})-d_b -\lambda)(-d_h+ 2 A (\alpha  B+K)e^{-(\lambda+d_b)\tau}-\lambda)\\
              &-2 A (\alpha  B+K)(1-e^{-(\lambda+d_b)\tau})\alpha AHe^{-d_b \tau }
\end{aligned}
\end{equation}

We can see the characteristic equation \eqref{th1:ch} of the interior equilibrium $E_i={(B^*_i,H^*_I)}$ is very complicated and difficult to analysis. Thus, for convenience, we start with the simpler case by setting $\alpha=0$ which is our model \eqref{BH3}.\\

\begin{theorem}\label{th1:stability}
[\textit{Stability of Interior Equilibria}] If $\frac{r}{2d_he^{d_b\tau}\sqrt{K}}>1$, then Model \eqref{BH3} has two interior equilibria $E_i$ where $E_1$ is always unstable and $E_2$ is always  locally asymptotically stable.\\
\end{theorem}
\noindent\textbf{Notes:} Theorem \ref{th1:stability} indicates that the value of the maturation time $\tau$ has no effects on the stability of its interior equilibria $E_i, i=1,2$ for Model \eqref{BH3}. In the case that $\frac{r}{2d_he^{d_b\tau}\sqrt{K}}=1$, i.e., $d_h = \frac{re^{-d_b\tau}}{2\sqrt{K}}$, then Model \eqref{BH3} has an unique interior $E=(B^{*}, H^{*})=\left( \frac{r\left[1 - e^{-d_b\tau}\right]}{2d_b},\sqrt{K}\right)$. The following theorem provides results on the interior equilibrium stability of this critical case:\\

\begin{theorem}\label{th4:stabilityctrical}
[\textit{Unique Interior Equilibrium}] If $d_h = \frac{re^{-d_b\tau}}{2\sqrt{K}}$, Model \eqref{BH3} has a unique interior equilibrium $E=(B^{*}, H^{*})=\left( \frac{r\left(1 - e^{-d_b\tau}\right)}{2d_b},\sqrt{K}\right)$ which is always locally asymptotically stable for any delay $\tau>0$.\\
\end{theorem}

\noindent\textbf{Notes:} The detailed proof of Theorem \ref{th4:stabilityctrical} uses normal form theory in \cite{faria1995normal}, and is provided in the last section. Both Theorem \ref{th1:stability} and Theorem \ref{th4:stabilityctrical} implies that when $\alpha=0$, i.e., Model \eqref{BH3}, the colony can survive at $E_2=(B^{*}_2, H^{*}_2)$ if $\frac{r}{2d_he^{d_b\tau}\sqrt{K}}\geq1$ and initial conditions are in a proper range. \\

What if $\alpha >0$? Our simulations suggest that  $E_1$ is still unstable and $E_2$ is  locally stable.
Based on our analytical results and simulations,  we summarize the general dynamics of Model \eqref{BH} as follows:
\begin{enumerate}
\item The extinction equilibrium $E_e$ of Model \eqref{BH3} always exists and is always locally asymptotically stable.
\item If $\frac{r}{d_h} < 2e^{d_b\tau}\sqrt{K}$, Model \eqref{BH} has its global stability at the extinction equilibrium $E_e$.
\item If $\frac{d_b r-d_h^2e^{d_b \tau }[e^{d_b \tau }-1]\alpha}{2d_h d_b e^{d_b\tau}\sqrt{K}}\geq 1$, Model \eqref{BH} has two locally asymptotically stable equilibria: the extinction equilibrium $E_e$ and the interior equilibrium $E_2=(B^*_2,H^*_2)$.\\
\end{enumerate}

\begin{figure}[!ht]
 \centering
    \begin{subfigure}{0.24\textwidth}
        \includegraphics[width=\textwidth]{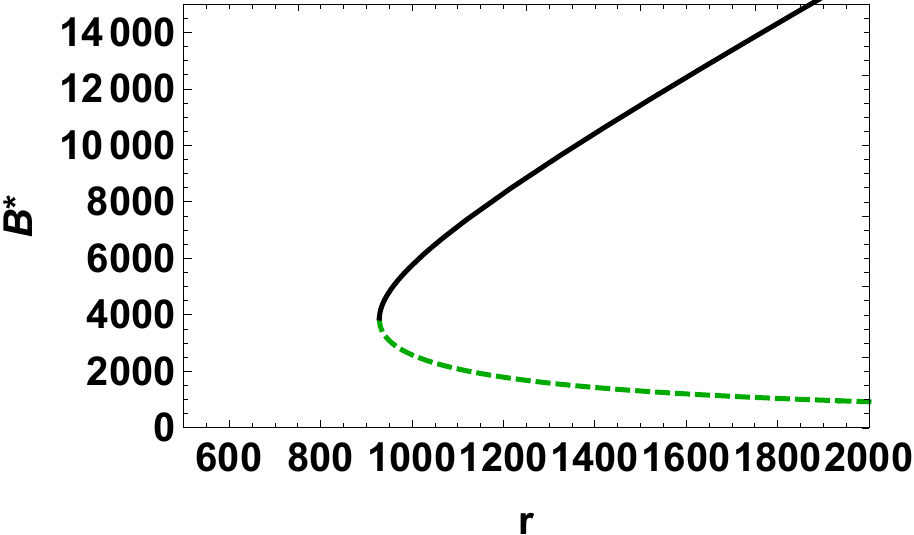}
\caption{Brood population and the queen egg-laying rate $r$ when $K=10^5, \alpha=3, d_b=0.1, d_h=0.17, \tau=21$.}
\label{fig:m1rbB}
    \end{subfigure}
    \begin{subfigure}{0.24\textwidth}
        \includegraphics[width=\textwidth]{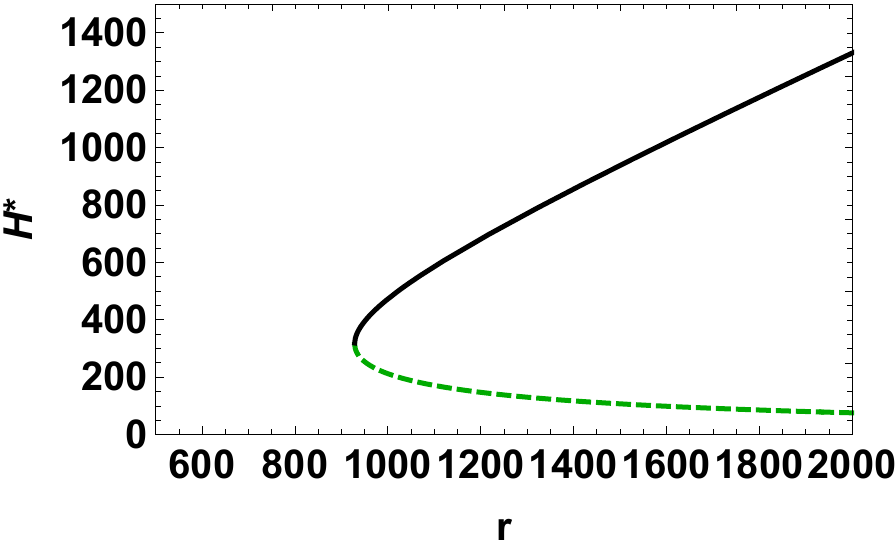}
\caption{Adult population and the queen egg-laying rate $r$ when $K=10^5, \alpha=3, d_b=0.1, d_h=0.17, \tau=21$.}
\label{fig:m1rbH}
    \end{subfigure}
    \begin{subfigure}{0.24\textwidth}
        \includegraphics[width=\textwidth]{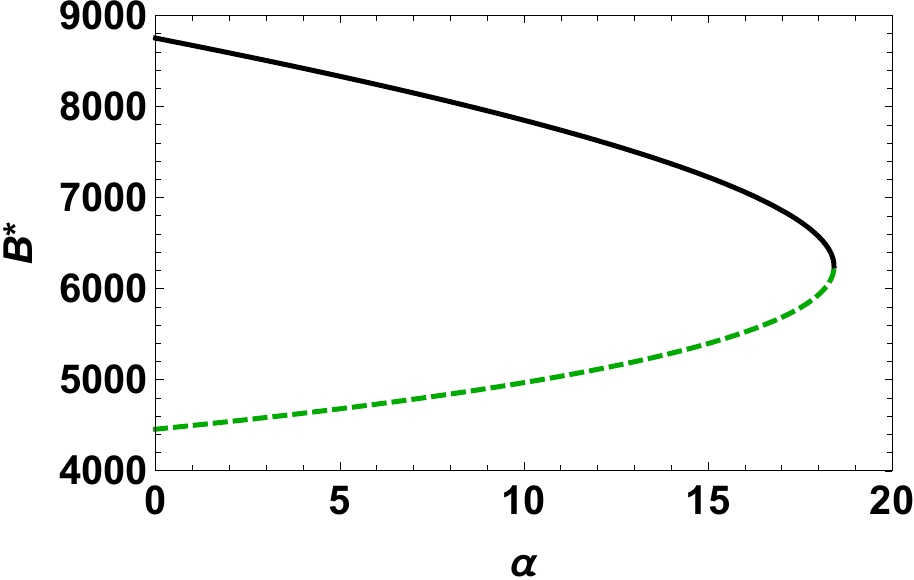}
\caption{Brood population and the brood regulation effect $\alpha$ when $K=10^6, r=1400, d_b=0.09, d_h=0.1, \tau=21$.}
\label{fig:m1abB}
    \end{subfigure}
    \begin{subfigure}{0.24\textwidth}
        \includegraphics[width=\textwidth]{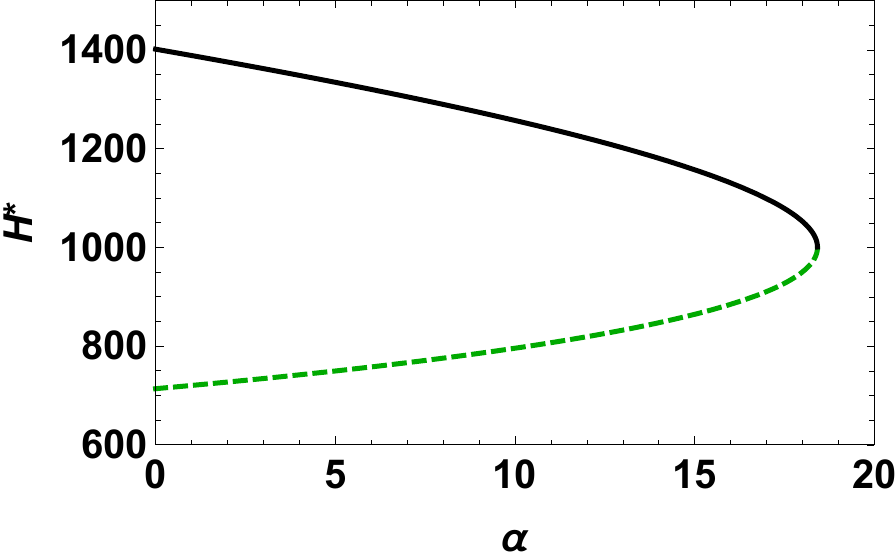}
\caption{Adult population and the brood regulation effect $\alpha$ when $K=10^6, r=1400, d_b=0.09, d_h=0.1, \tau=21$.}
\label{fig:m1abH}
    \end{subfigure}
    
    \begin{subfigure}{0.24\textwidth}
        \includegraphics[width=\textwidth]{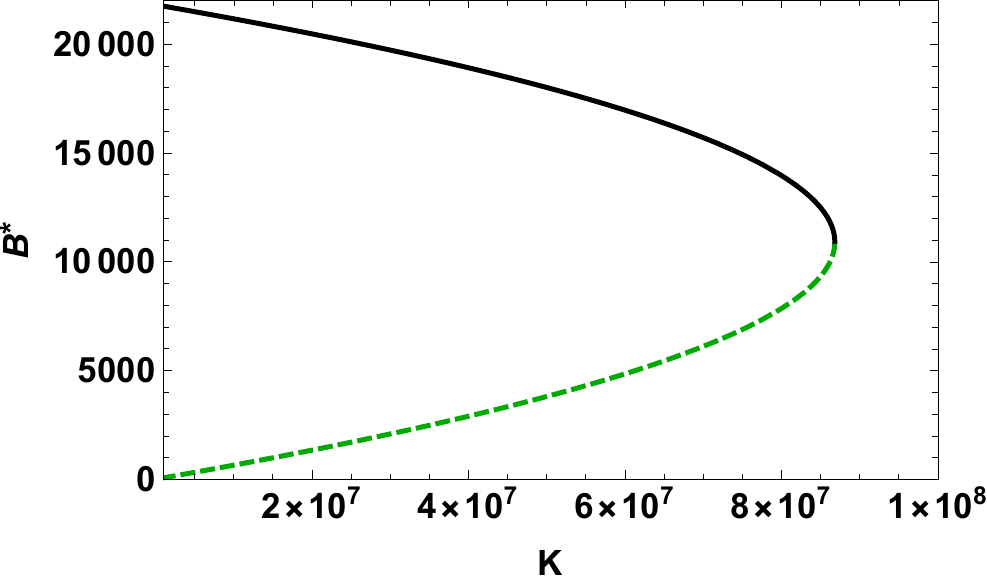}
\caption{Brood population and the square of half max of colony size $K$ when $r=1400, d_b=0.03, d_h=0.04, \alpha =3, \tau=21$ .}
\label{fig:m1kbB}
    \end{subfigure}  
      \begin{subfigure}{0.24\textwidth}
        \includegraphics[width=\textwidth]{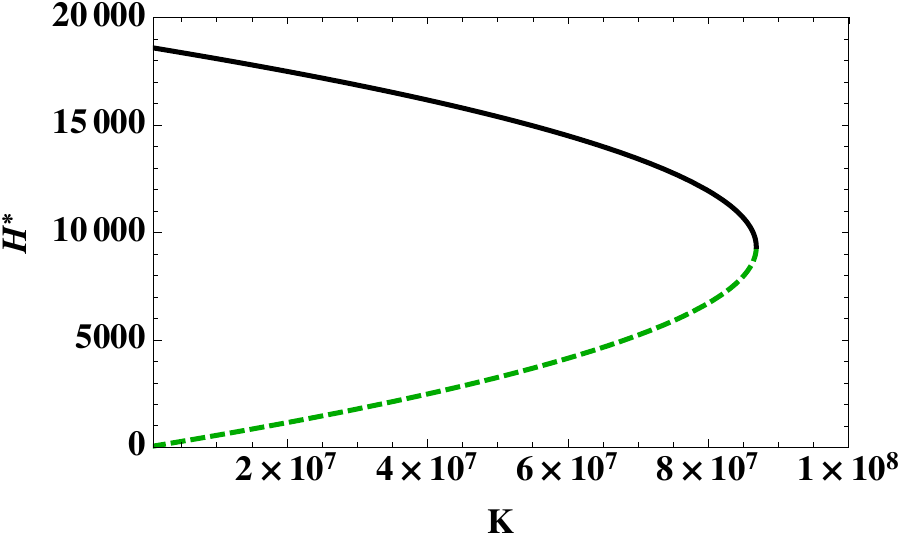}
\caption{Adult population and the square of half max of colony size $K$ when $r=1400, d_b=0.03, d_h=0.04, \alpha =3, \tau=21$ .}
\label{fig:m1kbH}
    \end{subfigure}  
         \begin{subfigure}{0.24\textwidth}
        \includegraphics[width=\textwidth]{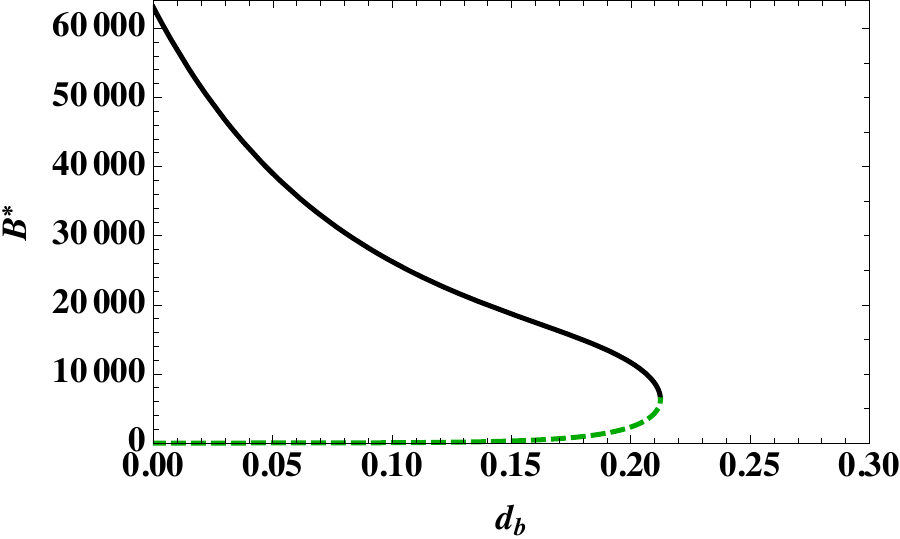}
\caption{Brood population and the death rate of the brood $d_b$ when $K=10^5, r=3000, \alpha =3, d_h=0.05, \tau=21$}
\label{fig:m1dbB}
    \end{subfigure}
    \begin{subfigure}{0.24\textwidth}
        \includegraphics[width=\textwidth]{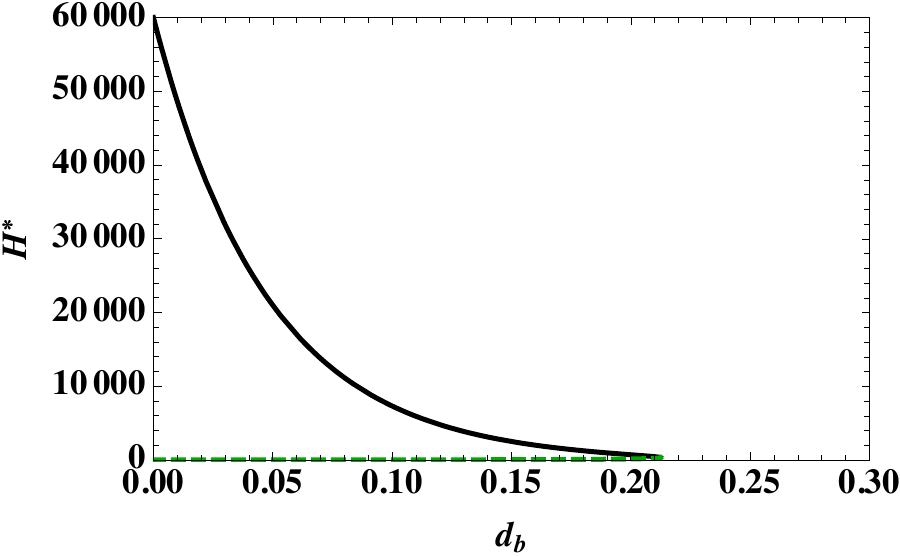}
\caption{Adult population and the death rate of the brood $d_b$ when $K=10^5, r=3000, \alpha =3, d_h=0.05, \tau=21$}
\label{fig:m1dbH}
    \end{subfigure}
    
     \begin{subfigure}{0.24\textwidth}
        \includegraphics[width=\textwidth]{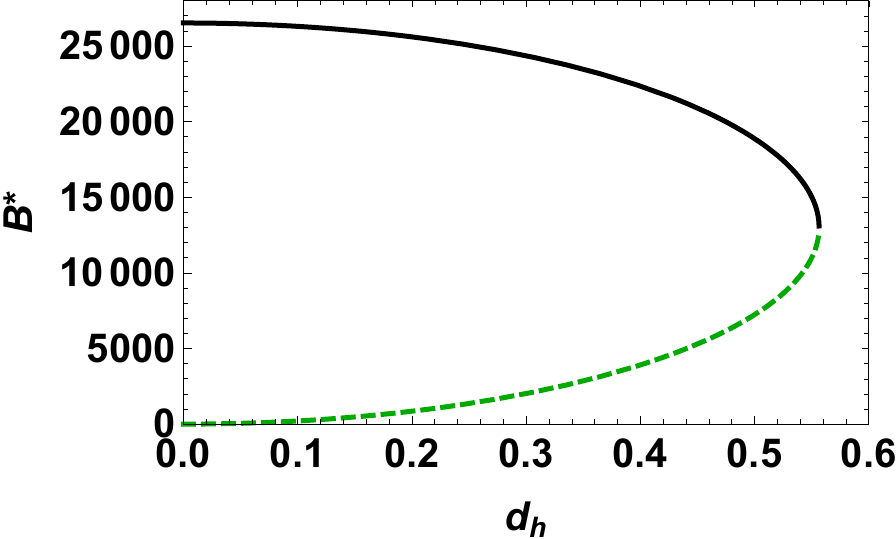}
\caption{Brood population and the death rate of the adult $d_h$ when $K=1*10^4, r=1400, d_b=0.01, \alpha =3, \tau=21$.}
\label{fig:m1dhB}
    \end{subfigure}     
         \begin{subfigure}{0.24\textwidth}
        \includegraphics[width=\textwidth]{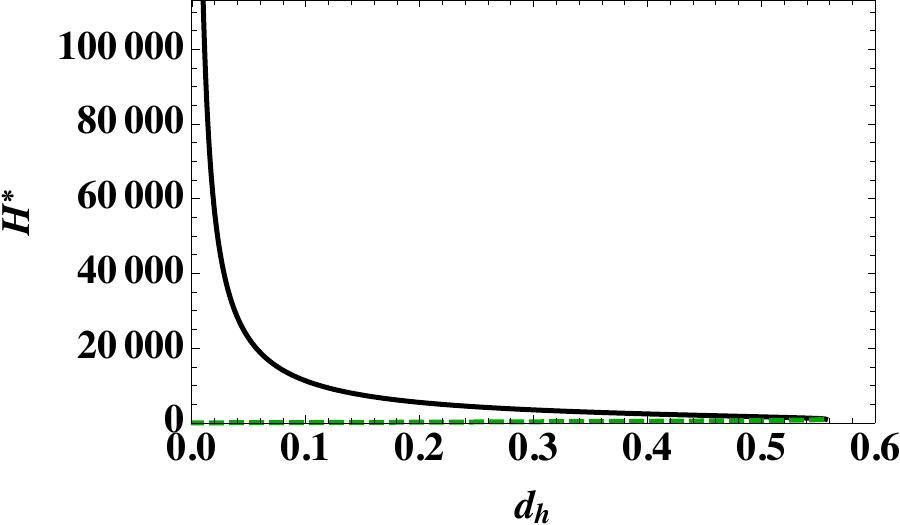}
\caption{Adult population and the death rate of the adult $d_h$ when $K=1*10^4, r=1400, d_b=0.01, \alpha =3, \tau=21$.}
\label{fig:m1dhH}
    \end{subfigure}
    
   \caption{Bifurcation diagrams of Model \eqref{BH} with the interior equilibrium $E_1=(B_1^*,H_1^*)$ in black and $E_2=(B_2^*,H_2^*)$ in green where the solid curve indicates stable, and the dash curve indicates saddle.}
    \label{fig:bif1}
 \end{figure}

Figure \ref{fig:bif1} shows bifurcation diagrams of Model (\ref{BH}) regarding (a) the queen egg-laying rate ($r$) (see Figure \ref{fig:m1rbB}\&\ref{fig:m1rbH}); (b) the brood regulation effects on reproduction $\alpha$ (see Figure \ref{fig:m1abB}\&\ref{fig:m1abH}); (c) the half-saturation coefficient $K$ (see Figure \ref{fig:m1kbB}\&\ref{fig:m1abH}) (d) the mortality of brood $d_b$ (see Figure \ref{fig:m1dbB}\&\ref{fig:m1dbH}); and (e) the mortality of adult $d_h$ (see Figure \ref{fig:m1dhB}\&\ref{fig:m1dhH}). Those bifurcation diagrams indicates that (1) the colony survival requires the large value of the queen egg-laying rate ($r$) which leads to the increased brood and adult population as it increases; (2) the large values of $\alpha$, the half-saturation coefficient $K$,  or any mortality rate $d_b$ or $d_h$ can lead to the colony collapsing, and both brood and adult population are decreasing with respect to these parameter values; and (3) increasing the value of the adult population mortality can lead to the dramatic decreasing of the adult population. \\

\textbf{The another modeling approach with age structure for the honeybee colony:} In our model derivation section, we proposed the model \eqref{BH2} below assuming that the adult $H(t)$ matures from the survived brood population at time $t-\tau$.
\begin{eqnarray*}
\begin{array}{lcl}
\frac{dB}{dt}&=&\frac{rH(t)^2}{K+H(t)^2}-d_b B- e^{- d_b \tau} B(t-\tau)   \\
\frac{dH}{dt}&=& e^{- d_b \tau} B(t-\tau)-d_h H(t)
\end{array}
\end{eqnarray*}
The model above is motivated from the compartmental ODE model in the literature \cite{tang2002density}. We aim to compare the dynamics of Model \eqref{BH2} to the model \eqref{BH} to address the importance of deriving a proper biological model with age structure. \\

 First, we notice that the extinction equilibrium $E_e=(0,0)$ always exists as for the model \eqref{BH}. However, $E_e$ can go through stability switching that leads to an oscillatory solution around $E_e$ for Model \eqref{BH2}

 \begin{theorem}\label{BaiBH2-00Stability}
[\textit{Extinction equilibria dynamics}] Model \eqref{BH2} always has the extinction equilibrium $E_e=(0,0)$.
\begin{enumerate}
  \item If $d_b\ge \frac{\sqrt{2}}{2}$, then $E_e=(0,0)$ is asymptotically stable for all $\tau\ge 0$.
  \item If $0<d_b< \frac{\sqrt{2}}{2}$, then $E_e=(0,0)$ is asymptotically stable for $\tau\in (0,\tau_0)$ or $\tau\ge \tau_1$, while unstable for $\tau\in (\tau_0, \tau_1)$, where $\tau_k=\frac{\theta+k\pi}{w}, k=0,1$, $\theta=\pi-\arctan\left(\frac{w}{d_b}\right)$ and $w=(e^{-2d_b \tau}-d_b^2)^{\frac{1}{2}}$.
\end{enumerate}
\end{theorem}

\begin{figure}[ht]
    \centering
    \includegraphics[width=0.4\textwidth]{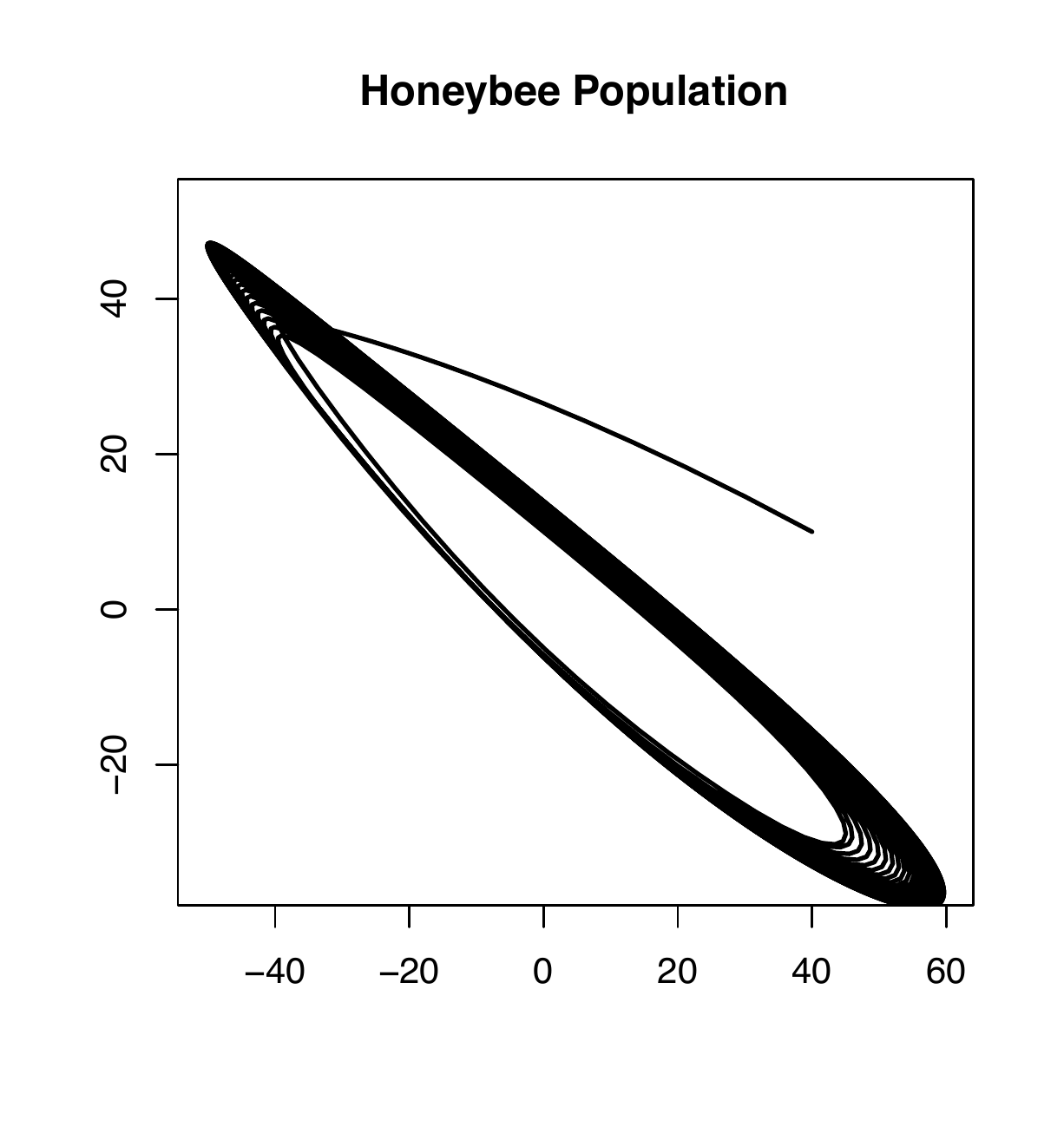}
    \caption{Phase plane of honeybee brood and adult  population of  with $r=1000, K=1*10^6, d_b=0.1, d_h=.17, \tau=18$ when Model (\ref{BH2}) has an unstable $E_e$.}
    \label{fig:E00}
\end{figure}

\noindent\textbf{Notes:} Theorem \ref{BaiBH2-00Stability} suggests that the smaller value of the brood mortality can destablize the colony dynamics. In addition, it implies that
Model \eqref{BH2} is not positive invariant as the extinction equilibrium $E_e=(0,0)$ could have stability switches that lead to oscillatory solution around $E_e$. See Figure \ref{fig:E00} when Model \eqref{BH2} exists a limit cycle of population around $E_e=(0,0)$. \\

 Let $(B,H)$ be an interior equilibrium of Model \eqref{BH2}, then it satisfies that the following two equations:

\begin{eqnarray*}
0&=&\frac{rH^2}{K+H^2}-d_b B- e^{- d_b \tau} B \label{B2}  \\
0&=& e^{- d_b \tau} B-d_h H \label{H2}
\end{eqnarray*}
which gives  the brood population at the equilibrium $B^* =d_he^{d_b\tau}H^*$ and by solving
$\frac{r H}{K+H^2}-d_bd_he^{d_b\tau}-d_h=0$ we could solve
\begin{align*}
H^*_i=\frac{r\pm\sqrt{r^2-4 K \left(d_b d_h e^{d_b \tau }+d_h\right)^2}}{2 \left(d_b d_h e^{d_b \tau }+ d_h\right)},\, i=1,2.
\end{align*}

Now define the characteristic equation of the interior equilibrium $(B^*, H^*)$ of Model \eqref{BH2} as follows:

\begin{align}
    C(\lambda,\tau)&=( -d_h -\lambda)(-d_b-e^{-d_b\tau}e^{\lambda\tau}-\lambda)-\frac{2rKH^*}{(K+(H^*)^2)^2}e^{-(d_b+\lambda)\tau}\\
    &=\lambda^2+(d_b+d_h)\lambda+(\lambda+d_h-\frac{2rKH^*}{(K+(H^*)^2)^2})e^{-(\lambda+d_b)\tau}+d_hd_b=0 \label{characteristic}
\end{align}

\begin{theorem}\label{BaiBH2-inEStability}
[\textit{Interior Equilibrium Dynamics}]  Let $r>2d_h\sqrt{K}(1+d_b)$ and $\tau^*=\frac{1}{d_b}\ln\left(\frac{1}{d_b}\left(\frac{r}{2d_h\sqrt{K}}-1\right)\right)$.
If $\tau\in[0,\tau^*)$,
Model (\ref{BH2}) has two positive interior equilibrium 
$$E_i=(B_i^*, H_i^*)=(d_he^{d_b\tau}H_i^*,H_i^*), i=1,2$$  
which $H_1^*<H_2^*$.
And $E_1=(B_1^*, H_1^*)$ is always unstable in $[0,\tau^*)$, $E_2=(B_2^*, H_2^*)$ is always stable or occurs stability switching by following cases:
 
 \begin{mycases}
    \item If $d_b\ge 1$ or $0<d_b<1$, $d_b^2+d_h^2\ge 1$ and $2d_h\sqrt{K}(1+d_b)<r\le\frac{2d_h\sqrt{K}(1+d_b)^2}{\sqrt{(1+d_b)^2-4d_b^2}}$, $E_2$ is locally asymptotically stable for all $\tau\in[0,\tau^*)$.
    \item If $0<d_b<1$, $d_b^2+d_h^2\ge 1$ and $r>\frac{2d_h\sqrt{K}(1+d_b)^2}{\sqrt{(1+d_b)^2-4d_b^2}}$, then the stability of $E_2$ switches just once from stable to unstable as $\tau$ increases in $[0,\tau^*)$.
    \item If $d_b^2+d_h^2< 1$ and $2d_h\sqrt{K}(1+d_b)<r\le\frac{2d_h\sqrt{K}(1+d_b)^2}{\sqrt{(1+d_b)^2-4d_b^2}}$, the stability of $E_2$ can change a finite number of times at most as $\tau$ is increased $\tau\in[0, \tau^*)$, and eventually it becomes unstable.
    \item If $d_b^2+d_h^2< 1$ and $r>\frac{2d_h\sqrt{K}(1+d_b)^2}{\sqrt{(1+d_b)^2-4d_b^2}}$, the stability of $E_2$ switches at least once in $[0,\tau^*)$ from stable to unstable.
\end{mycases}

\end{theorem}

\begin{figure}[ht]
 \centering
    \begin{subfigure}{0.35\textwidth}
        \includegraphics[width=\textwidth]{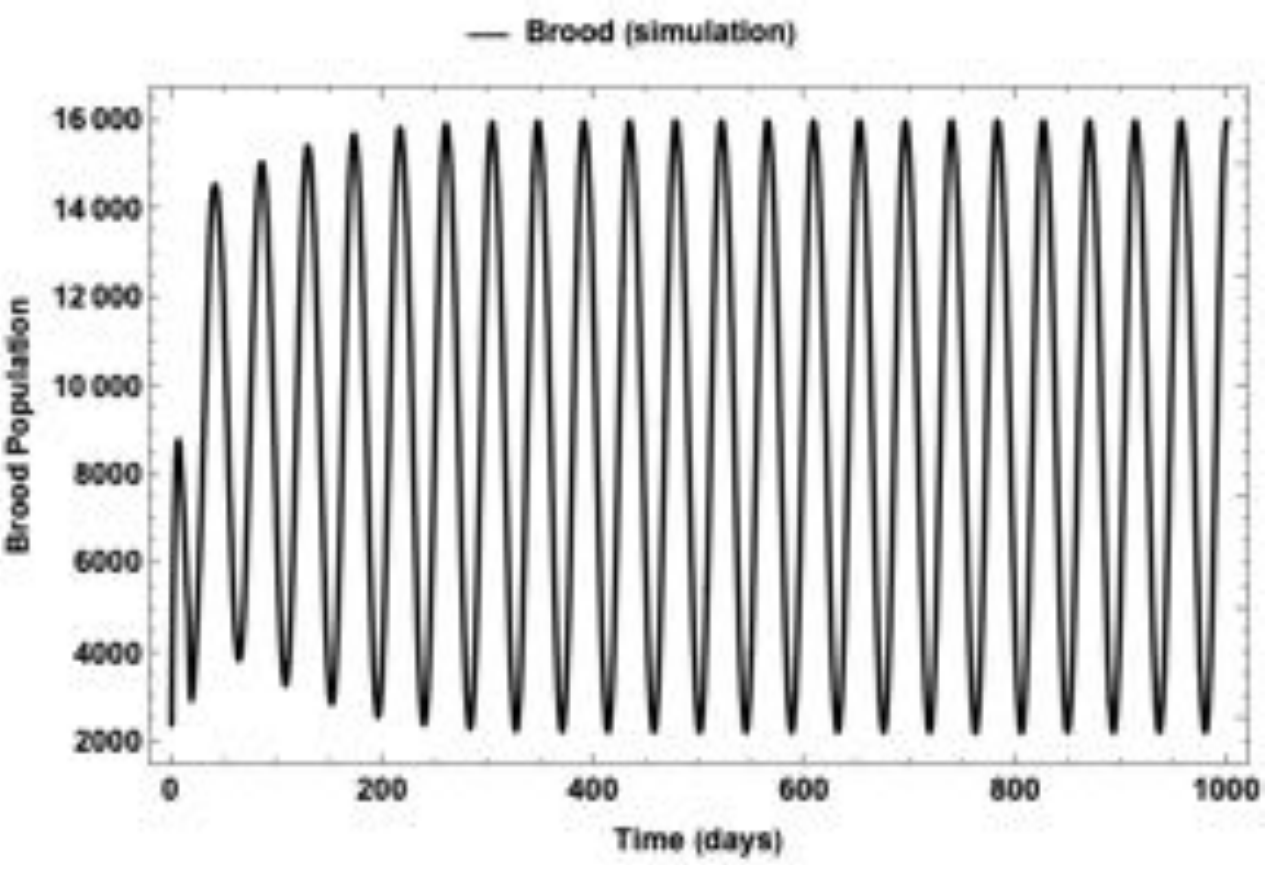}
\caption{Brood population in $\tau=16$}
    \end{subfigure}
    \begin{subfigure}{0.35\textwidth}
        \includegraphics[width=\textwidth]{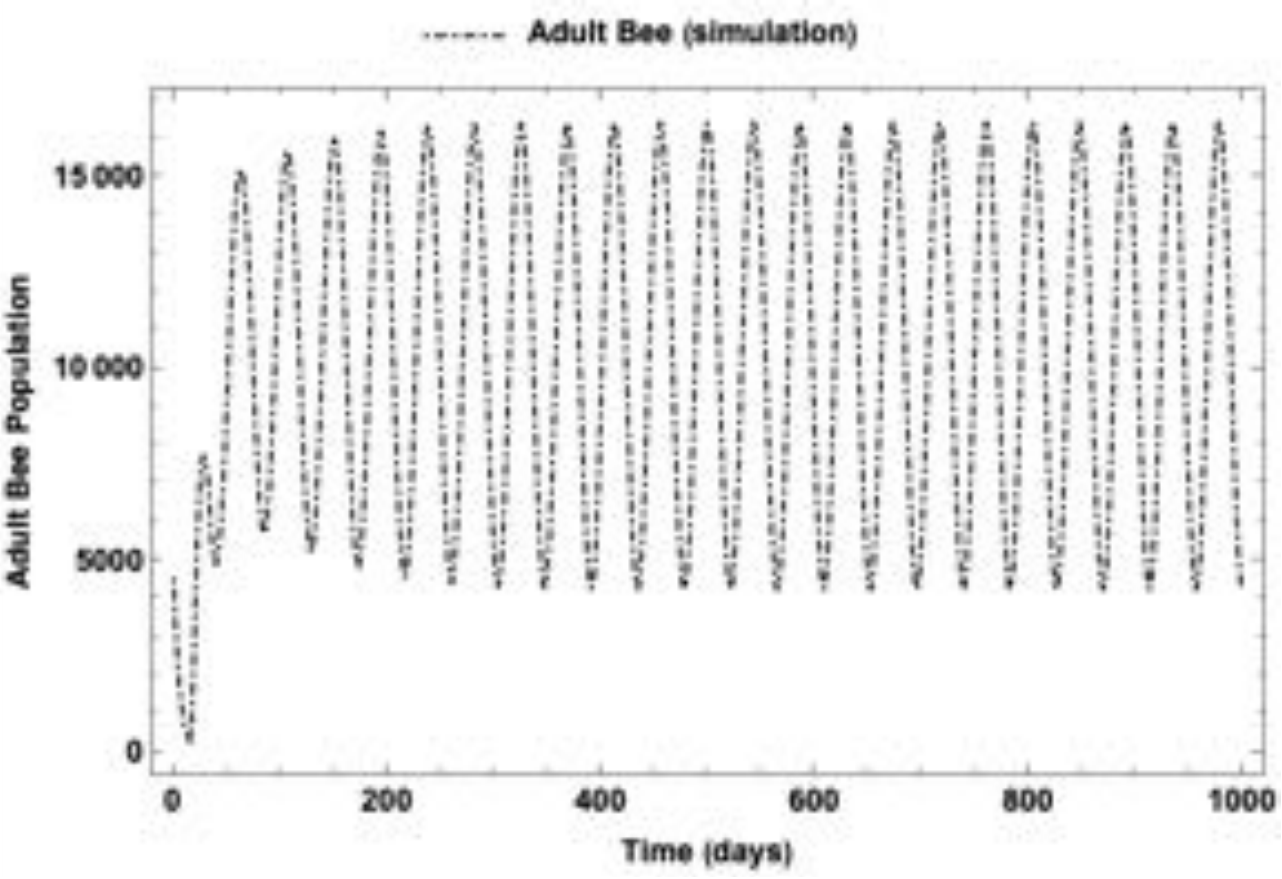}
\caption{Adult population in $\tau=16$}
    \end{subfigure}
     \vfill
    \begin{subfigure}{0.35\textwidth}
        \includegraphics[width=\textwidth]{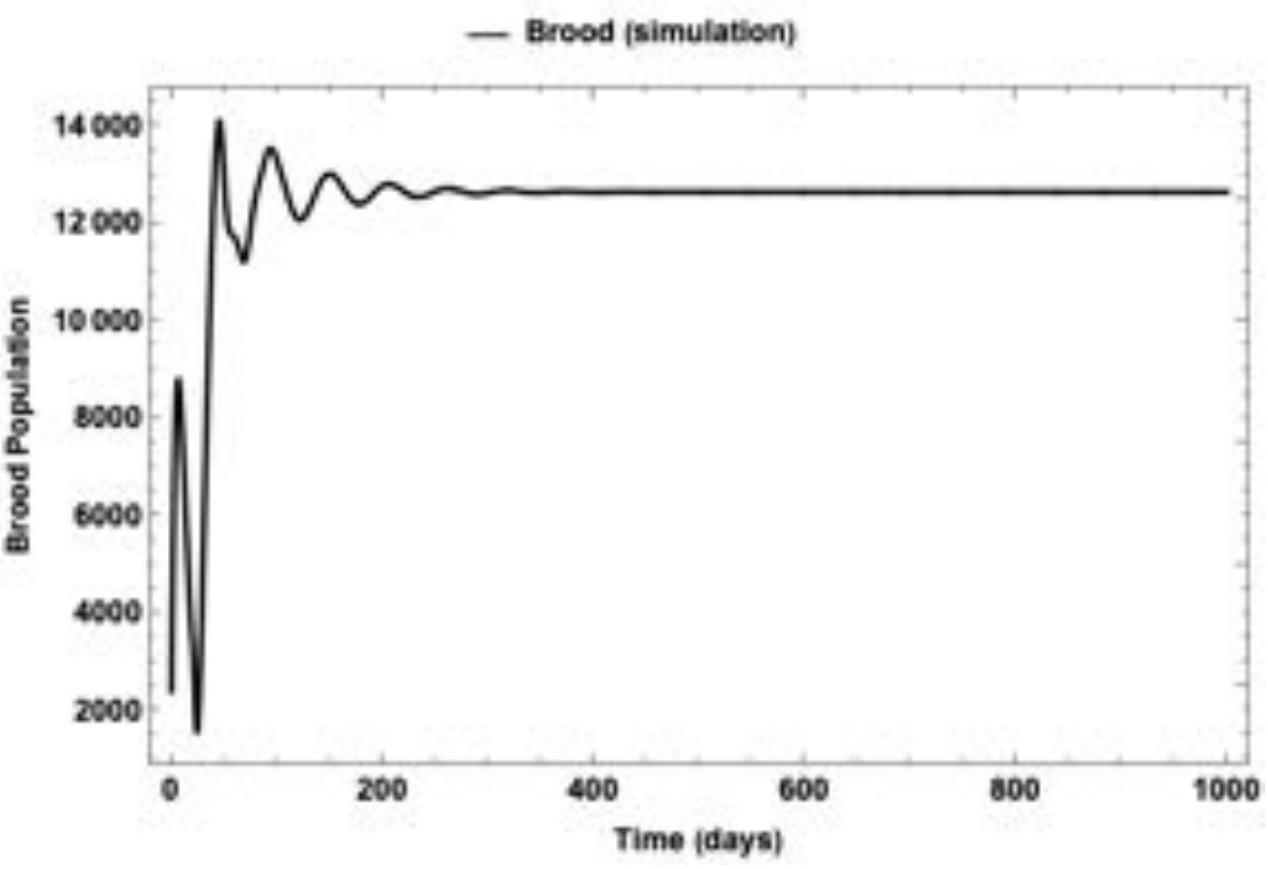}
\caption{Brood population in $\tau=21$}
    \end{subfigure}
    \begin{subfigure}{0.35\textwidth}
        \includegraphics[width=\textwidth]{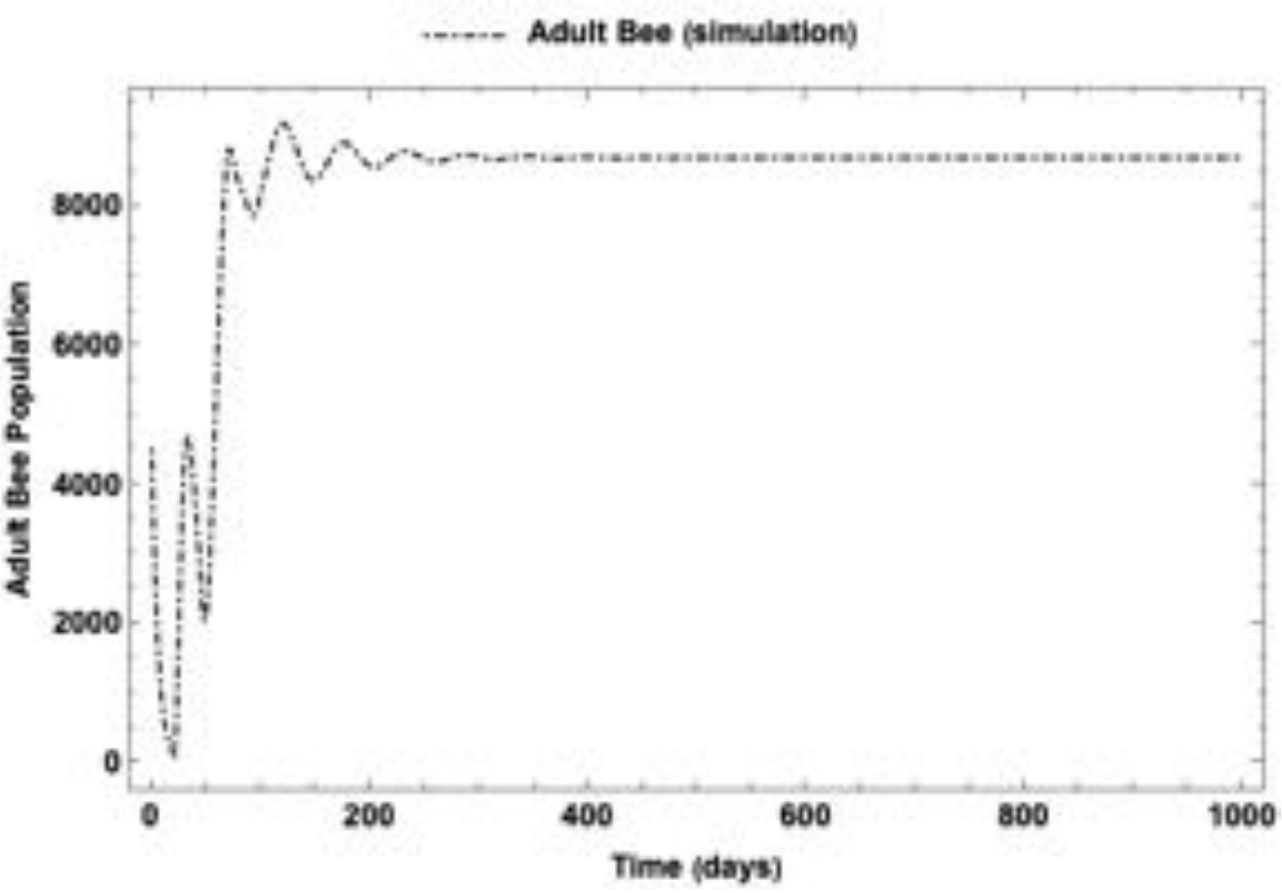}
\caption{Adult population in $\tau=21$}
    \end{subfigure}
    \caption{Time series of the brood (solid) and adult (dot dashed) bee when $r = 3000; d_h = 0.178; d_b = 0.1; K = 5,000,000$; {$B(\theta)=2400;H(\theta)=4500, \theta\in[-\tau,0]$.}}
    \label{fig:tau}
 \end{figure}

\begin{figure}[ht]
 \centering
    \begin{subfigure}{0.35\textwidth}
        \includegraphics[width=\textwidth]{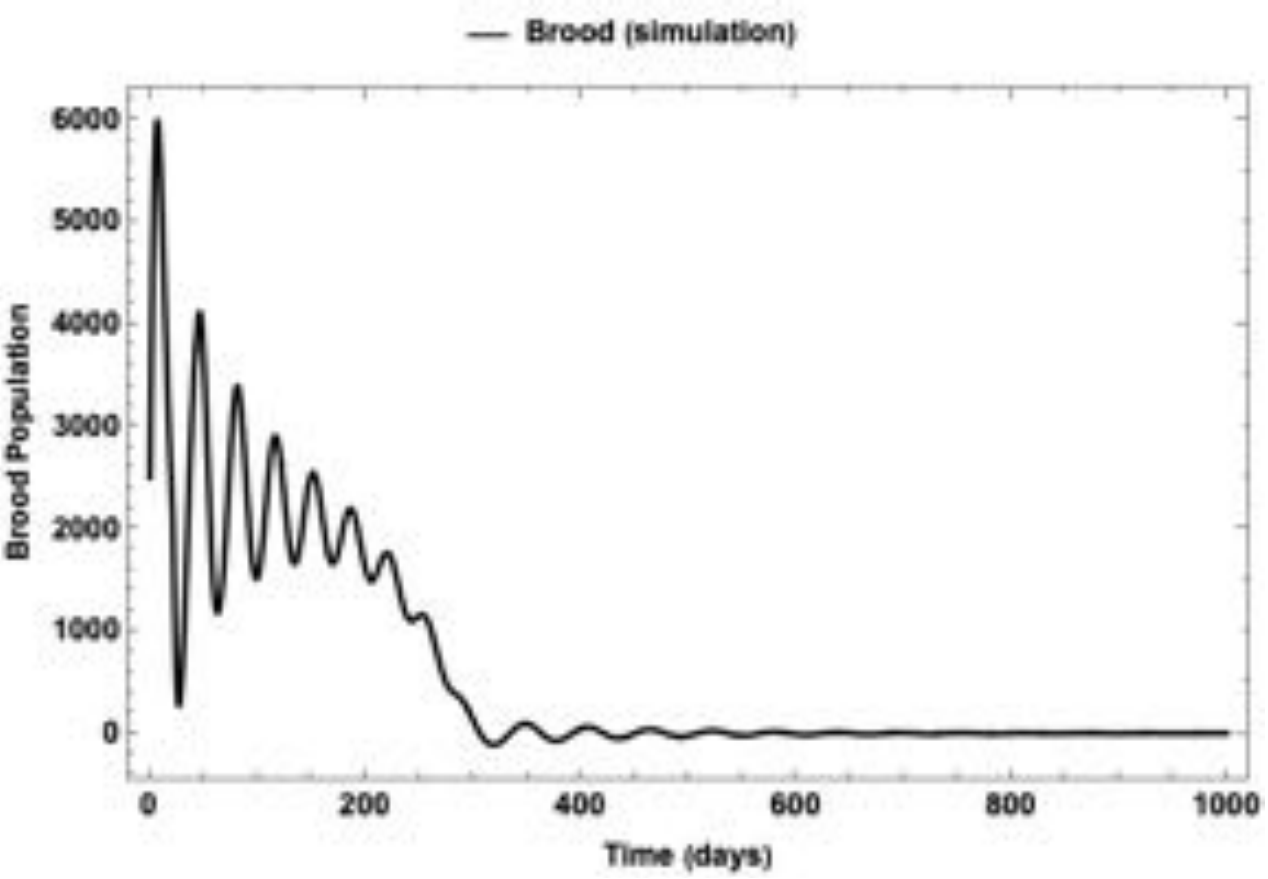}
\caption{Brood population for $B(0)=2500;H(0)=6000$}
    \end{subfigure}
    \begin{subfigure}{0.35\textwidth}
        \includegraphics[width=\textwidth]{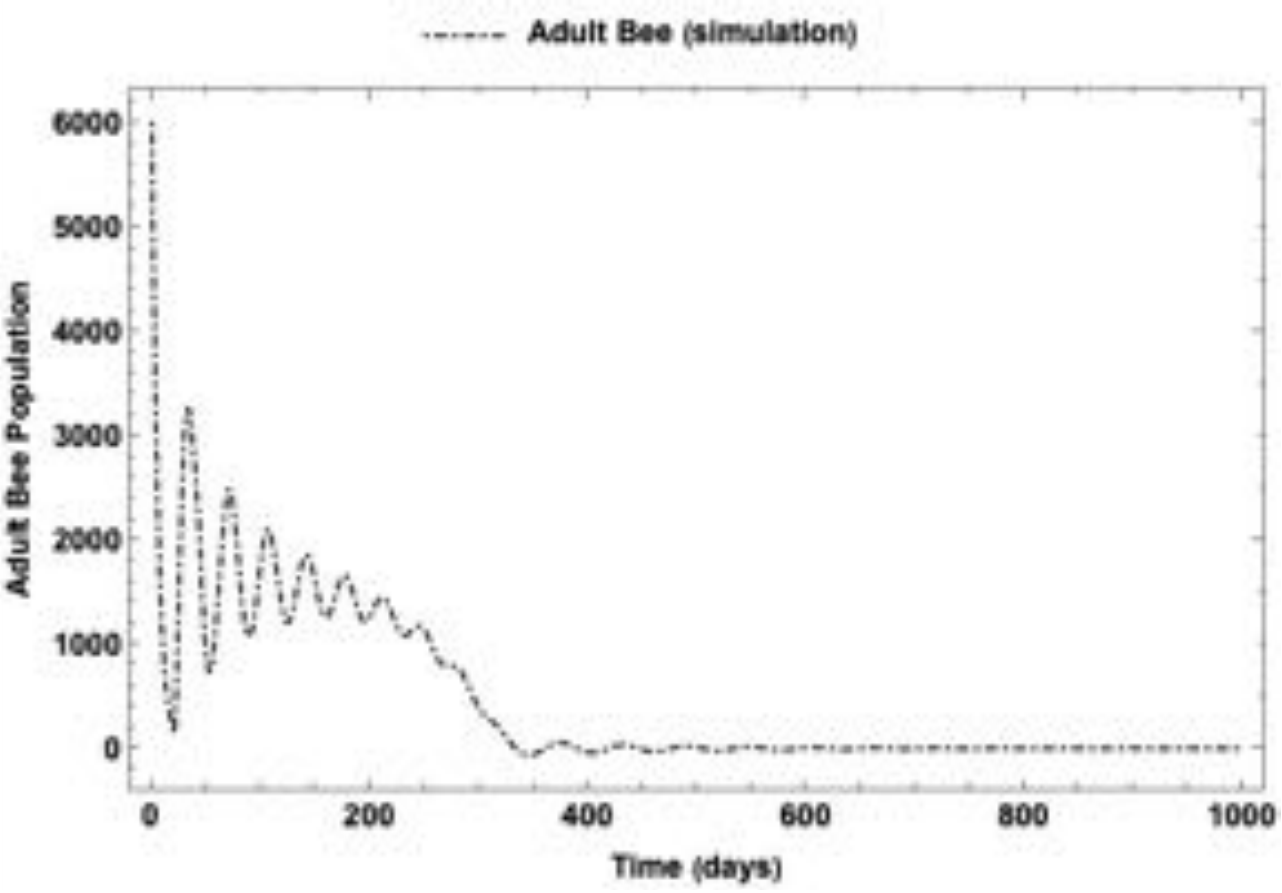}
\caption{Adult population for $B(0)=2500;H(0)=6000$}
    \end{subfigure}
     \vfill
    \begin{subfigure}{0.35\textwidth}
        \includegraphics[width=\textwidth]{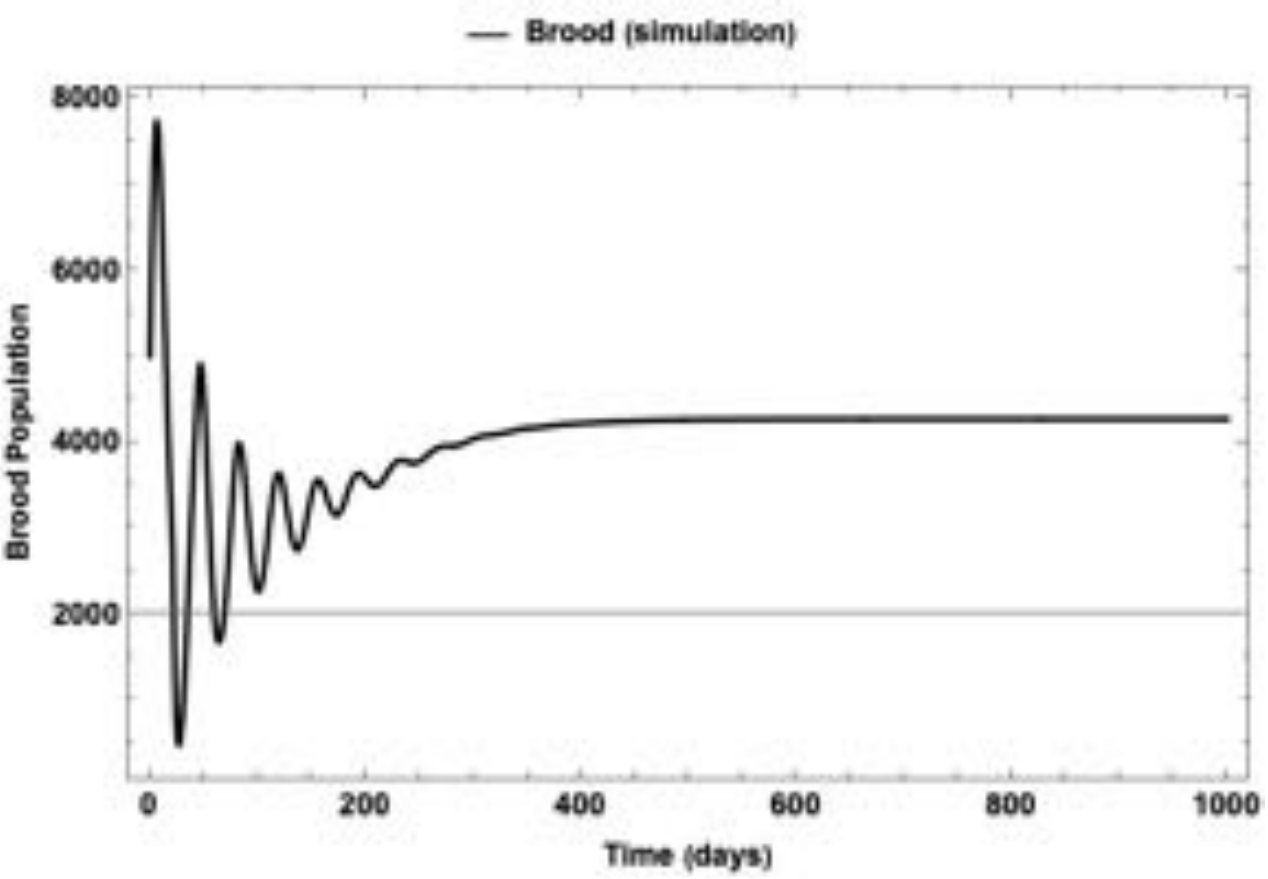}
\caption{Brood population for $B(0)=5000;H(0)=7,000$}
    \end{subfigure}
    \begin{subfigure}{0.35\textwidth}
        \includegraphics[width=\textwidth]{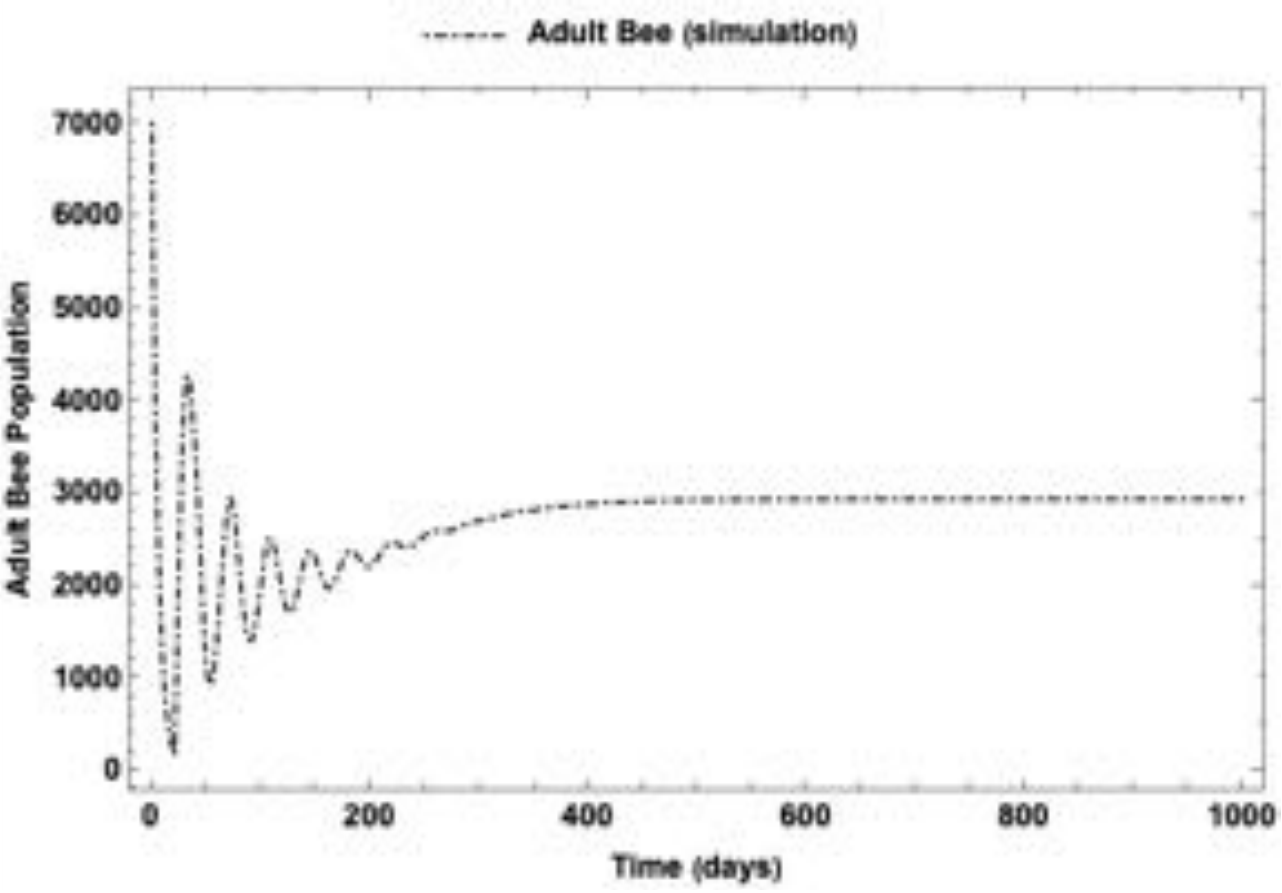}
\caption{Adult population for $B(0)=5000;H(0)=7,000$}
    \end{subfigure}
    \caption{Time series of the brood (solid) and adult (dot dashed) bee using $r = 1500; d_h = 0.178; d_b = 0.1; K = 5,000,000;\tau=21.$; \red{$B(\theta)=B(0)$ and $H(\theta)=H(0)$, $\theta\in[-\tau,0]$.} }
    \label{fig:initial}
 \end{figure}

\noindent\textbf{Notes:} Theorem \ref{BaiBH2-inEStability} indicates that: (1) The large value of mortality in brood and/or adult bees with the intermediate value of the egg laying rate $r$ can have the simple equilibrium dynamics; and (2) The relative small values of mortality in brood and/or adult bees with the large value of the egg laying rate $r$ can destabilize the colony dynamics that lead to stability switching in the interior equilibrium  $E_2$. For example, Figure \ref{fig:tau} provides an example when Model \eqref{BH2} can have stability switches at its interior attractor $E_2$ as $\tau$ changes: Model \eqref{BH2} has local stability when $\tau=21$ while its has oscillatory solution when $\tau=16$.  Figure \ref{fig:initial} also indicates the importances of initial condition that may lead to the survival or collapsing of the colony.\\

\noindent\textbf{Comparisons between Model \eqref{BH} and Model \eqref{BH2}:}  Both models can have up to three equilibria with always the existence of the extinction equilibrium $E_e$. However, the maturation time $\tau$ has no effects on the stability of the equilibrium of Model \eqref{BH} while it could lead to stability switches for  Model \eqref{BH2}. The consequence is that Model \eqref{BH2} is not positive invariant and could have a oscillatory solution around the extinction equilibrium $E_e$ and the interior equilibrium $E_2$. \\

To continue exploring how we should model population dynamics of honeybee with the proper age structure so that we could have a better understanding of important factors  contributing the colony survival, we perform bifurcation diagrams on both Model \eqref{BH} and Model \eqref{BH2}. Figure \ref{fig:bif1} shows bifurcation diagrams of Model (\ref{BH}) regarding (a) the queen egg-laying rate ($r$) (see Figure \ref{fig:m1rbB} \& \ref{fig:m1rbH}); (b) the brood regulation effects on reproduction $\alpha$ (see Figure \ref{fig:m1abB} \& \ref{fig:m1abH}); (c) the half-saturation coefficient $K$ (see Figure \ref{fig:m1kbB} \& \ref{fig:m1kbH}) (d) the mortality of brood $d_b$ (see Figure \ref{fig:m1dbB} \& \ref{fig:m1dbH}); and (e) the mortality of adult $d_h$ (see Figure \ref{fig:m1dhB} \& \ref{fig:m1dhH}).\\

\begin{figure}[!ht]
 \centering
    \begin{subfigure}{0.24\textwidth}
        \includegraphics[width=\textwidth]{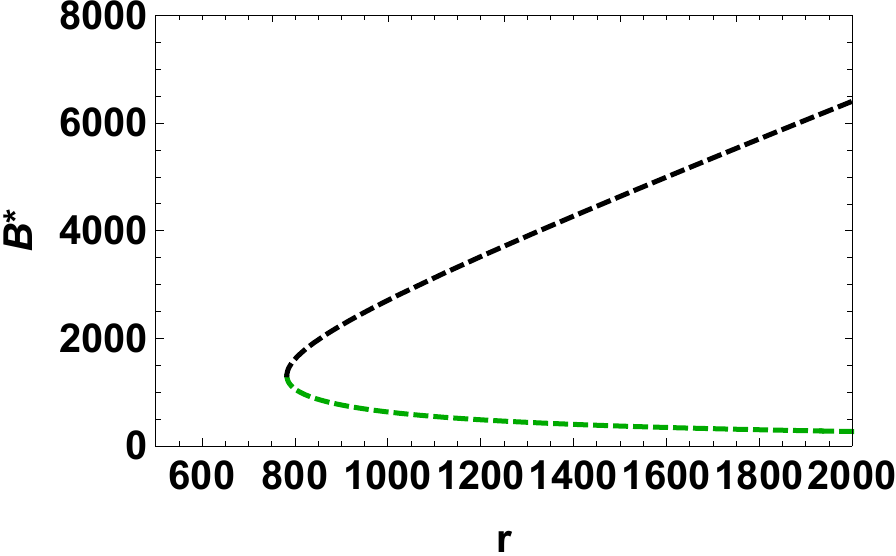}
\caption{Brood population and the egg-laying rate $r$ when $K=9*10^6, d_b=0.07, d_h=0.1, \tau=21$.}
\label{fig:m2rbu}
    \end{subfigure}
    \begin{subfigure}{0.24\textwidth}
        \includegraphics[width=\textwidth]{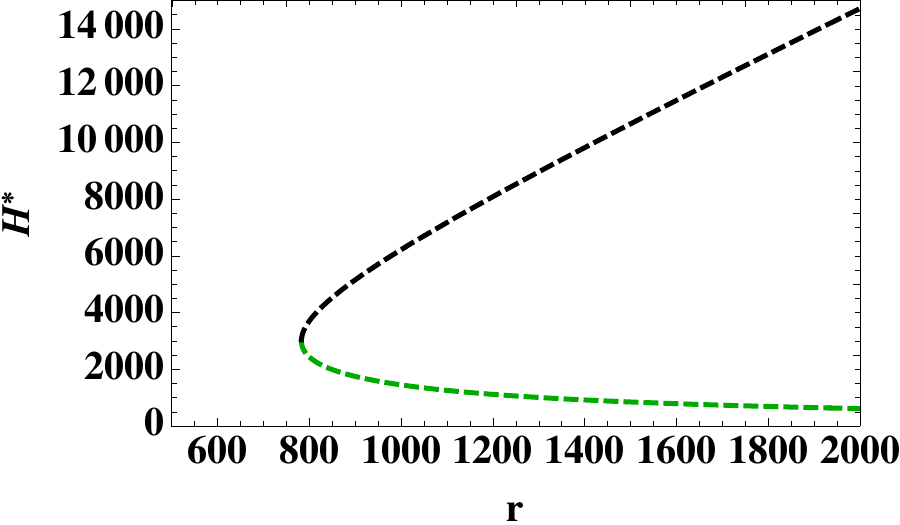}
\caption{Adult population and the egg-laying rate $r$ when $K=9*10^6, d_b=0.07, d_h=0.1, \tau=21$.}
\label{fig:m2rhu}
    \end{subfigure}
    \begin{subfigure}{0.24\textwidth}
        \includegraphics[width=\textwidth]{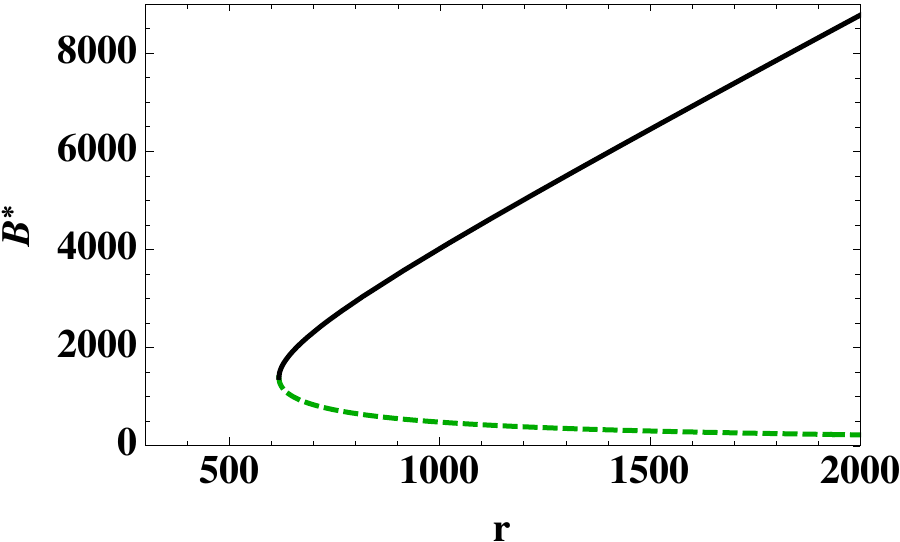}
\caption{Brood population and the egg-laying rate $r$ when $K=10^6, d_b=0.1, d_h=0.17, \tau=21$.}
\label{fig:m2rbs}
    \end{subfigure}
    \begin{subfigure}{0.24\textwidth}
        \includegraphics[width=\textwidth]{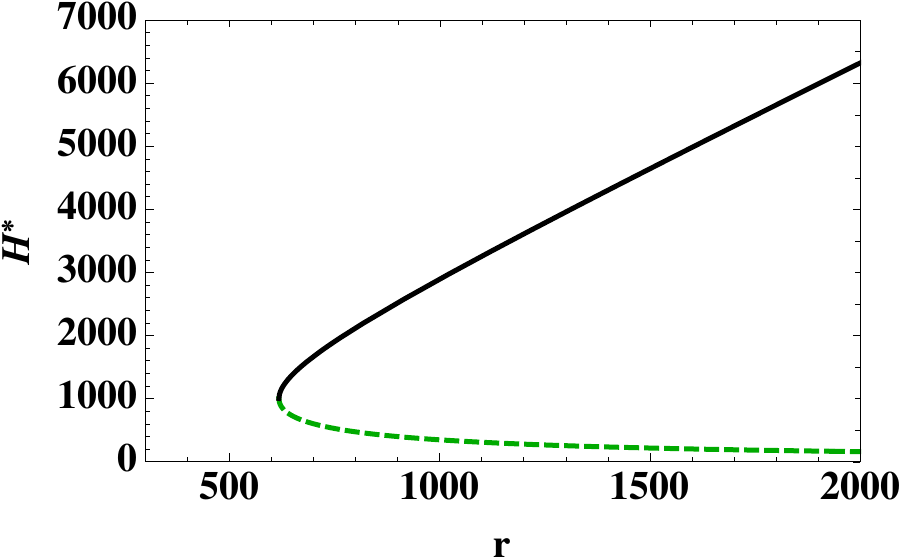}
\caption{Adult population and the egg-laying rate $r$ when $K=10^6, d_b=0.1, d_h=0.17, \tau=21$.}
\label{fig:m2rhs}
    \end{subfigure}
    
    \begin{subfigure}{0.24\textwidth}
        \includegraphics[width=\textwidth]{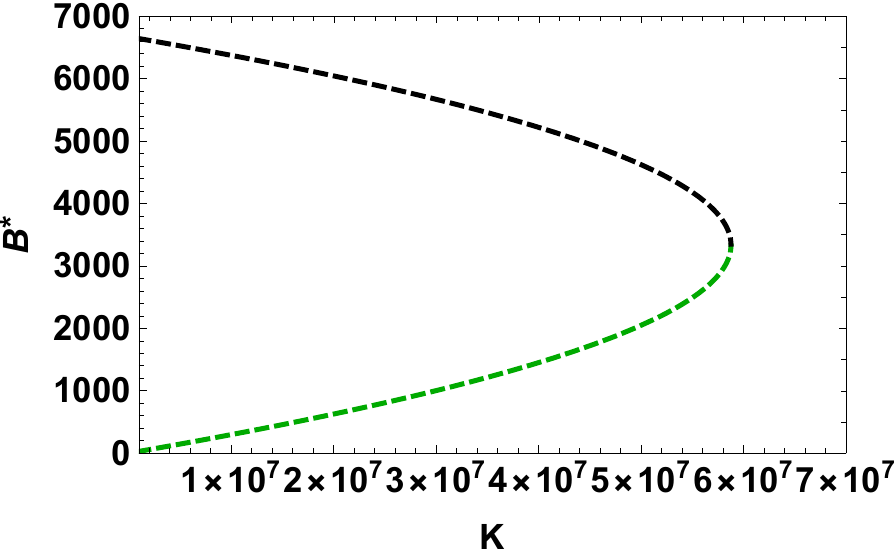} 
\caption{Brood population and the square of half max of colony size $K$ when $r=2000, d_b=0.07, d_h=0.1,\tau=21$.}
\label{fig:m2kbu}
    \end{subfigure}   
    \begin{subfigure}{0.24\textwidth}
        \includegraphics[width=\textwidth]{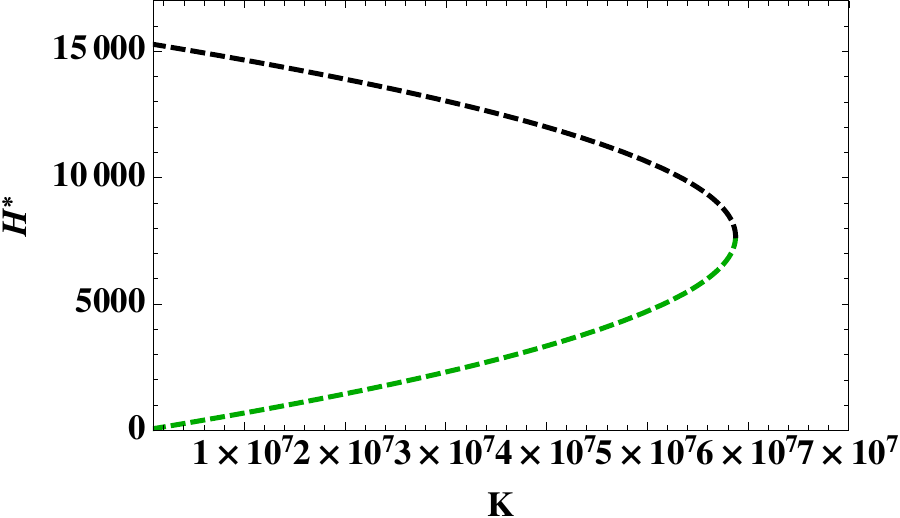} 
\caption{Adult population and the square of half max of colony size $K$ when $r=2000, d_b=0.07, d_h=0.1,\tau=21$.}
\label{fig:m2khu}
    \end{subfigure}   
         \begin{subfigure}{0.24\textwidth}
        \includegraphics[width=\textwidth]{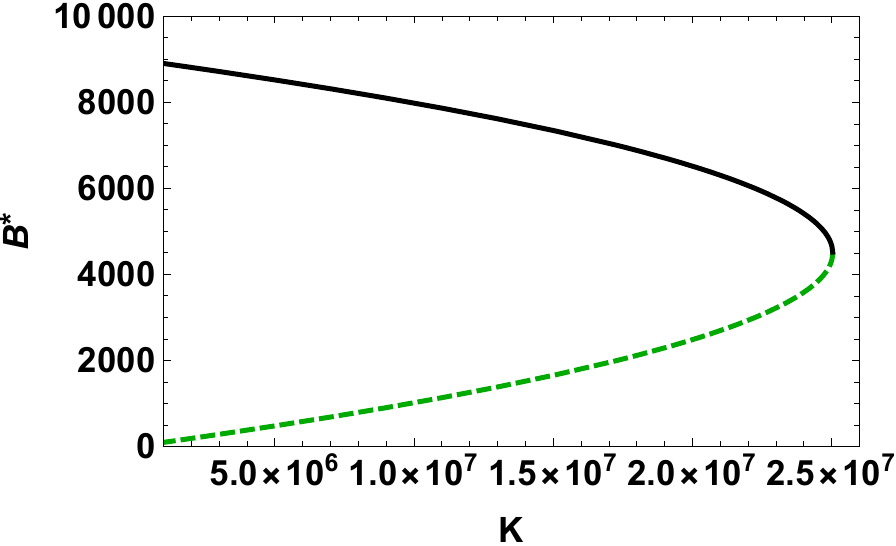}
\caption{Brood population and the square of half max of colony size $K$ when $ r=2000, d_b=0.1, d_h=0.11, \tau=21$.}
\label{fig:m2kbs}
    \end{subfigure}
     \begin{subfigure}{0.24\textwidth}
        \includegraphics[width=\textwidth]{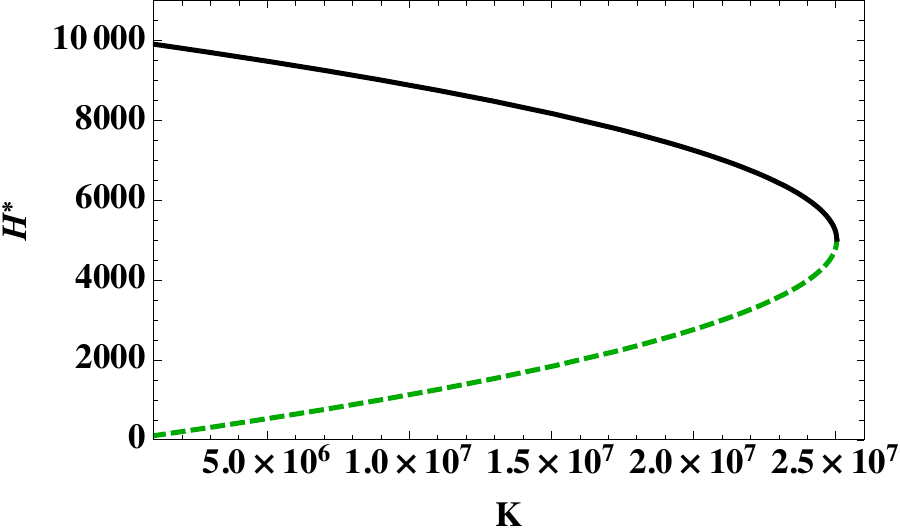}
\caption{Adult population and the square of half max of colony size $K$ when $ r=2000, d_b=0.1, d_h=0.11, \tau=21$.}
\label{fig:m2khs}
    \end{subfigure}
    
    \begin{subfigure}{0.24\textwidth}
        \includegraphics[width=\textwidth]{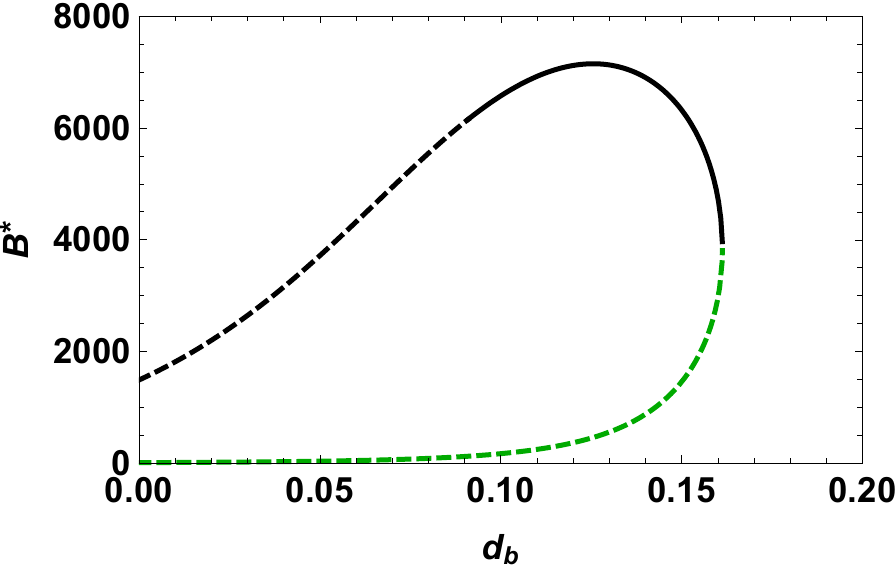}
\caption{Brood population and the death rate of the brood $d_b$ when $K=10^6, r=1500, d_h=0.13, \tau=21$.}
\label{fig:m2dbB}
    \end{subfigure}
    \begin{subfigure}{0.24\textwidth}
        \includegraphics[width=\textwidth]{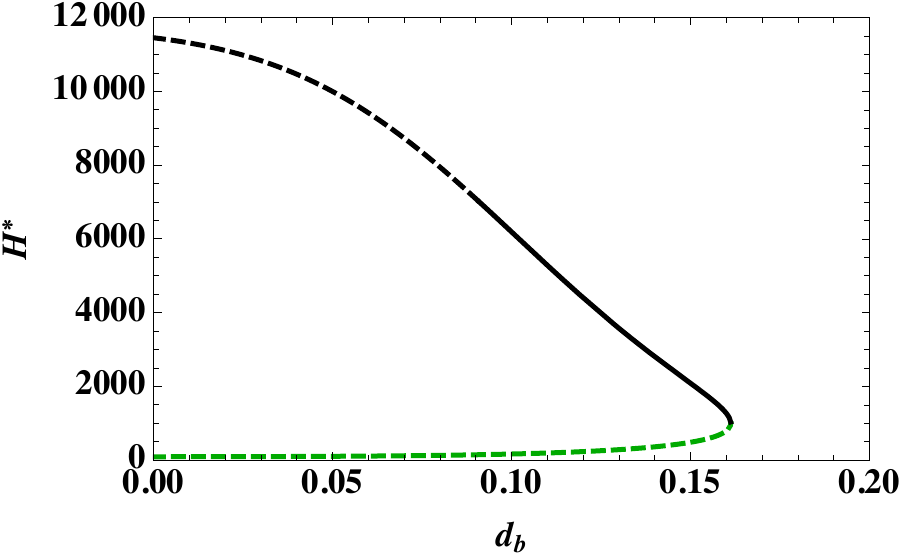}
\caption{Adult population and the death rate of the brood $d_b$ when $K=10^6, r=1500, d_h=0.13, \tau=21$.}
\label{fig:m2dbH}
    \end{subfigure}
    \begin{subfigure}{0.24\textwidth}
        \includegraphics[width=\textwidth]{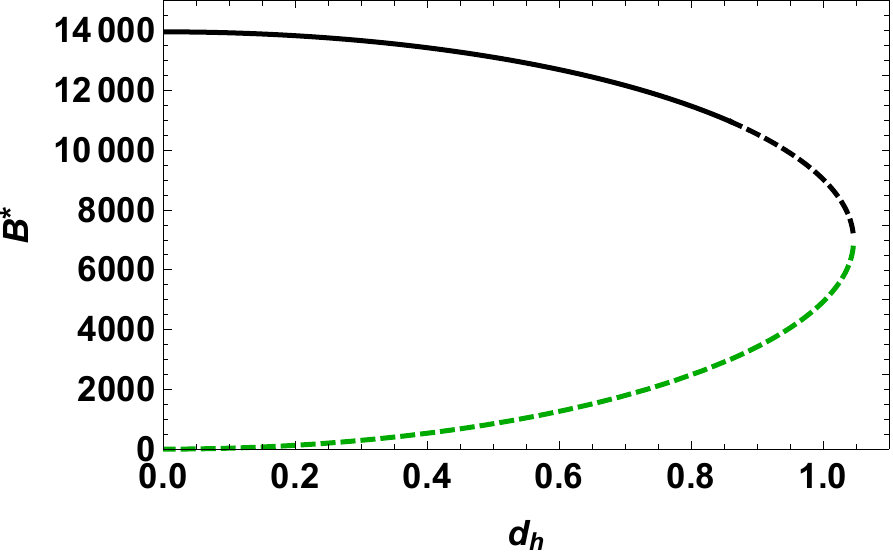}
\caption{Brood population and the death rate of the adult $d_h$ when $K=10^4, r=3000, d_b=0.2, \tau=21$.}
\label{fig:m2dhB}
    \end{subfigure}
    \begin{subfigure}{0.24\textwidth}
        \includegraphics[width=\textwidth]{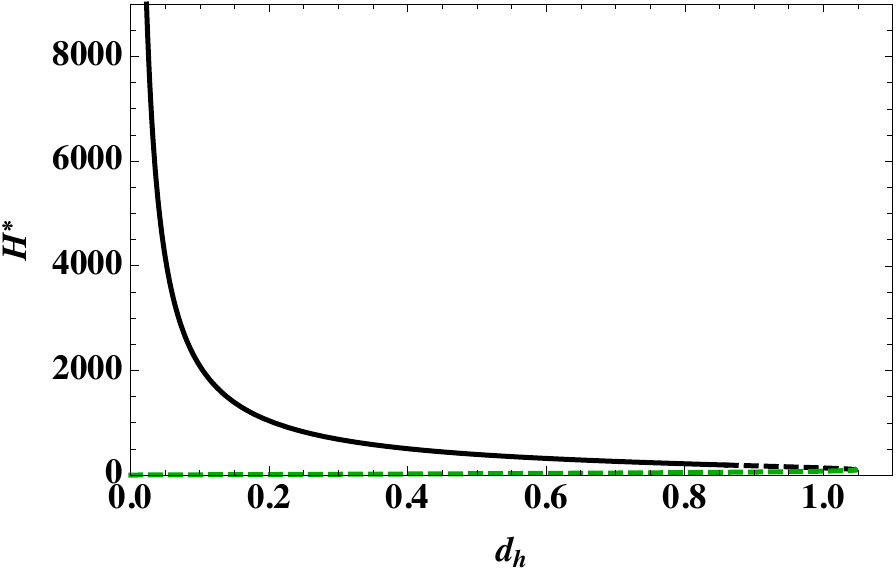}
\caption{Adult population and the death rate of the adult $d_h$ when $K=10^4, r=3000, d_b=0.2, \tau=21$.}
\label{fig:m2dhH}
    \end{subfigure}
    
    \caption{Bifurcation diagrams of interior equilibrium $E_1=(B_1^*,H_1^*)$ (black) and $E_2=(B_2^*,H_2^*)$ (green) for Model \eqref{BH2}: solid curve indicates stable, and dash curve indicates unstable}
    \label{fig:bif2}
 \end{figure}

 Figure \ref{fig:bif2} shows bifurcation diagrams of Model (\ref{BH2}) regarding (a) the queen egg-laying rate ($r$) (see Figure \ref{fig:m2rbu} \& \ref{fig:m2rhs});  (b) the half-saturation coefficient $K$ (see Figure \ref{fig:m2kbu} \& \ref{fig:m2khs}) (c) the mortality of brood $d_b$ (see Figure \ref{fig:m2dbB} \& \ref{fig:m2dbH}); and (d) the mortality of adult $d_h$ (see Figure \ref{fig:m2dhB} \& \ref{fig:m2dhH}). The biggest differences of those bifurcation diagrams between Model (\ref{BH}) and Model (\ref{BH2}) are: (1) The survival equilibrium ($E_2$) can become destabilized if we decrease the value of the brood mortality $d_b$ and/or increase the adult mortality;  (2) the brood population may have its maximum point when the mortality of the brood $d_b$ is in a proper range: In Figure \ref{fig:m2dbB}, it shows that the interesting pattern on how the brood population changes with its mortality rate; and (3) Model \eqref{BH} has only equilibrium dynamics either at the extinction $E_e$ or the interior equilibrium $E_2$. { Honeybee population data shown in Figure \ref{data} seems to exhibit seasonality.  By comparing dynamics of Model \eqref{BH} with Model \eqref{BH2}, we know that only Model \eqref{BH2}  has  oscillatory solutions. Does it mean that Model \eqref{BH2} is better than Model \eqref{BH} as it has oscillatory solutions?} \\

\section{Data and Seasonality}

\begin{figure}[ht]
 \centering
    \begin{subfigure}{0.4\textwidth}
        \includegraphics[width=\textwidth]{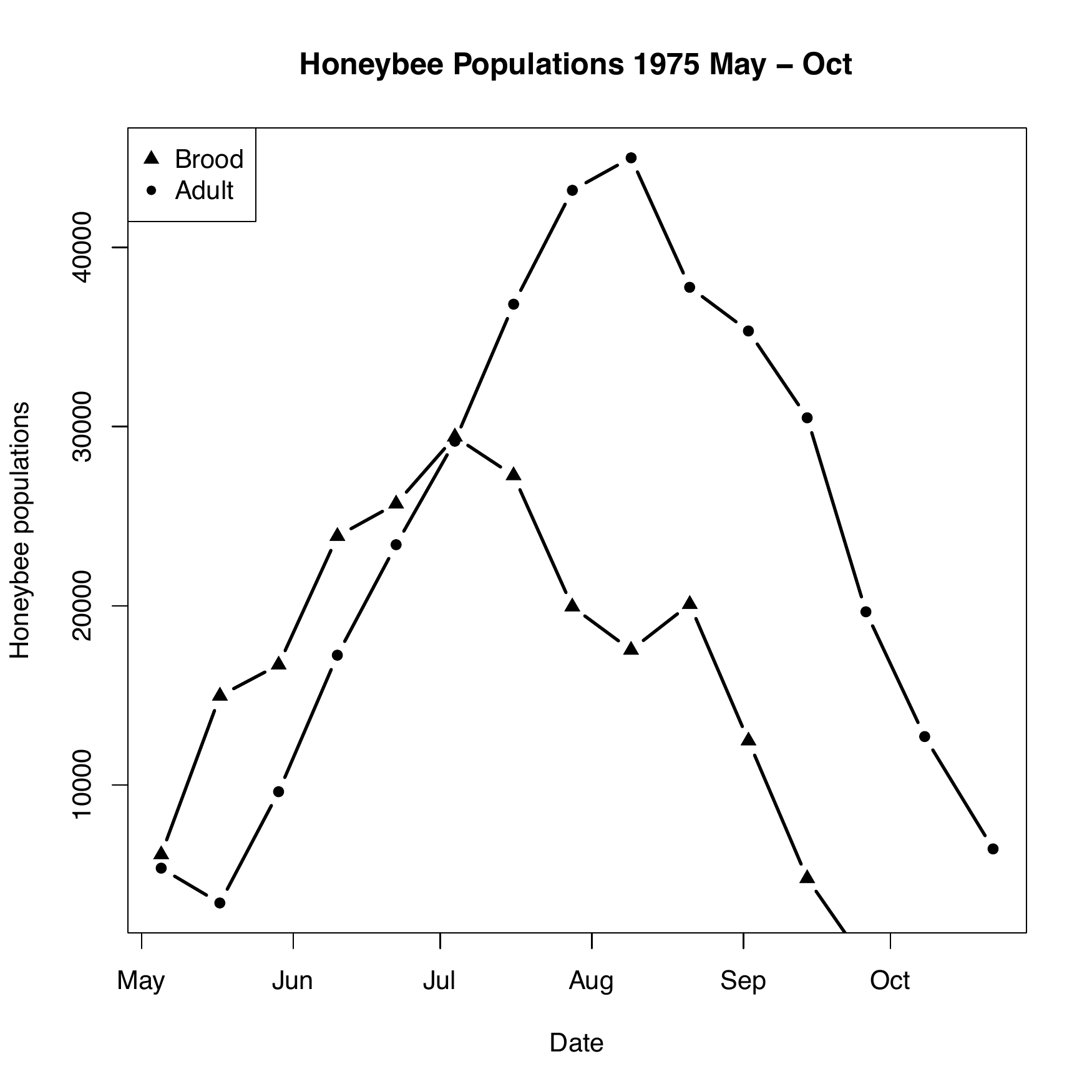}
\caption{Honeybee population in 1975}
    \end{subfigure}
    \begin{subfigure}{0.4\textwidth}
        \includegraphics[width=\textwidth]{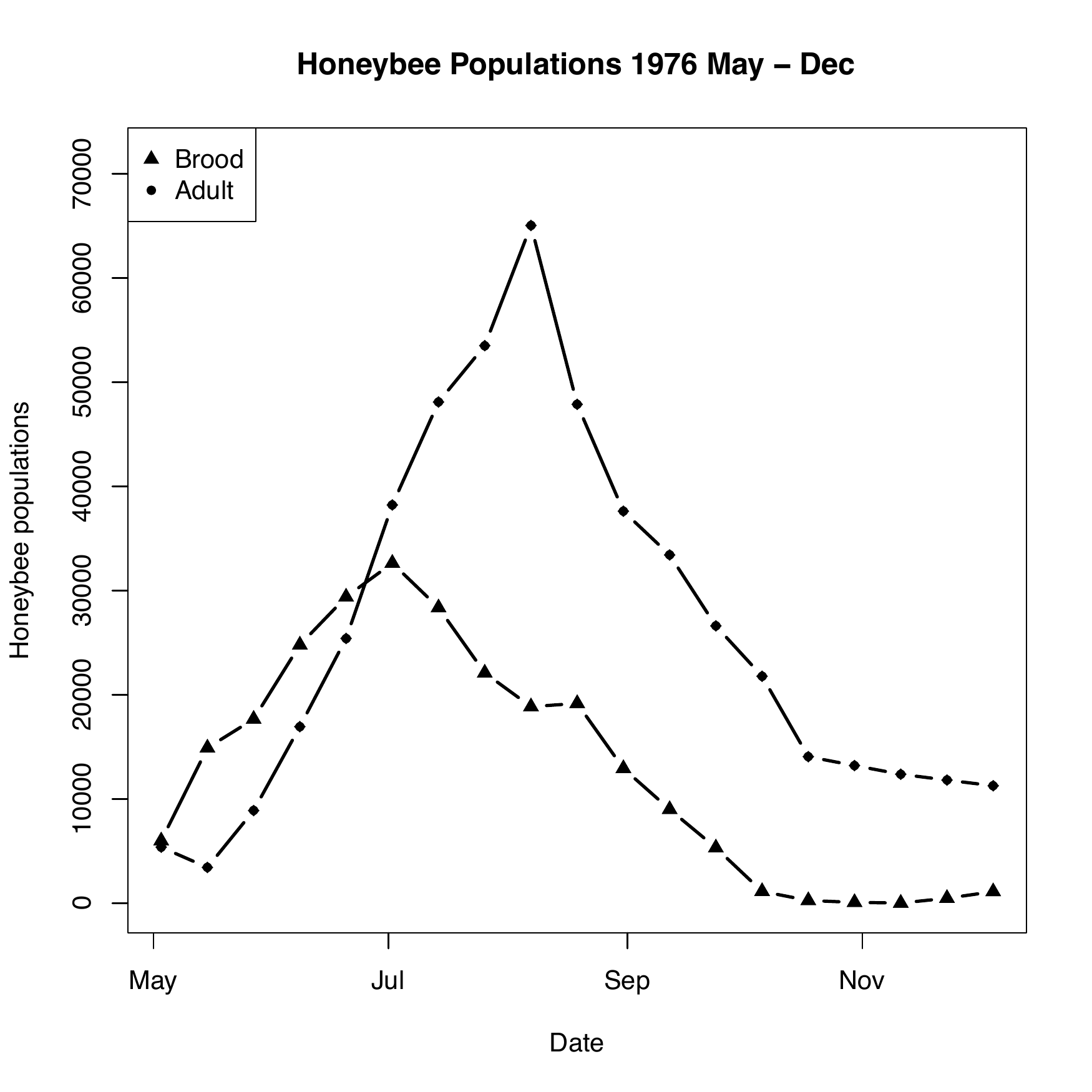}
\caption{Honeybee population in 1976}
    \end{subfigure}
	\caption{\label{fig:data} The observed population data for  honeybee colony in 1975 (left) and 1976 (right), respectively. The triangle line corresponds to brood population (eggs, larva and pupa), while the circle line corresponds to adults.}\label{data}
\end{figure}

In this section, we use the honeybee population data collected by James Harris (1980) \cite{harris1980population} to do parameter estimations and model validations.
The honeybee population data of two colonies: May 5, 1975 to Oct 22, 1975  and  May 3, 1976 to Dec 5, 1976,  is shown in Figure \ref{data} in the left (1975) and right (1976) side, respectively: The brood $B$ (the sum of egg, larvae, and pupa) population is shown by triangle dots while the adult $H$ population is represented by point dots. Based on Figure \ref{data}, the initial population of brood  is $B_0=6125$ and adult $H_0=5362$ for 1975 while $B_0=5982$ and $H_0=5362$ for 1976. Figure \ref{data} shows seasonality. Mathematical analysis provided in our previous section indicates that Model \eqref{BH2} can have oscillatory solutions under certain conditions while Model \eqref{BH} only exhibits simple equilibrium dynamics.\\

The questions are: (1) Is Model \eqref{BH2} better than Model \eqref{BH} because it shows oscillatory solutions? Or (2) Is seasonality showed in honeybee population data (see Figure \ref{data}) caused by the external factors such as resource?\\

{To address the questions above, we first assume that  the queen lays egg is seasonal due to resource constraints.  The literature work suggests that food, temperature, weather and oviposition place would affect the queen \cite{bodenheimer1937studies,khoury2011quantitative,degrandi1989beepop}, thence her egg-laying rate is not constant, and assumed tohave the following expression:}
    \begin{eqnarray}
     r(t) &=& r_0*(1 + \cos(\frac{2\pi(t -\psi)}{\gamma})) \label{seasonalitya};\\
     r(t-\tau) &=& r_0*(1 + \cos(\frac{2\pi(t -\tau-\psi)}{\gamma})) .
     \label{seasonalityb}
    \end{eqnarray}where $\gamma$ indicates the length of seasonality; $\tau$ indicates the time length of the juvenile period; $\psi$ indicates the time of the maximum laying rate; and $r_0$ indicates the baseline egg-laying rate.\\


{Then we perform parameter estimations and model validations based on data shown in Figure \ref{data}: We implement the Monte Carlo parameter sweep method as our fitting method to the honeybee population data to attain parameter estimates \cite{cowan1998statistical}. 
Essentially, we randomly sample hypotheses for the parameters following negative binomial regression with appropriate ranges (see Table.\ref{tb1}). For each observed value, we defined the negative binomial probability density function. The mean ($\mu$) is set by the corresponding predictive value, and variance is $\mu+\alpha*\mu^2$, which is $\alpha=k^{-1}$. $k$ indicates dispersion parameter, which we set range [0.5,2] \cite{piegorsch1990maximum, lloyd2007maximum}. Then we calculate the likelihood for negative binomial regression model to get better estimate for parameters \cite{lawless1987negative,ismail2007handling}. Afterwards, we performed data fitting on the above model (model number) and compare the results..}\\

We first assume that the egg laying rate $r$ is constant. {All fittings are set by constant history functions with $B(\theta)=6125$ and $H(\theta)=5362$ for 1975, and $B(\theta)=5982$ and $H(\theta)=5362$ for 1976, for all $\theta\in[-\tau,0]$.}

\begin{figure}[ht]
 \centering
 \begin{subfigure}{0.45\textwidth}
        \includegraphics[width=\textwidth]{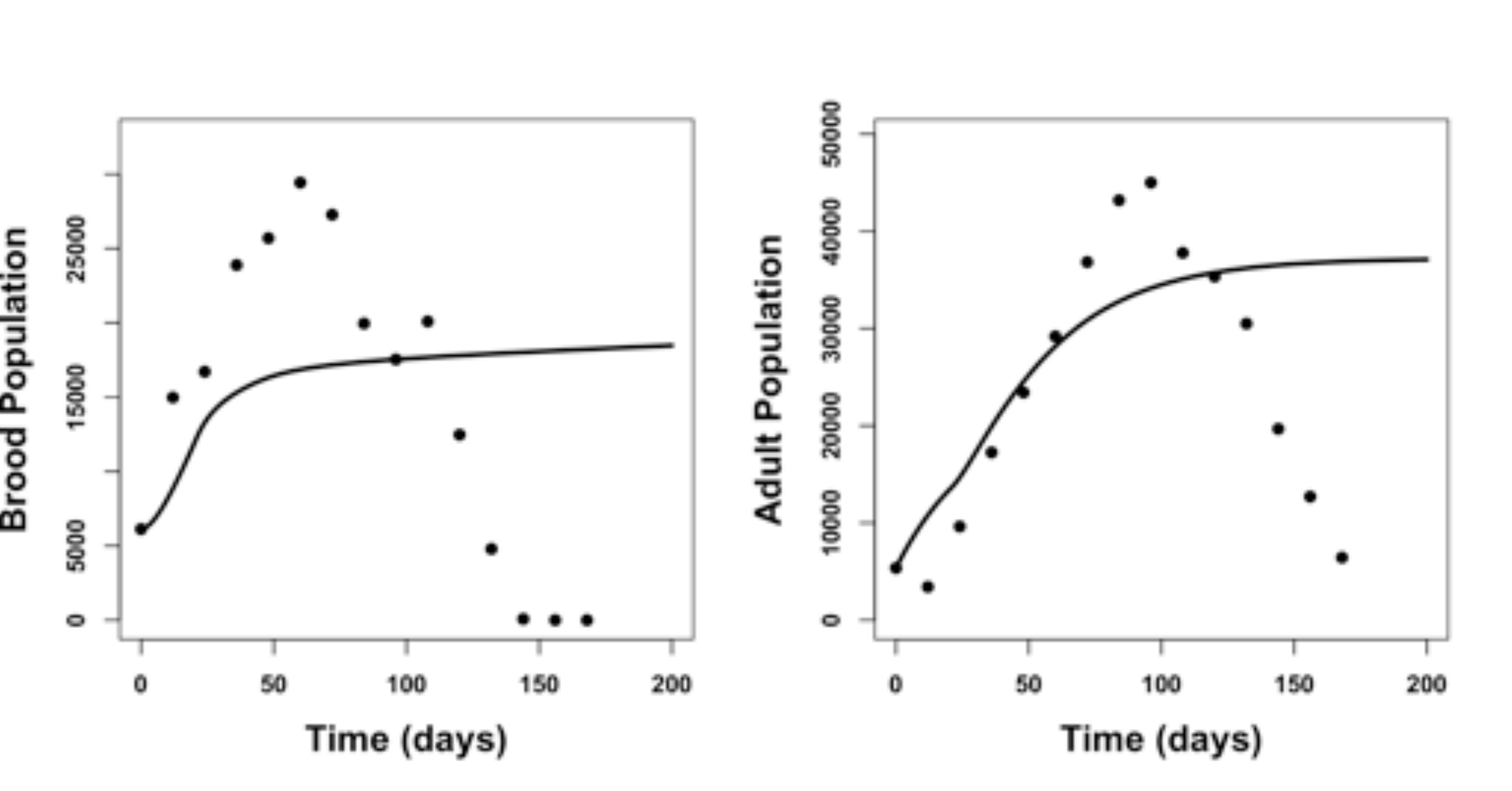}
\caption{Model (\ref{BH}) (1975), $r = 1237; d_h = 0.033; d_b = 0.001; K = 20,574,000; \alpha = 16.9.$}
\label{fig: 1975_without_1}
    \end{subfigure}
    \begin{subfigure}{0.45\textwidth}
        \includegraphics[width=\textwidth]{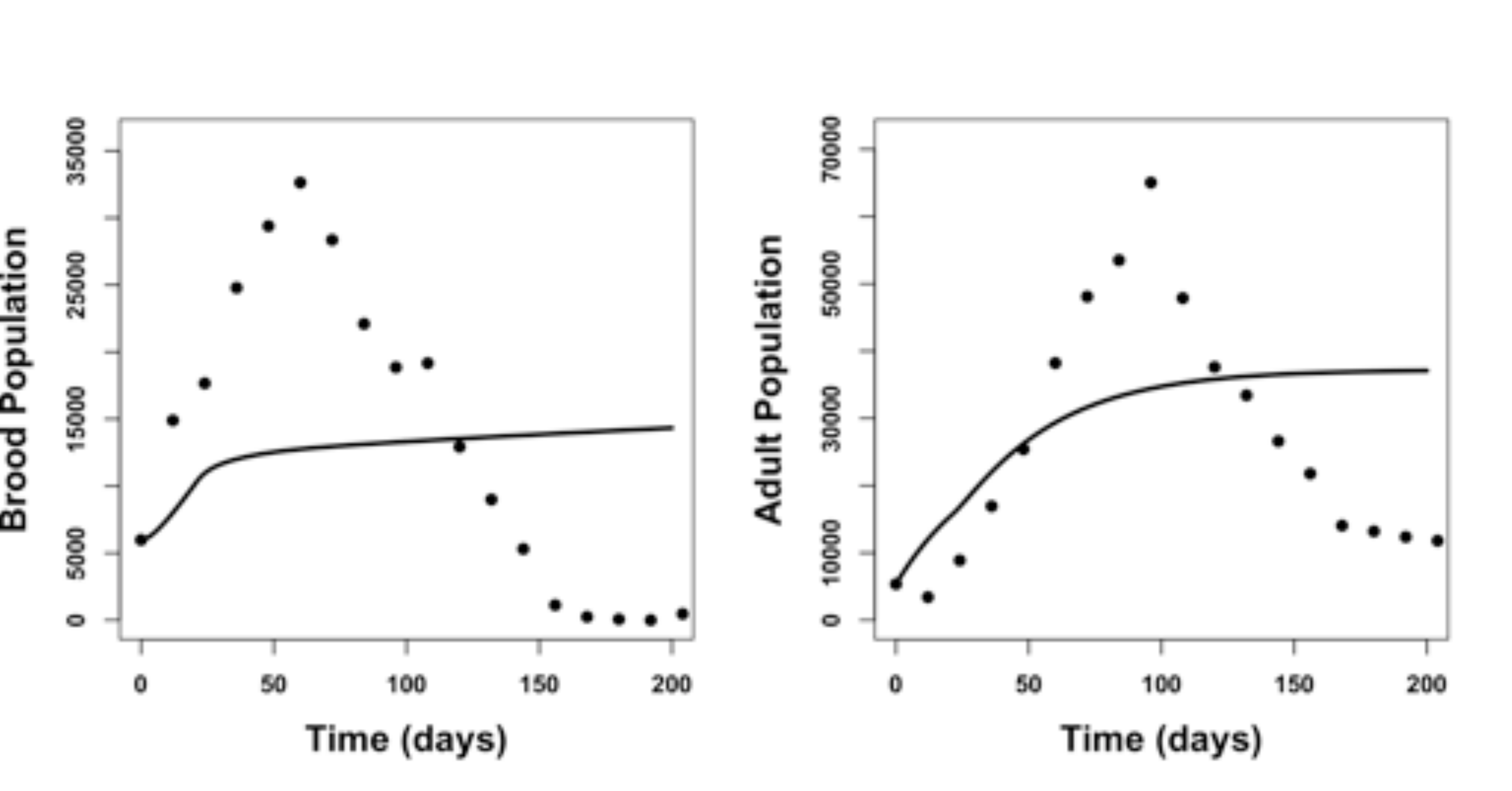}
\caption{Model (\ref{BH}) (1976), $r = 1149; d_h = 0.03; d_b = 0.001; K = 10,653,000; \alpha = 4.$}
\label{fig: 1976_without_1}
    \end{subfigure}
   \vfill
    \begin{subfigure}{0.45\textwidth}
        \includegraphics[width=\textwidth]{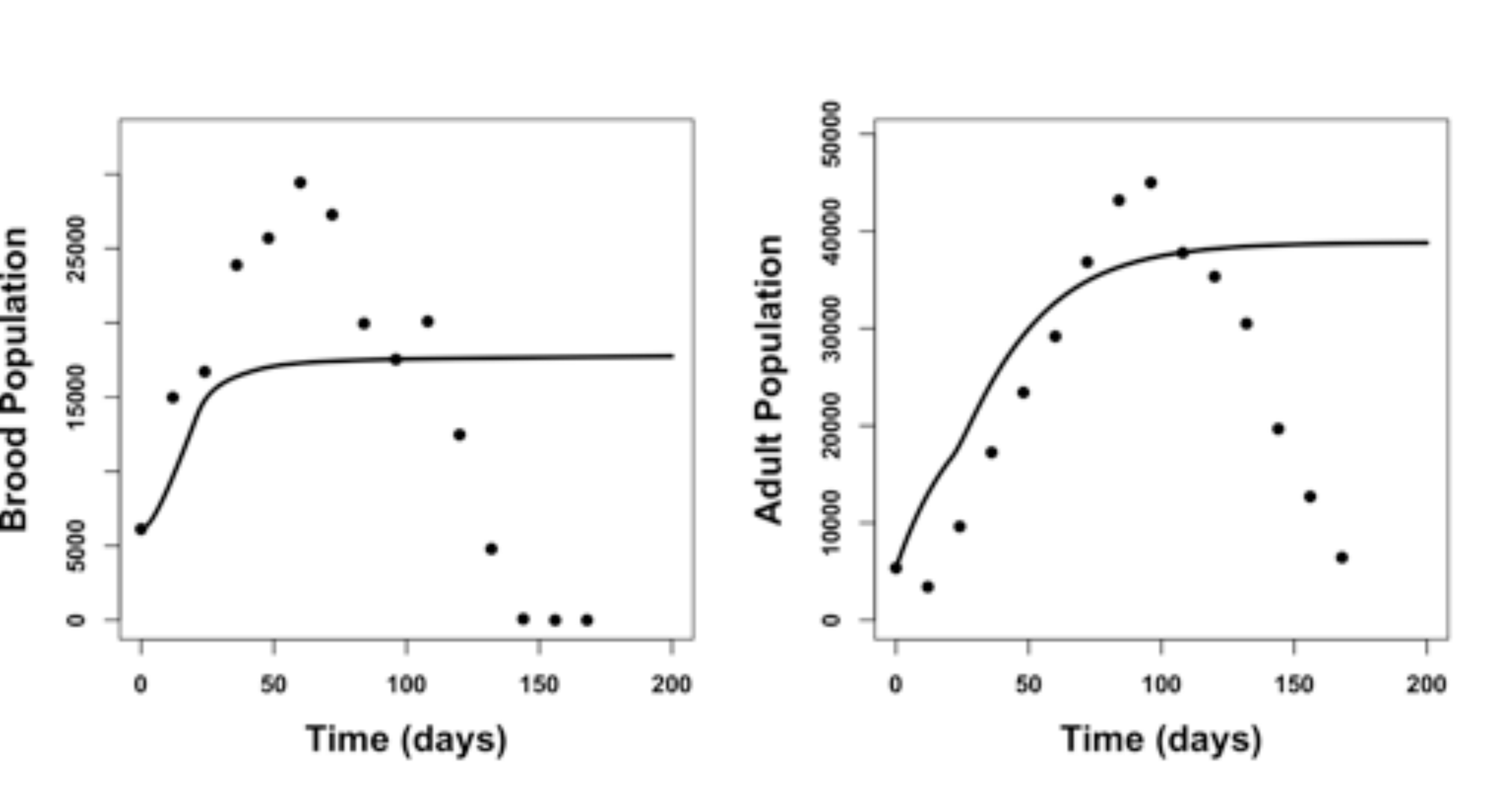}
\caption{Model (\ref{BH3}) (1975), $r = 1573; d_h = 0.04; d_b = 0.0001; K = 15,716,000.$}
\label{fig: 1975_without_2}
    \end{subfigure}
    \begin{subfigure}{0.45\textwidth}
        \includegraphics[width=\textwidth]{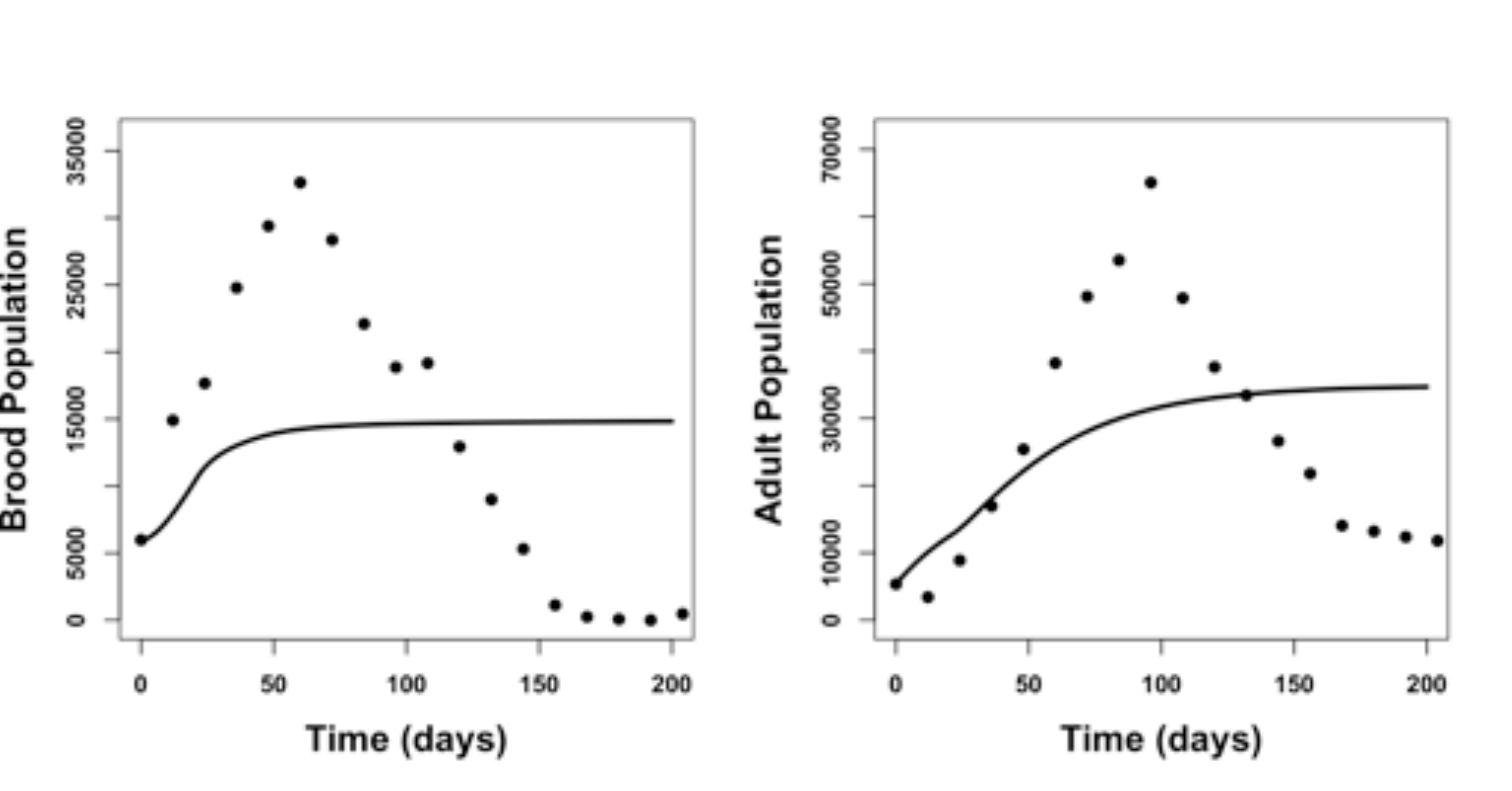}
\caption{Model (\ref{BH3}) (1976), $r = 1065; d_h = 0.03; d_b = 0.0001; K = 19,600,000.$}
\label{fig: 1976_without_2}
    \end{subfigure}
     \vfill
    \begin{subfigure}{0.45\textwidth}
        \includegraphics[width=\textwidth]{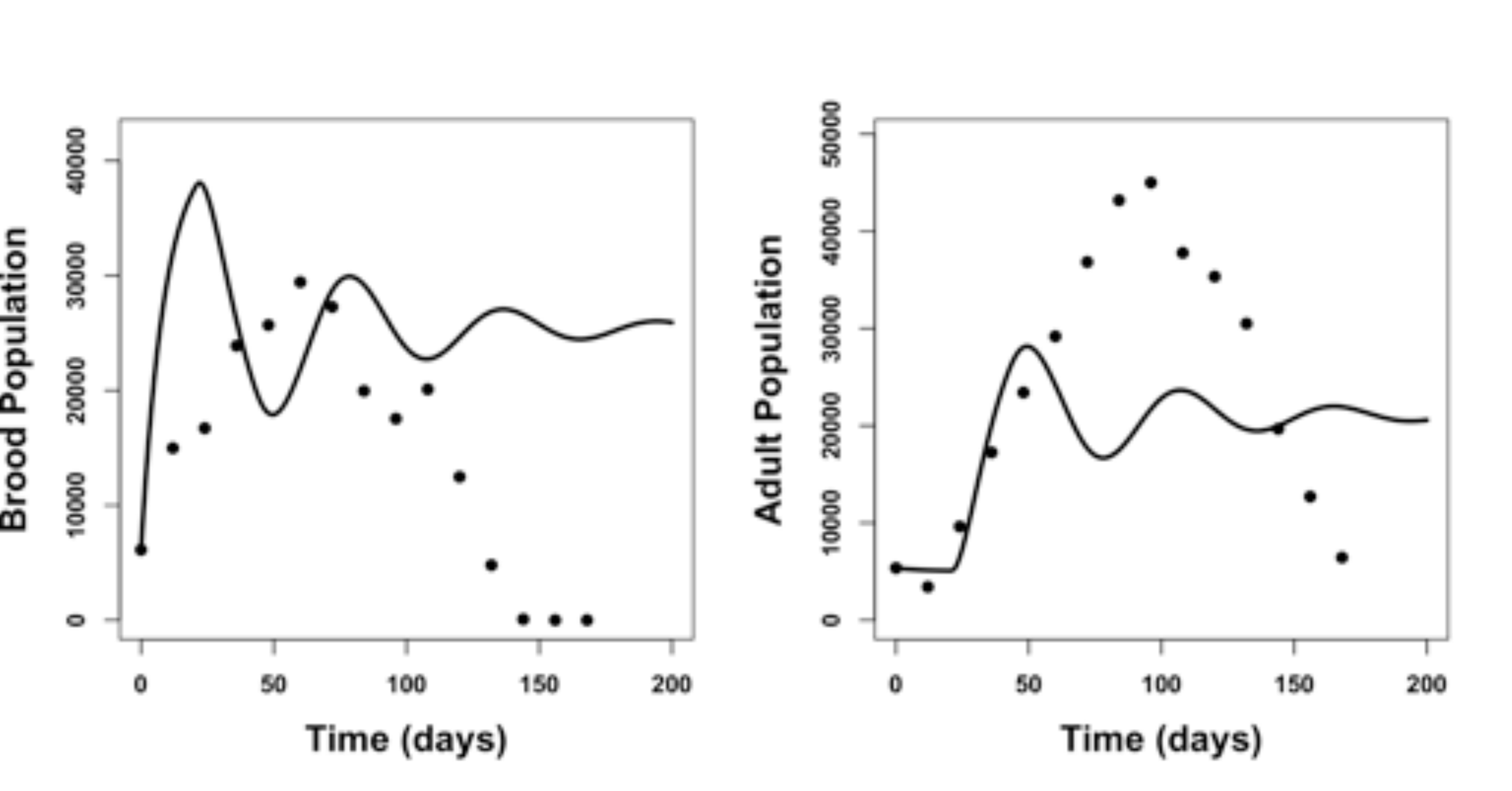}
\caption{Model (\ref{BH2}) (1975), $r = 5333; d_h = 0.12; d_b = 0.11; K = 867,000.$}
\label{fig: 1975_without_3}
    \end{subfigure}
     \begin{subfigure}{0.45\textwidth}
        \includegraphics[width=\textwidth]{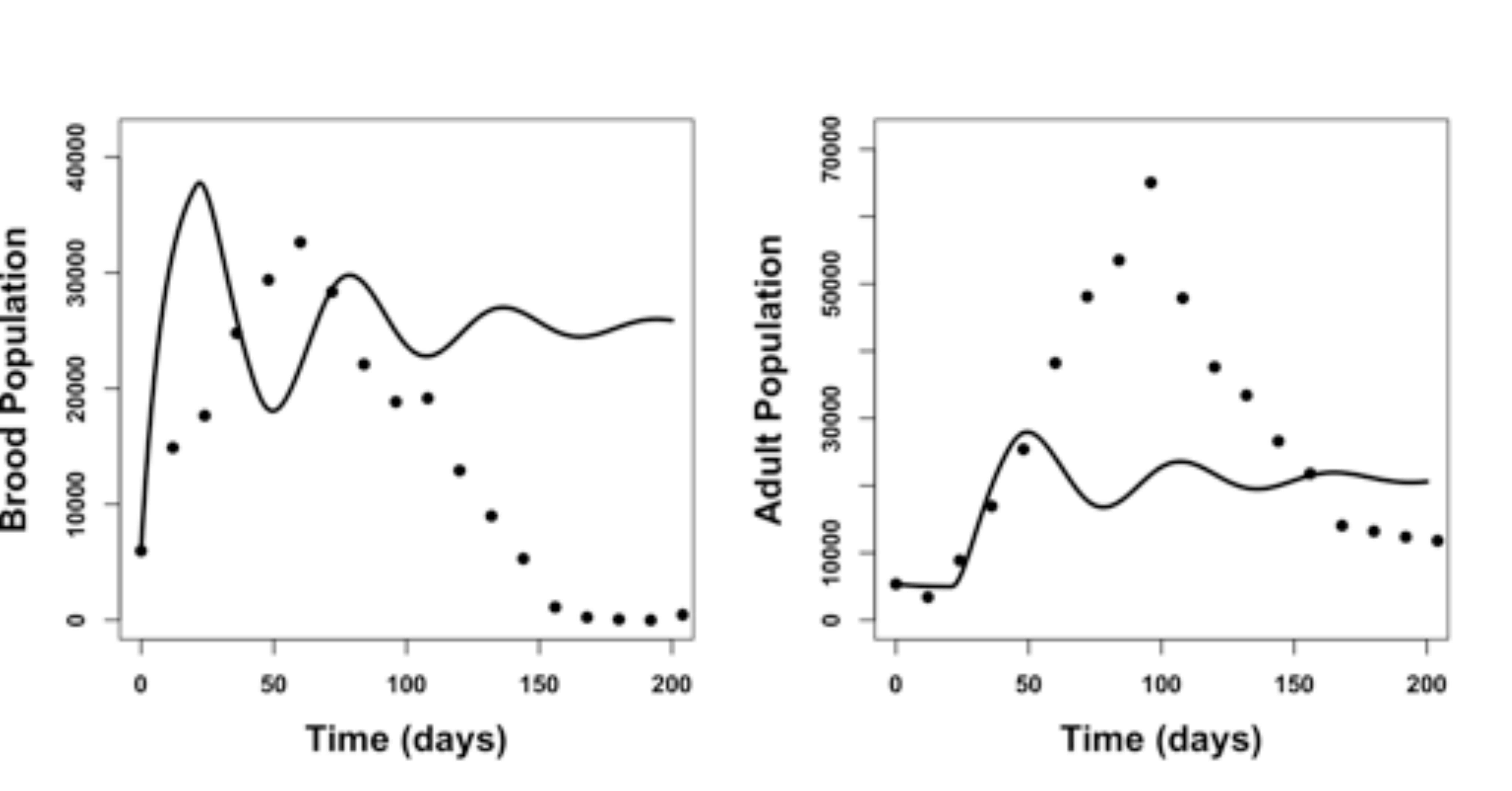}
\caption{Model (\ref{BH2}) (1976), $r = 5333; d_h = 0.12; d_b = 0.11; K = 1,088,000.$}
\label{fig: 1976_without_3}
    \end{subfigure}
    \caption{Data fitting without seasonality for Harris honeybees data in 1975 (a, c, e) and 1976 (b, d, f) with $\tau=21$. Black dots indicate Harris data, and black curve indicates our model. }
    \label{fig:without}
 \end{figure}

\begin{enumerate}
\item Fitting Model \eqref{BH}: the best fit is shown in Figure (\ref{fig: 1975_without_1} \& \ref{fig: 1976_without_1}). If we use the estimated parameters: $r = 1237; d_h = 0.033; d_b = 0.001; K = 20,574,000; \alpha = 16.9$, then $E_2=(25299, 36124.5)$ in 1975; and $r = 1149; d_h = 0.03; d_b = 0.001; K = 10,653,000; \alpha = 4$, then $E_2=(23693.5, 37215.3)$ in 1976. Model \eqref{BH} approaches to its plateau under those two fittings.

\item Fitting Model \eqref{BH3}: the best fit is shown in Figure (\ref{fig: 1975_without_2} \& \ref{fig: 1976_without_2}). If we use the estimated parameters: $r = 1573; d_h = 0.04; d_b = 0.0001; K = 15,716,000$, then $E_2=(32658.1, 38837.8)$ in 1975; and $r = 1065; d_h = 0.03; d_b = 0.0001; K = 19,600,000$, then $E_2=(21987, 34863.3)$ in 1976. Model \eqref{BH3} approaches to its plateau under those two fittings.

\item Fitting Model \eqref{BH2}: the best fit is shown in Figure (\ref{fig: 1975_without_3} \& \ref{fig: 1976_without_3}). If we use the estimated parameters: $r = 5333; d_h = 0.12; d_b = 0.11; K = 867,000$, then $E_2=(25435.1, 21039.3)$ in 1975; and $r = 5333; d_h = 0.12; d_b = 0.11; K = 1,088,000$, then $E_2=(25422.3, 21028.8)$ in 1976. Model \eqref{BH2} approaches to its steady state through damping oscillations.\\ 
\end{enumerate}
\renewcommand{\arraystretch}{1.5}

\begin{figure}[ht]
 \centering
 \begin{subfigure}{0.45\textwidth}
        \includegraphics[width=\textwidth]{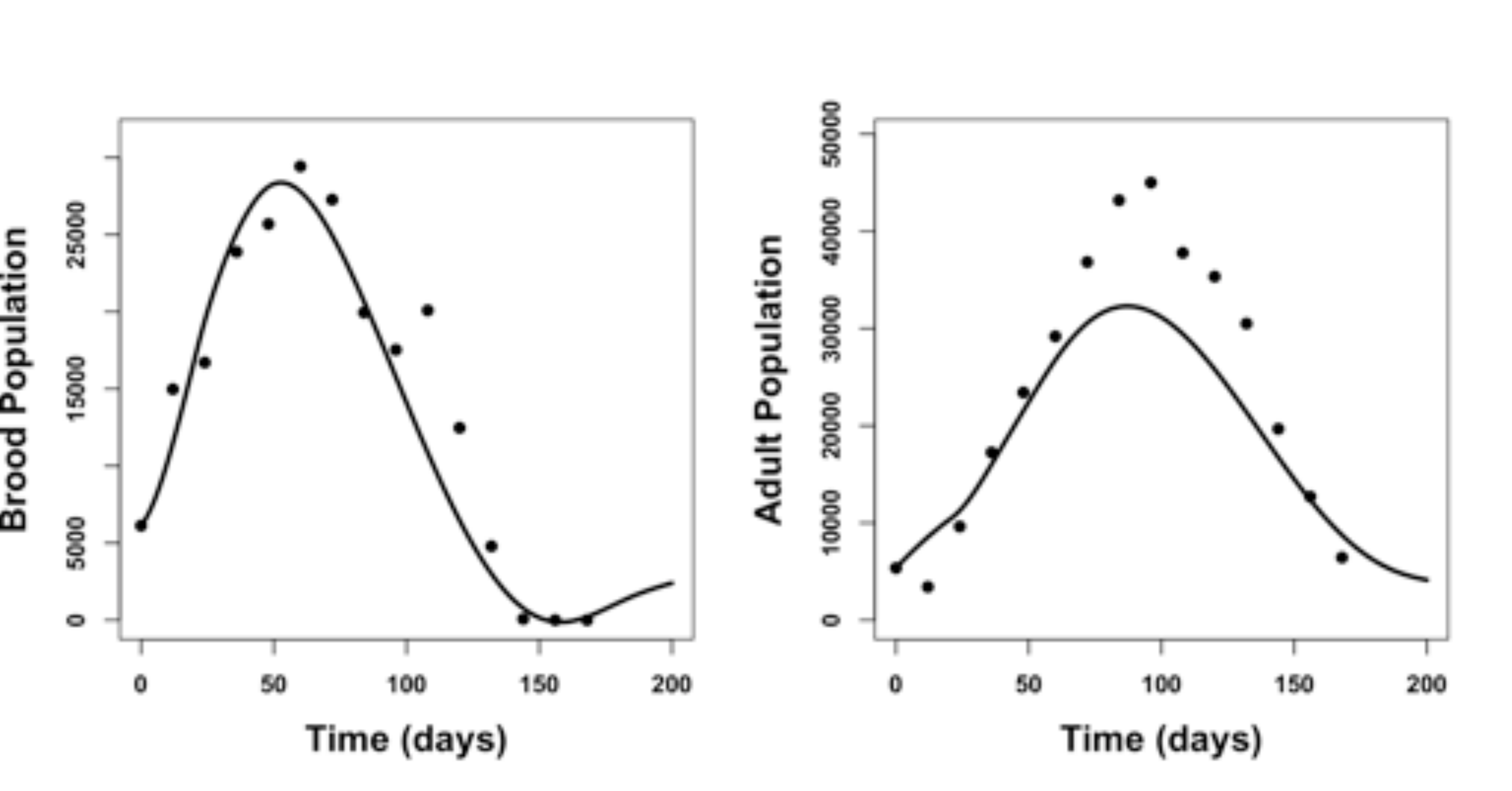}
\caption{Model (\ref{BH}) (1975), $r_0 = 1193; d_h = 0.03; d_b = 0.02; K = 56,963,000; \alpha = 4; \gamma = 273; \psi = 12.$}\label{fig: 1975_with_1}
    \end{subfigure}
    \begin{subfigure}{0.45\textwidth}
        \includegraphics[width=\textwidth]{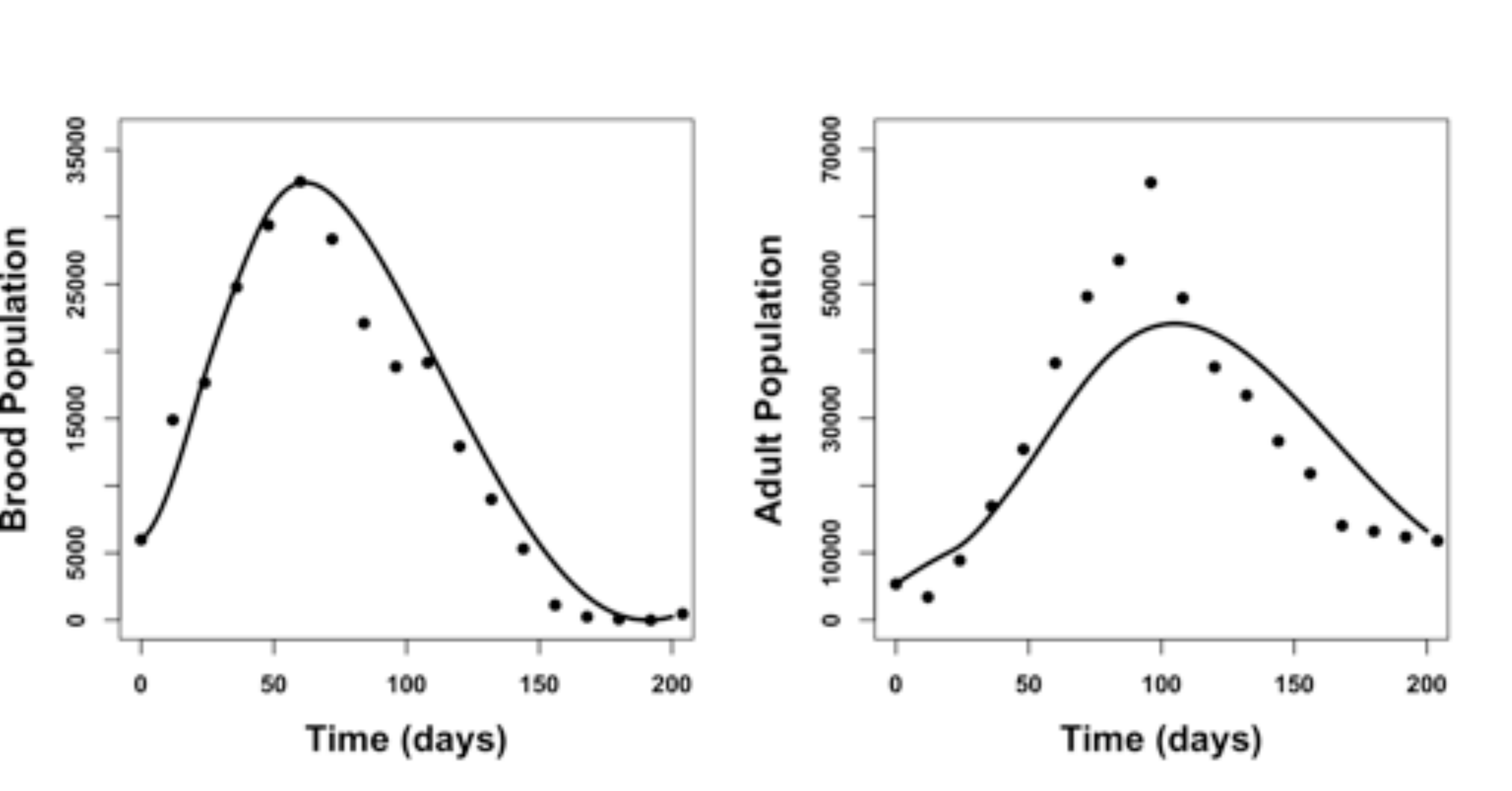}
\caption{Model (\ref{BH}) (1976), $r_0 = 1319; d_h = 0.023; d_b = 0.02; K = 84,933,000; \alpha = 3.2; \gamma = 338; \psi = 12.$}\label{fig: 1976_with_1}
    \end{subfigure}
     \vfill
    \begin{subfigure}{0.45\textwidth}
        \includegraphics[width=\textwidth]{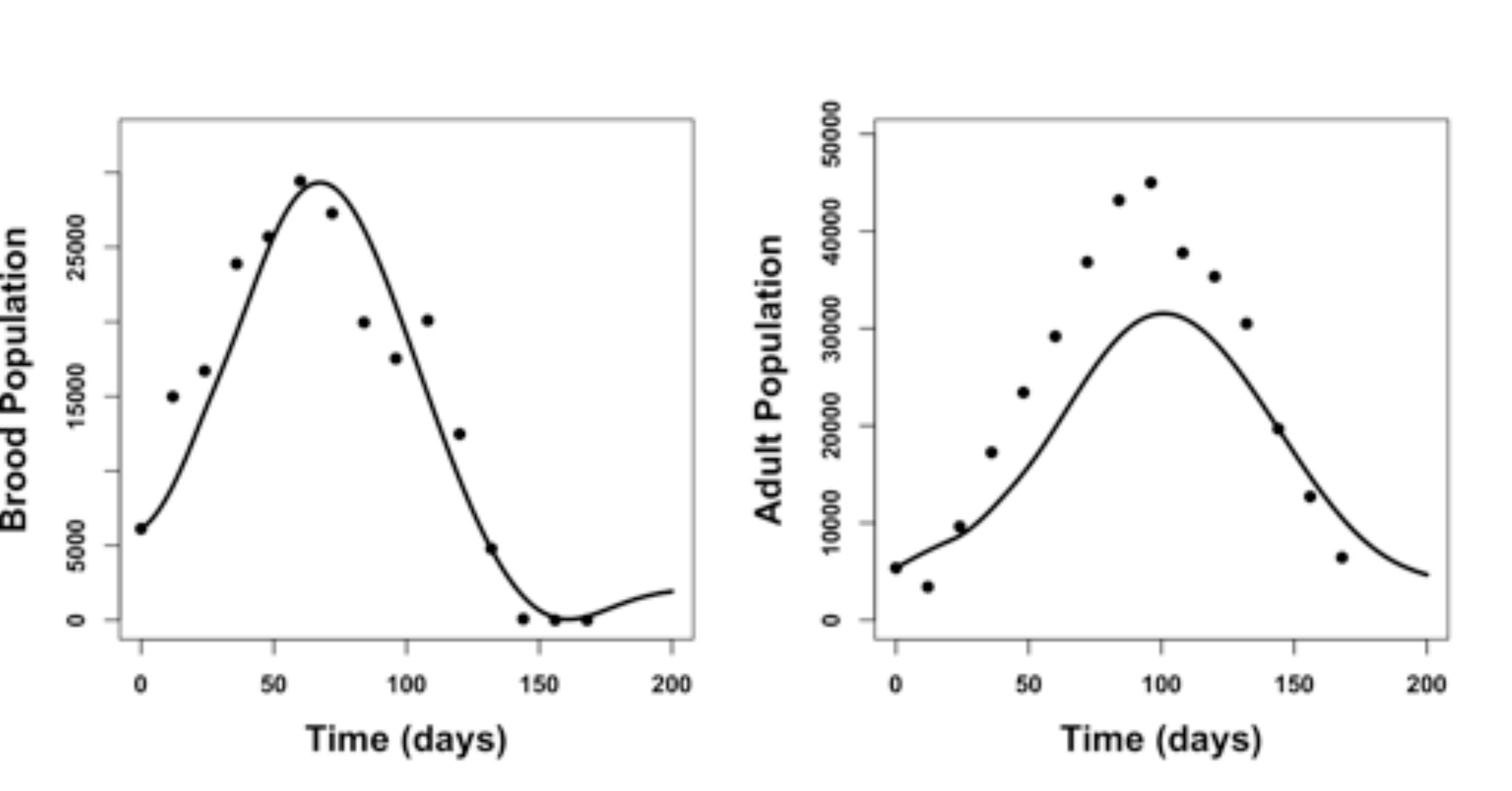}
\caption{Model \ref{BH3} (1975),  $r_0 = 1644; d_h = 0.03; d_b = 0.02; K = 139,137,000; \gamma = 277; \psi = 12.$}\label{fig: 1975_with_2}
    \end{subfigure}
    \begin{subfigure}{0.45\textwidth}
        \includegraphics[width=\textwidth]{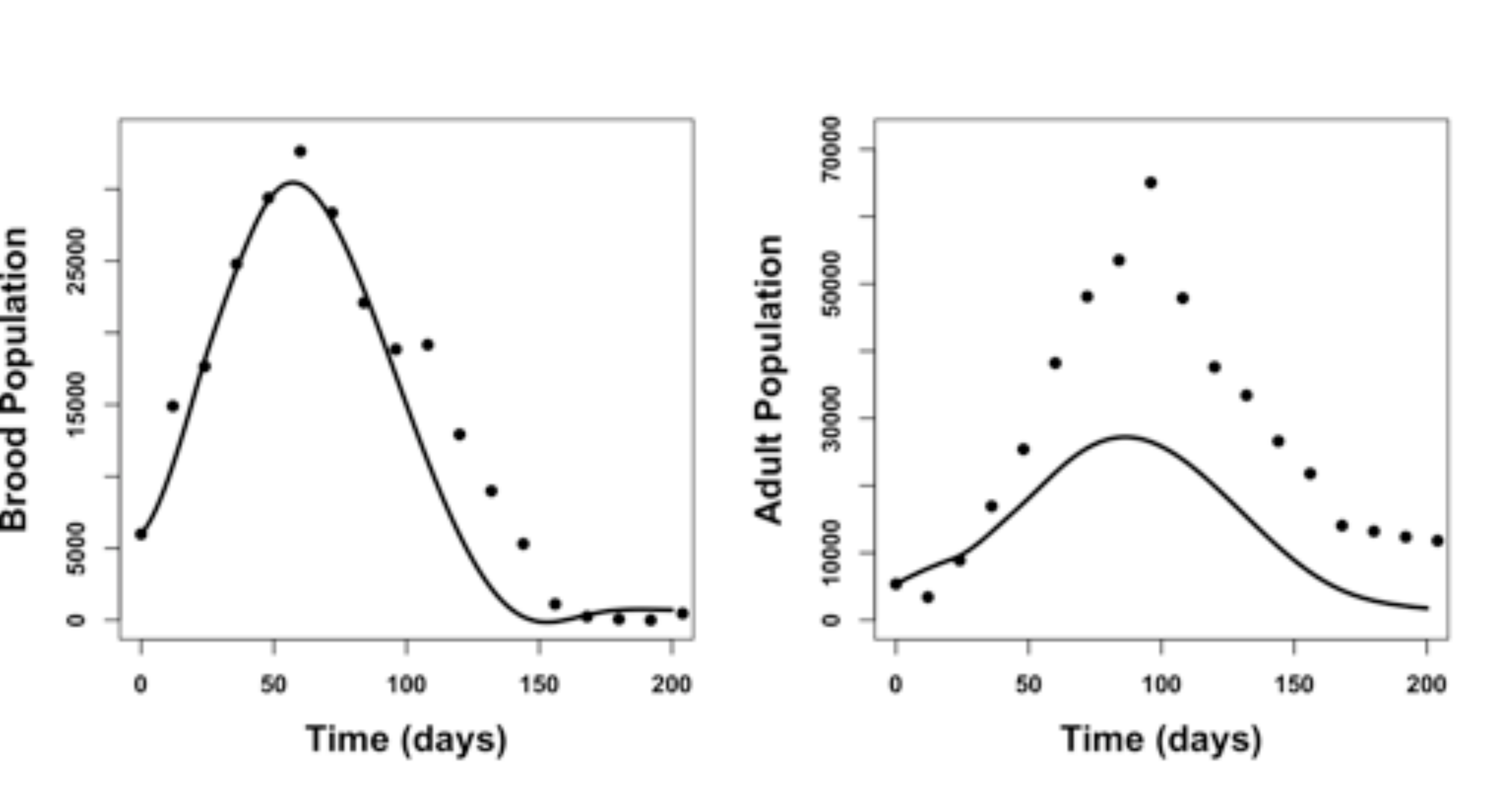}
\caption{Model (\ref{BH3}) (1976),  $r_0 = 1477; d_h = 0.04; d_b = 0.02; K = 81,048,000; \gamma = 261; \psi = 12.$}\label{fig: 1976_with_2}
    \end{subfigure}
     \vfill
     \begin{subfigure}{0.45\textwidth}
        \includegraphics[width=\textwidth]{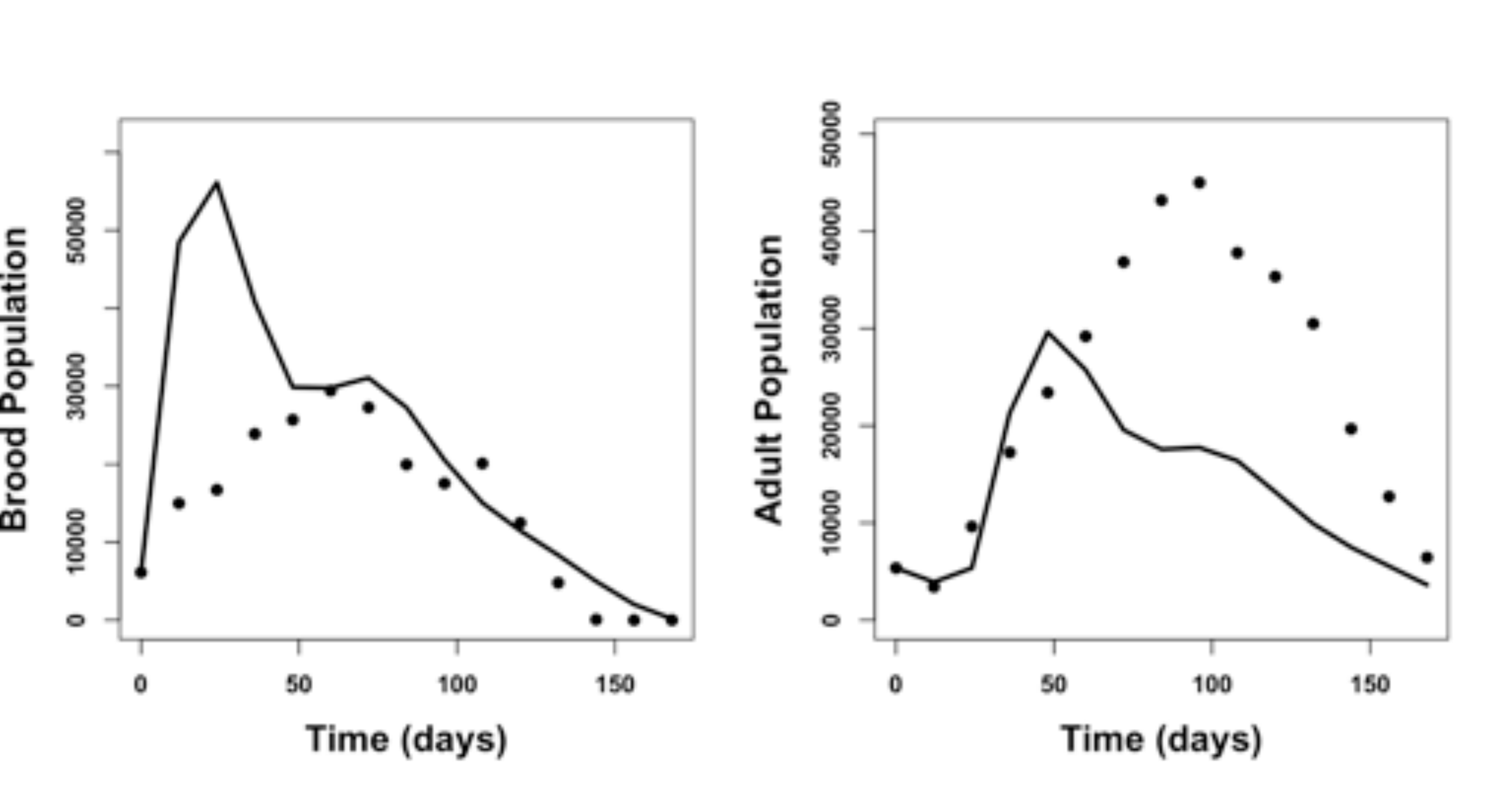}
\caption{Model (\ref{BH2}) (1975), $r_0 = 4024; d_h = 0.13; d_b = 0.12; K = 89,000; \gamma = 350; \psi = 12.$}\label{fig: 1975_with_3}
    \end{subfigure}
    \begin{subfigure}{0.45\textwidth}
        \includegraphics[width=\textwidth]{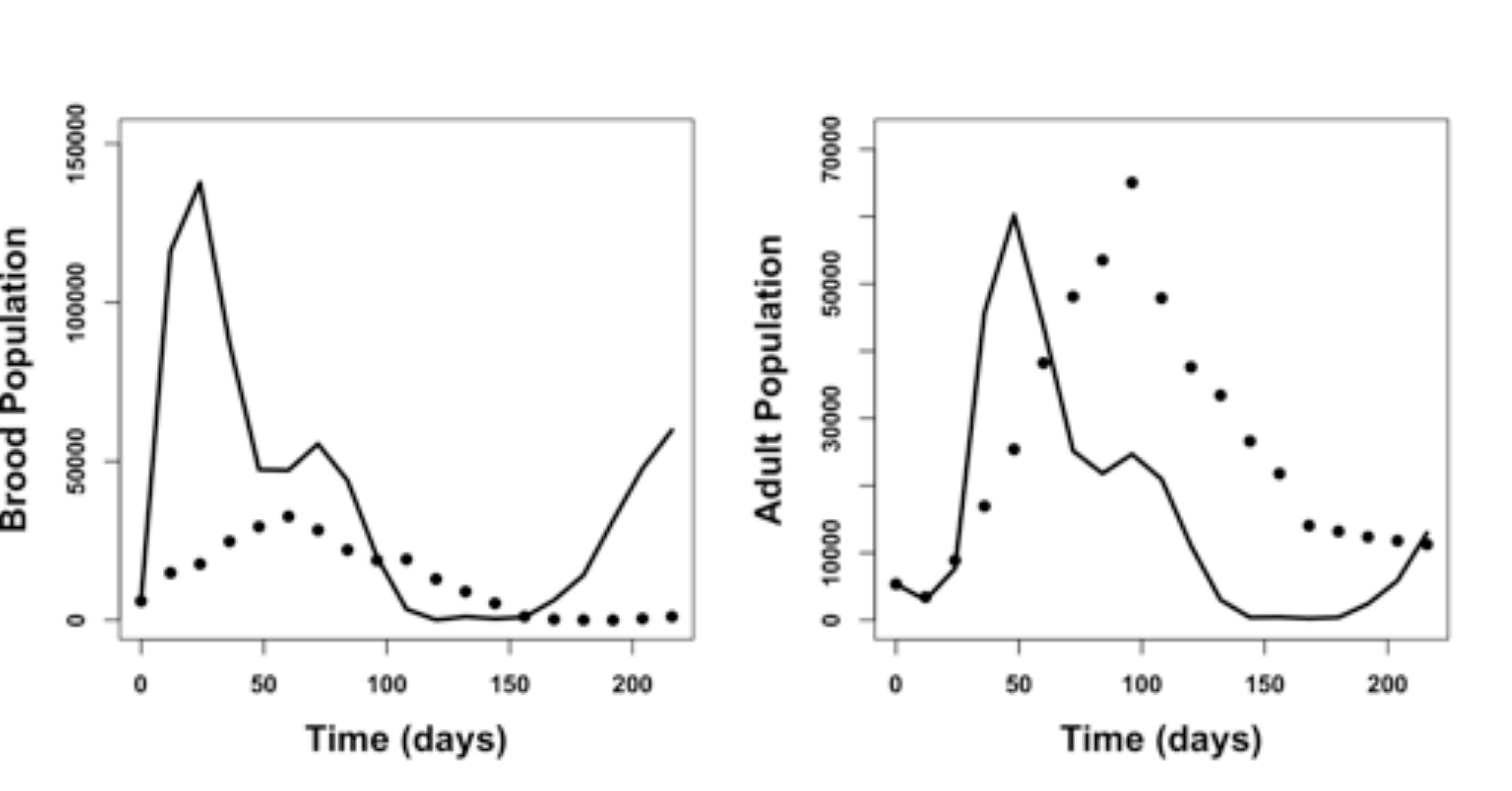}
\caption{Model (\ref{BH2}) (1976), $r_0 = 9171; d_h = 0.19; d_b = 0.12; K = 20,000; \gamma = 261; \psi = 12.$}\label{fig: 1976_with_3}
    \end{subfigure}
    \caption{Data fitting with the seasonality equation in $r$ for Harris honeybees data in 1975 (a, c, e) and 1976 (b, d, f) with $\tau=21$. Black dots indicate Harris data, and black curve indicates our model. }
    \label{fig:with}
 \end{figure}

The fittings shown in Figure \ref{fig:without} suggest that the assumption of the queen egg laying being constant is not realistic enough. Thus  we assume that the egg laying rate is  a periodic function $r(t)=r_0*(1 + \cos(\frac{2\pi(t -\psi)}{\gamma}))$.  {All fittings are set by constant history functions, i.e., $B(\theta)=6125$ and $H(\theta)=5362$ for 1975, and $B(\theta)=5982$ and $H(\theta)=5362$ for 1976, for all $\theta\in[-\tau,0]$}
\begin{enumerate}
\item Fitting Model \eqref{BH}: the best fit is shown in Figure (\ref{fig: 1975_with_1} \& \ref{fig: 1976_with_1}). If we use the estimated parameters: $r_0 = 1193; d_h = 0.03; d_b = 0.02; K = 56,963,000; \alpha = 4; \gamma = 273; \psi = 12$ in 1975; and $r_0 = 1319; d_h = 0.023; d_b = 0.02; K = 84,933,000; \alpha = 3.2; \gamma = 338; \psi = 12$ in 1976. Both data fittings of Model \eqref{BH} have periodic solutions.

\item Fitting Model \eqref{BH3}: the best fit is shown in Figure (\ref{fig: 1975_with_2} \& \ref{fig: 1976_with_2}). If we use the estimated parameters: $r_0 = 1644; d_h = 0.03; d_b = 0.02; K = 139,137,000; \gamma = 277; \psi = 12$ in 1975; and $r_0 = 1477; d_h = 0.04; d_b = 0.02; K = 81,048,000; \gamma = 261; \psi = 12$ in 1976. Both data fittings of Model \eqref{BH3} have periodic solutions.

\item Fitting Model \eqref{BH2}: the best fit is shown in Figure (\ref{fig: 1975_with_3} \& \ref{fig: 1976_with_3}). If we use the estimated parameters: $r_0 = 5333; d_h = 0.12; d_b = 0.11; K = 867,000; \gamma = 350; \psi = 12$ in 1975; and $r_0 = 9171; d_h = 0.19; d_b = 0.12; K = 20,000; \gamma = 261; \psi = 12$ in 1976.Both data fittings of Model \eqref{BH2} have periodic solutions that would lead to negative solutions as Model \eqref{BH2} is not positively invariant.
\end{enumerate}


The data fittings assuming that $ r(t) = r_0*(1 + \cos(\frac{2\pi(t -\psi)}{\gamma}))$ have better outcomes than the previous ones (Figure \ref{fig:without}) by assuming $r$ being constant. By comparison through the negative log likelihood method \cite{ismail2007handling}, 
we could deduce that Model (\ref{BH}) has the best fittings in both scenarios: $r$ being constant in Figure (\ref{fig:without}) and $r$ being periodic in Figure \ref{fig:with}. Thus, based on both theoretical work and model validation, we could conclude that even though Model \eqref{BH2} could have oscillations in its solution, Model \eqref{BH} with the egg laying rate $r$ being periodic (supported by the best fitting based on data, see Figure \ref{fig:with}) should be a better model for us to explore the important factors contributing to the healthy of honeybee colones.\\

\renewcommand{\arraystretch}{1.5}

{\noindent\textbf{Effects of seasonality:} Here we perform two scenarios that seasonality may promote (see Figure \ref{simulation-season}) or suppress (see Figure \ref{simulation-season2}) the survival of honeybee colony, respectively. Figure \ref{fig:m1B} \& \ref{fig:m1H} are simulations without seasonality by taking $r=1200, K=5.4*10^6, d_b=0.01, d_h=0.05, \alpha=10$ with a constant history function $B(\theta)=300;H(\theta)=200$ for all $\theta\in[-\tau, 0]$, which show that honeybee colony collapses. Figure \ref{fig:m1sB} \& \ref{fig:m1sH} has seasonality by taking $r= r_0*(1 + \cos(\frac{2\pi(t -45)}{365}))$ whose average is $r_0=1200$, which show that honeybee colony survives.\\
\begin{figure}[ht]
 \centering
    \begin{subfigure}{0.24\textwidth}
        \includegraphics[width=\textwidth]{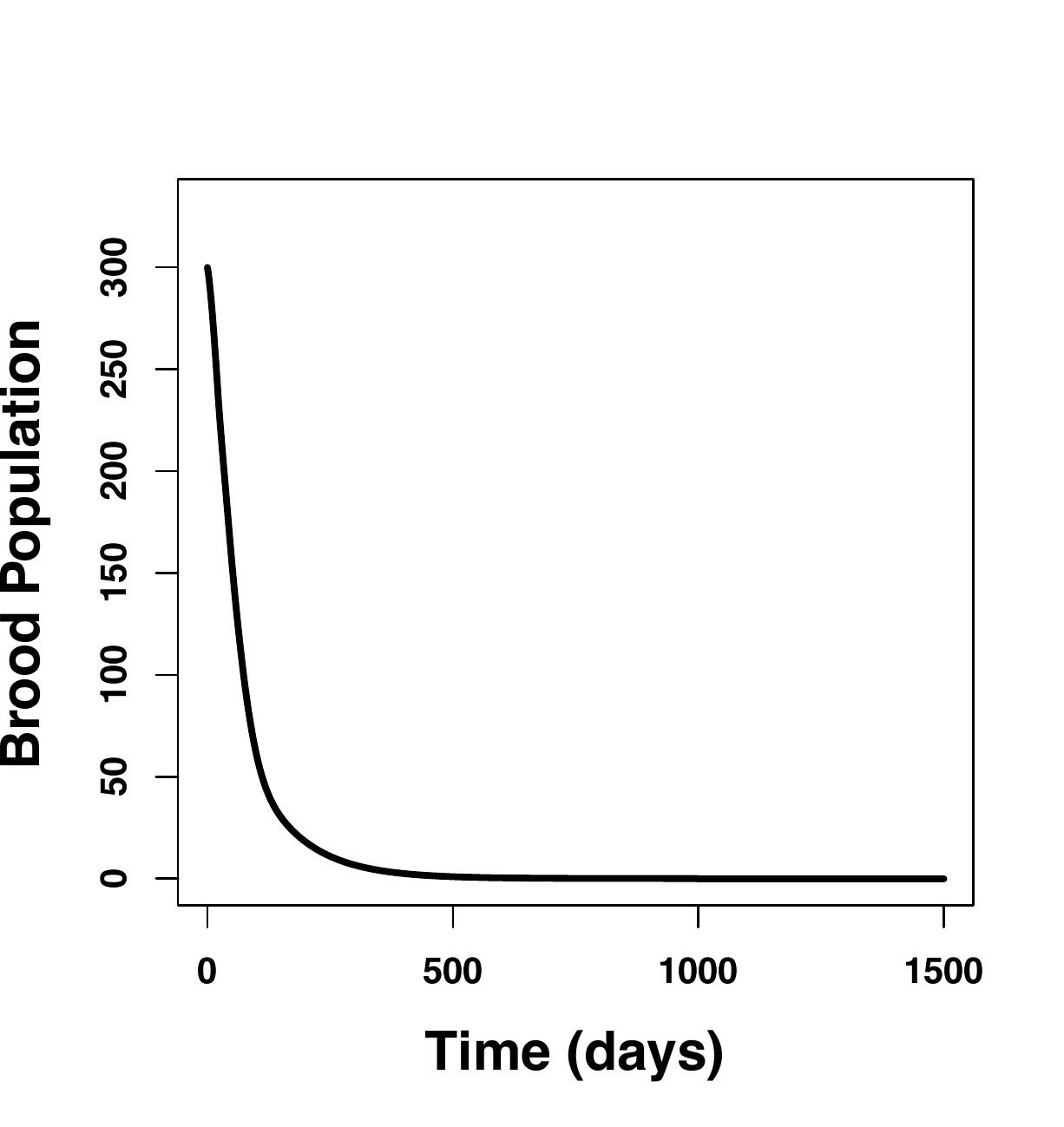}
\caption{Brood population of Model (\ref{BH}) without seasonality}
\label{fig:m1B}
    \end{subfigure}
    \begin{subfigure}{0.24\textwidth}
        \includegraphics[width=\textwidth]{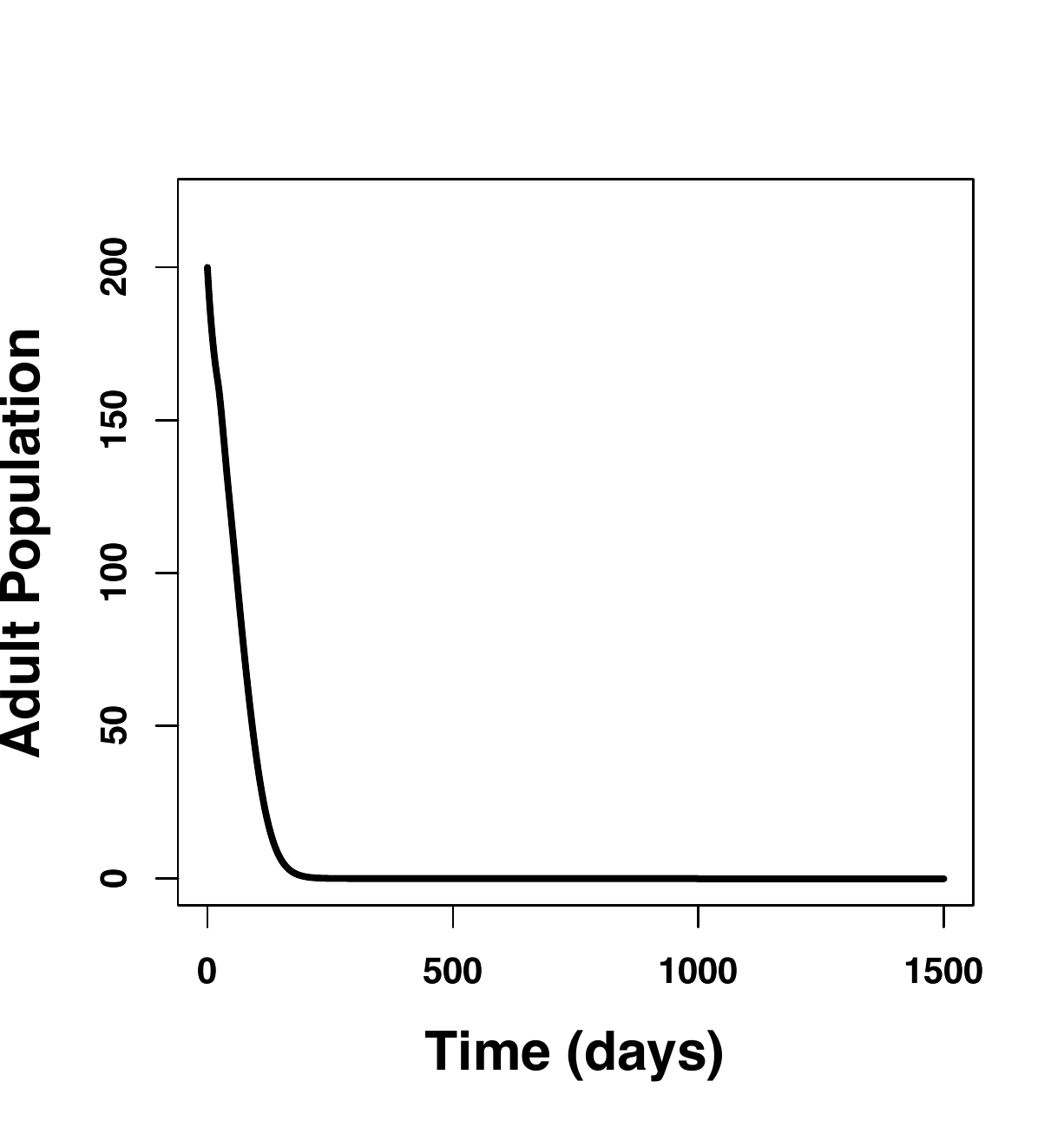}
\caption{Adult population of Model (\ref{BH}) without seasonality}
\label{fig:m1H}
    \end{subfigure}
    \begin{subfigure}{0.24\textwidth}
        \includegraphics[width=\textwidth]{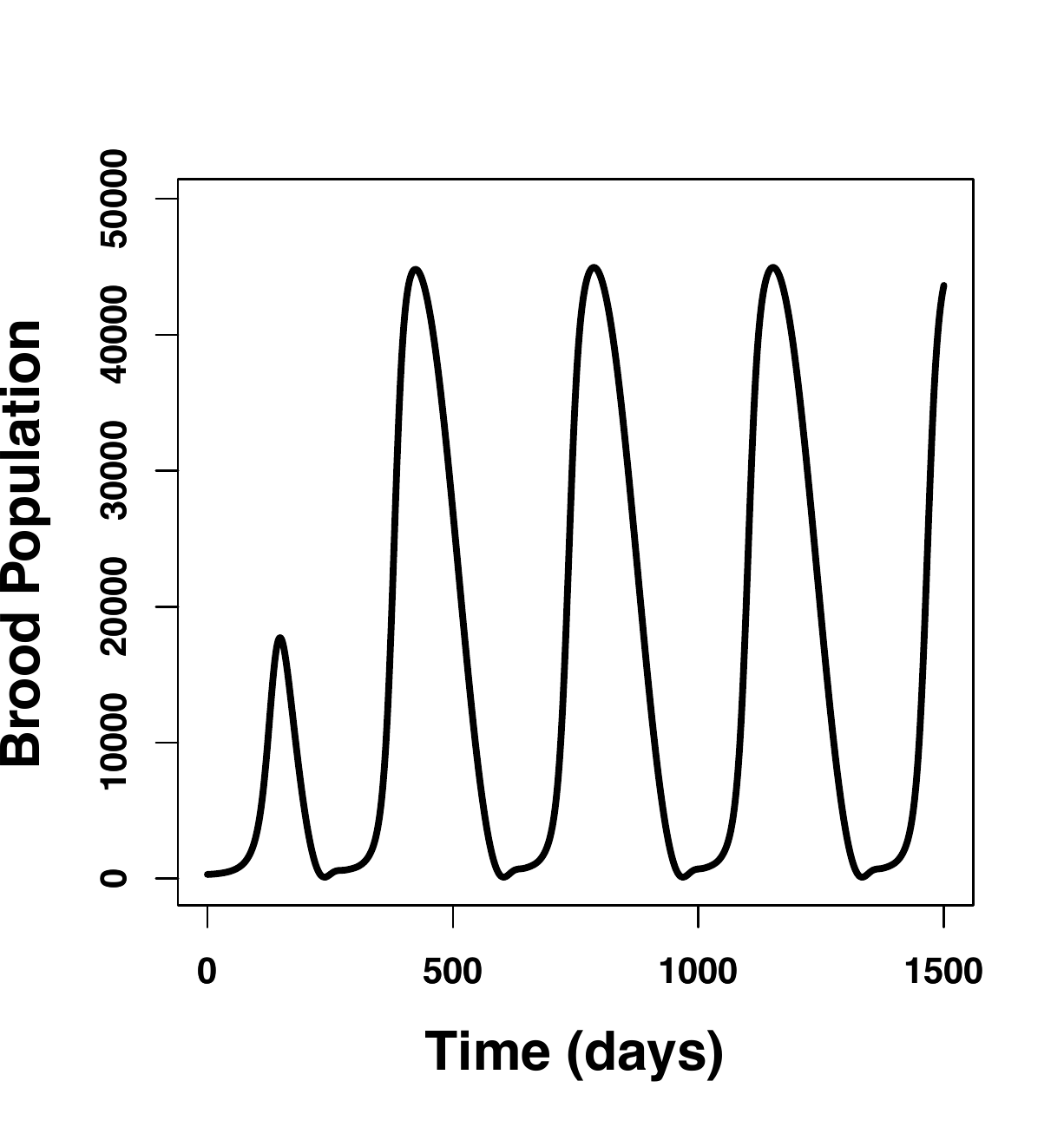}
\caption{Brood population of Model (\ref{BH}) with seasonality}
\label{fig:m1sB}
    \end{subfigure}
    \begin{subfigure}{0.24\textwidth}
        \includegraphics[width=\textwidth]{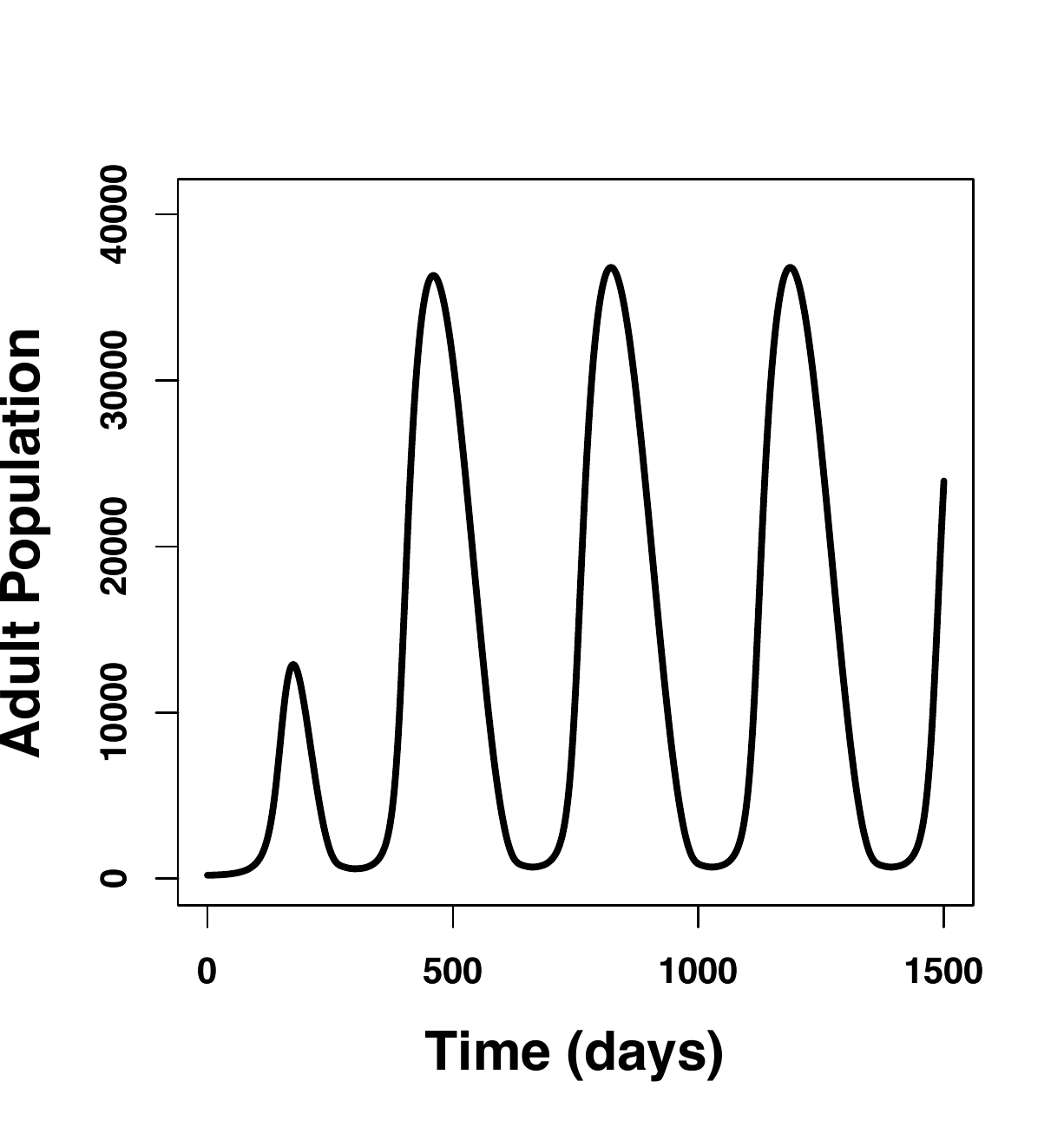}
\caption{Adult population of Model (\ref{BH}) with seasonality}
\label{fig:m1sH}
    \end{subfigure}
	\caption{Colony dynamic of simulation for Model (\ref{BH}) collapses without seasonality while survives with seasonality: $r=1200, K=5.4*10^6, d_b=0.01, d_h=0.05, \alpha=10,  \gamma=365, \psi=45;$ \red{$B(\theta)=300$ and $H(\theta)=200$, $\theta\in[-\tau,0]$ }.}\label{simulation-season}
\end{figure}   

{On the other hand, Figure \ref{simulation-season2} shows that seasonability can make honeybee colony collapse. Figures \ref{fig:m1Bs} \& \ref{fig:m1Hs} has no seasonality by  taking $r=1200, K=1*10^6, d_b=0.06, d_h=0.11, \alpha=10$ with a constant history function $B(\theta)=6125;H(\theta)=5362$ for all $\theta\in[-\tau, 0]$, which shows that honeybee colony could survive. While Figure \ref{fig:m1sBex} \& \ref{fig:m1sHex} has seasonality by taking  $r= r_0*(1 + \cos(\frac{2\pi(t -45)}{365}))$  with $r_0=1200$. In this case, we can see that seasonality may suppress the survival of honeybee colony.\\ }

  \begin{figure}[ht]
 \centering  
    \begin{subfigure}{0.24\textwidth}
        \includegraphics[width=\textwidth]{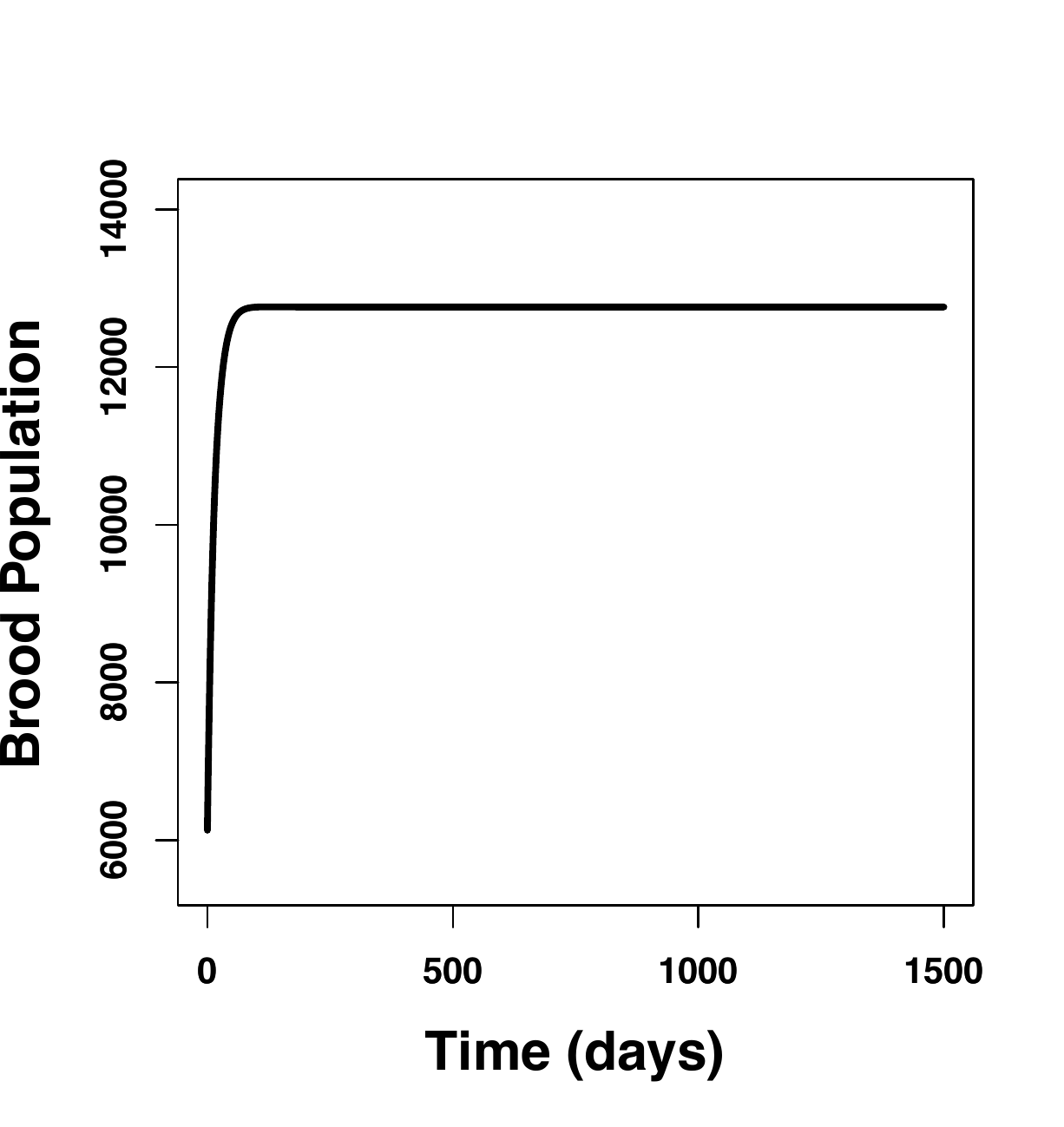}
\caption{Brood population of Model (\ref{BH}) without seasonality}
\label{fig:m1Bs}
    \end{subfigure}
    \begin{subfigure}{0.24\textwidth}
        \includegraphics[width=\textwidth]{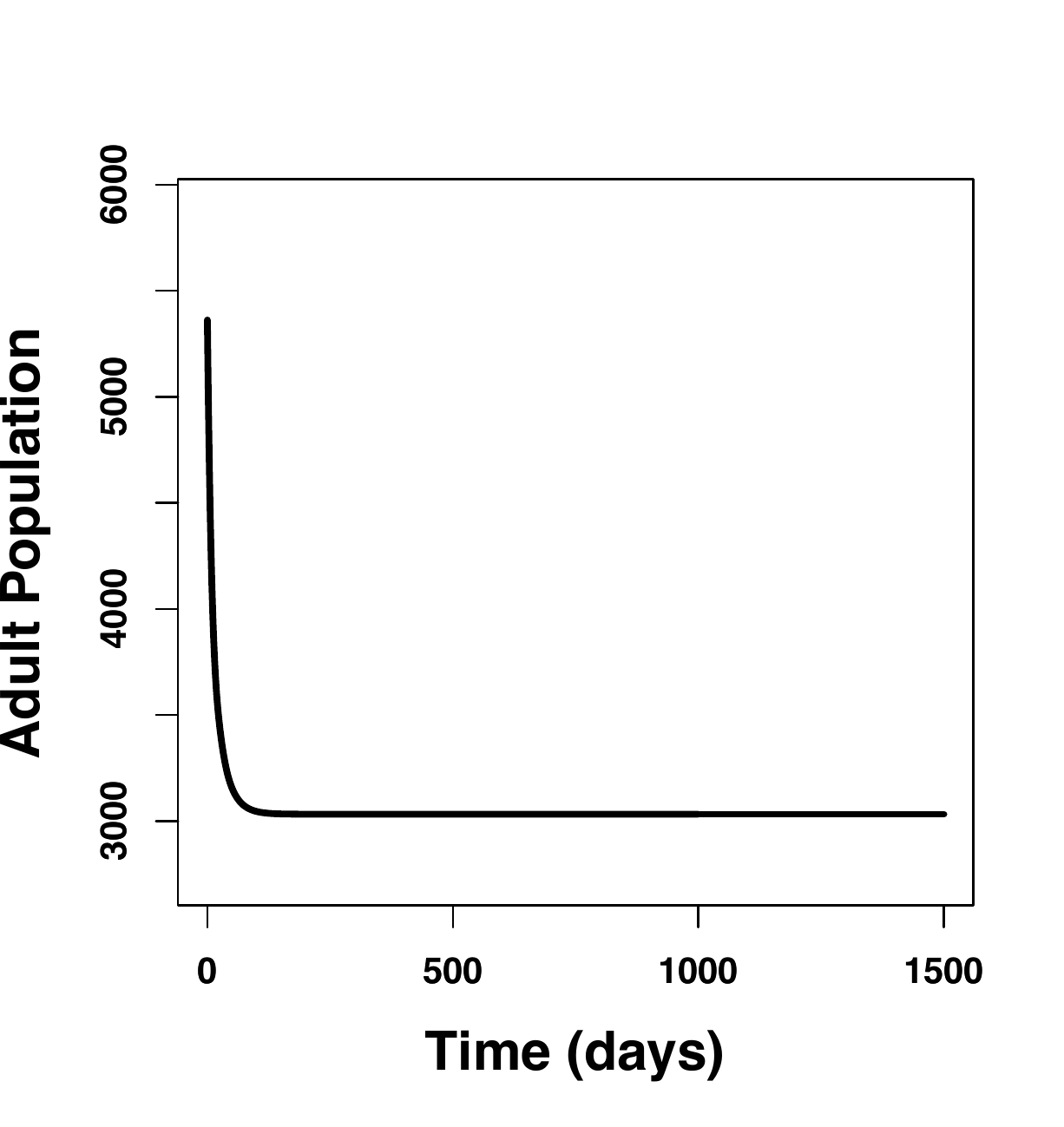}
\caption{Adult population of Model (\ref{BH}) without seasonality}
\label{fig:m1Hs}
    \end{subfigure}
    \begin{subfigure}{0.24\textwidth}
        \includegraphics[width=\textwidth]{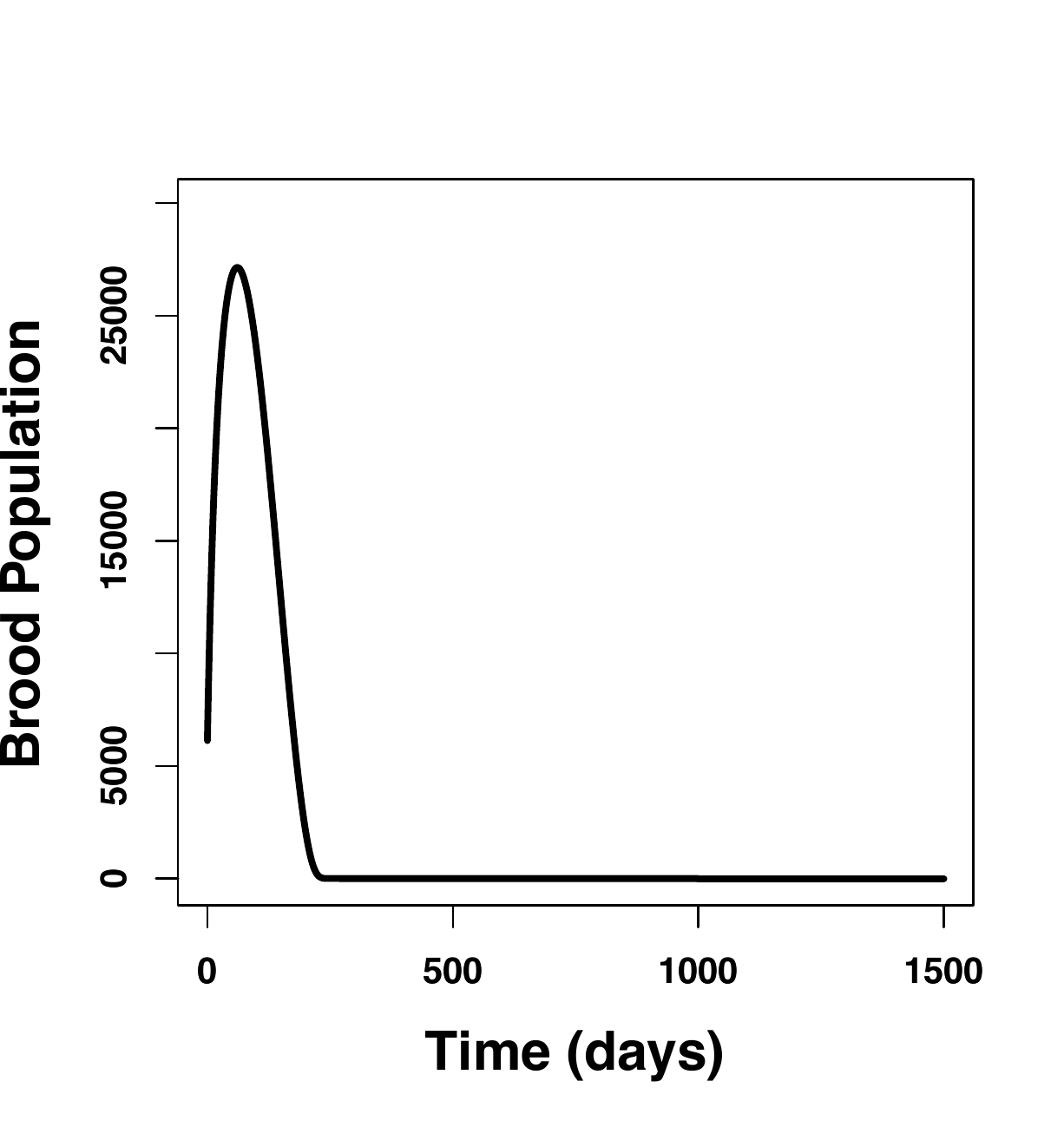}
\caption{Brood population of Model (\ref{BH}) with seasonality}
\label{fig:m1sBex}
    \end{subfigure}
    \begin{subfigure}{0.24\textwidth}
        \includegraphics[width=\textwidth]{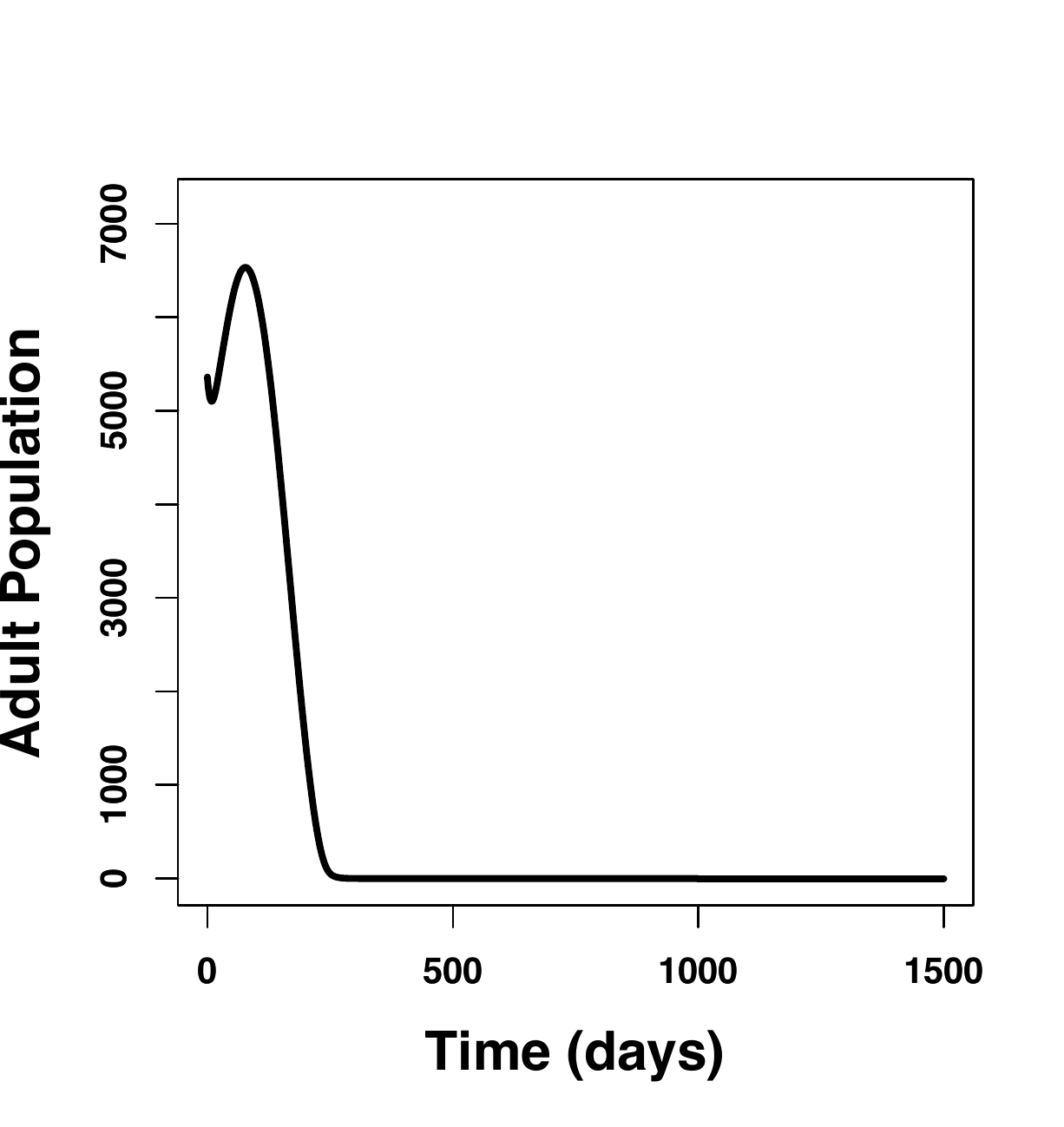}
\caption{Adult population of Model (\ref{BH}) with seasonality}
\label{fig:m1sHex}
    \end{subfigure}  
	\caption{Simulations for Model (\ref{BH}) survives without seasonality while collapes with seasonality: $r=1200, K=1*10^6, d_b=0.06, d_h=0.11, \alpha=10, \gamma=365, \psi=45;$ \red{$B(\theta)=300$ and $H(\theta)=200$, $\theta\in[-\tau,0]$ .}}\label{simulation-season2}
\end{figure}

\section{Conclusions}

Honeybees have dramatical decreased population over the long-term and each year \cite{smith2013pathogens}. As a result, great economic losses and the increase in the price of bee products have adversely affected the market \cite{smith2013pathogens, USDA2017}. The causes of the decline in the number of honeybees have been the great interests, whether they may directly link to human, environmental or disease \cite{perry2015rapid,oldroyd2007s,smith2013pathogens}. Some previous work always focus on foragers or recruitment, other works investigated external causes \cite{khoury2011quantitative,khoury2013modelling,russell2013dynamic,perry2015rapid,kang2015ecological,aronstein2012varroa}. In this study, we focus on modeling proper honeybee population age structure model with model validation using empirical data to obtain better biological understanding of the critical factors that could maintain healthy of honeybee colonies. \\


We propose two different models with age structure to explore the importance of proper modeling.  The first model \eqref{BH} has an assumption that the adult bees are surviving from eggs while the second model \eqref{BH2} assumes that adult bees are survived from brood stage rather than the egg stage. Our theoretical work (see Theorem \ref{th2:Ee}, \ref{th1:stability}, \ref{BaiBH2-00Stability} \& \ref{BaiBH2-inEStability}  )  implies that Model \eqref{BH} and Model \eqref{BH2} have huge differences in their dynamics. Specifically, Model \eqref{BH}  has only equilibrium dynamics and the maturation time doesn't affect its dynamics (see Theorem \ref{th1:stability} ) while Model \eqref{BH2} can be destabilized by the maturation time along with its life history parameter values (see Theorem \ref{BaiBH2-inEStability}) and Model \eqref{BH2}  is not positively invariant. Also our bifurcation diagrams (see Figure \ref{fig:bif1} \& \ref{fig:bif2})  confirmed such different dynamical outcomes. Both theoretical and bifurcation results indicates that the different assumptions can lead to different age structure models with dramatic dynamical outcomes. So which model would be more appealing and biological relevant? Can we say the second model \eqref{BH2} be better as it has oscillatory dynamics that could be supported by seasonality observed in data?\\

Given that the queen reproduction depends on seasonality \cite{seasonality,bodenheimer1937studies,khoury2011quantitative}, this suggests that it is paramount to include seasonality when modeling honeybee population dynamics. To address whether the seasonal pattern observed from data is due to the internal factor such as the maturation time  and/or other life history parameters (for example, Model \eqref{BH2} could be a better model for generating seasonal patterns from the internal factors) or the external factor such as the queen egg laying rate that is regulated by the temperature and the resource \cite{degrandi1989beepop,bodenheimer1937studies,coffey2007parasites}. We use data to validate Model \eqref{BH}, and Model \eqref{BH2} by assuming that the queen egg laying rate is constant and seasonal. Our validations on models without seasonality did not have a good fit by comparing to the corresponding models with seasonality. Among all models with seasonality, Model \eqref{BH} has the best fit (see Figure \ref{fig:with}). Our model validations with data suggest that the seasonal pattern observed from data is very likely due to the external factor such the temperature or available resources that may generate periodic dynamics in the queen egg laying rate while the internal factor such as the maturation time doesn't seem to be responsible seasonal pattern observed from data.\\


Both theoretical and numerical results including model validations suggest that Model \eqref{BH} with seasonality in the queen egg laying rate seems to be the most fit model for studying honeybee population dynamics with age structure. Theoretical results (Theorem \ref{th1:stability} ) and bifurcation diagrams (Figure \ref{fig:bif1}) imply that (1) the survival of honeybee colonies requires a large value of the queen egg-laying rate ($r$) and smaller values of the other life history parameter values in addition to the proper initial condition; (2) both brood and adult bee population is increasing with respect to the {egg-laying rate $r$ and is decreasing with respect to the regulation effects of brood $\alpha$, the square of half maximum of colony size at which brood survival rate $K$, and the mortality rates $d_b$, $d_h$}; and (3) seasonality may promote the survival of the honeybee colony (see Figure \ref{simulation-season}) but also may lead to the colony collapsing (see Figure \ref{fig:m1sBex}\&\ref{fig:m1sHex}). In summary, our work suggests that Model (\ref{BH}) with seasonality could be used for our future model that includes more external factors, such as diseases, parasite, food and human activities \cite{perry2015rapid,oldroyd2007s,smith2013pathogens,khoury2013modelling}. Our ongoing work has extended the current model \eqref{BH}  to include parasites.\\

\section{Proofs}

\subsection*{Proof of Theorem \ref{th1:pb}}
\begin{proof}
First, we look at the following equation that describes the population of adult bees:
\begin{equation}\label{H}
\begin{array}{lcl}
\frac{dH}{dt}&=& e^{- d_b \tau} \frac{rH(t-\tau) ^2}{K+H(t-\tau)^2+\alpha B(t-\tau)}-d_h H\\
\end{array}
\end{equation}
Since $H(t)$ is a nonnegative continuous function during the time $t\in [-\tau,0]$, the equation \eqref{H} implies that
$$\frac{dH}{dt}\geq -d_h H \mbox{ for time } t\in [0,\tau] \Rightarrow H(t)\geq H(0)e^{-d_h t} \geq 0  \mbox{ for time } t\in [0,\tau] .$$
By deduction on intervals $[(n-1)\tau, n\tau], n\geq 1]$, we could show that $H(t)\geq 0$.\\

By integration, we could set
\begin{equation}\label{Bt}
\begin{array}{lcl}
B(t)&=&\int_{t-\tau}^{t} \left[\frac{rH(s)^2}{K+H(s)^2+\alpha B(s)}e^{-d_b\left(t-s\right)}\right]ds
 - B(0)e^{-d_bt}\end{array}\end{equation}
 which gives
$\frac{dB}{dt}=\frac{rH(t)^2}{K+H(t)^2+\alpha B(t)}-d_b B- e^{- d_b \tau} \frac{rH(t-\tau) ^2}{K+H(t-\tau)^2+\alpha B(t-\tau)} $ and $B(0)=\int^0_{-\tau}\frac{rH^2(s)e^{d_bs}}{K+H^2(s)+\alpha B(s)}ds$. Thus, the equation \eqref{Bt} provides an explicit mathematical expression of $B(t)$ which is nonnegative for all time $t\geq0$. Thus, the state space $\mathbb X$ of the proposed model \eqref{BH} is positive invariant.\\

To show the boundedness of the model, define $V= B+H$, then we have
$$\begin{array}{lcl}
\frac{dV}{dt}= \frac{dB}{dt}+\frac{dH}{dt}
&=& \frac{rH^2}{K+H^2+\alpha B}-d_bB-d_hH\\
&\leq& r- \min\{d_b,d_h\}(B+H)=r-d_{\mbox{min}} V
\end{array}$$
with $d_{\mbox{min}} =\min\{d_b,d_h\}$. Consequently, we have
$$\limsup_{t\rightarrow\infty} V(t)=\limsup_{t\rightarrow\infty}(B(t)+H(t))\leq \frac{r}{d_{\mbox{min}}}$$which implies that Model \eqref{BH} is  bounded in $\mathbb X$.

\end{proof}

\subsection*{Proof of Theorem \ref{th2:Ee}}
\begin{proof}
We linearize the Model \ref{BH}:
{\footnotesize
\begin{equation}
\begin{aligned}
D  \left(  \left[ \begin{array}{c}
                 \dot{B}(t)\\
                 \dot{H}(t)
                \end{array}
            \right] \right)
            \Bigg\vert_{(B^*,H^*)}
                       &= \left[ \begin{array}{lcr}
               -\frac{\alpha  H^2 r}{\left(\alpha  B+H^2+K\right)^2}-d_b & \frac{2 H r (\alpha  B+K)}{\left(\alpha  B+H^2+K\right)^2}\\
               0 &  -d_h
               \end{array} \right]
	      \left[ \begin{array}{c}
	              B(t)\\
	              H(t)
	             \end{array} \right]\\
	    &+  \left[ \begin{array}{lcr}
		  \frac{\alpha  H^2 r e^{-d_b \tau }}{\left(\alpha  B+H^2+K\right)^2} & -\frac{2 H r e^{-d_b \tau } (\alpha  B+K)}{\left(\alpha  B+H^2+K\right)^2} \\
		    -\frac{\alpha  H^2 r e^{-d_b \tau }}{\left(\alpha  B+H^2+K\right)^2}& \frac{2 H r e^{-d_b \tau } (\alpha  B+K)}{\left(\alpha  B+H^2+K\right)^2}
		  \end{array} \right]
		  \left[ \begin{array}{c}
		     B(t-\tau)\\
	             H(t-\tau)
			\end{array} \right].
			\label{JE1}
	   \end{aligned}
	   \end{equation}
}

For extinction equilibrium, the matrix \eqref{JE1} gives:
\begin{align*}
D \left(  \left[ \begin{array}{c}
                 \dot{B}(t)\\
                 \dot{H}(t)
                \end{array}
            \right] \right)
              \Bigg\vert_{(0,0)}
                      = \left[ \begin{array}{lcr}
               -d_b &0\\
               0 & -d_h
              \end{array} \right]
	      \left[ \begin{array}{c}
	              B(t)\\
	              H(t)
	             \end{array} \right]
	    +  \left[ \begin{array}{lcr}
		  0 &0\\
		    0&0
		  \end{array} \right]
		  \left[ \begin{array}{c}
		     B(t-\tau)\\
	             H(t-\tau)
			\end{array} \right]
\end{align*}
 and from this we obtain the following eigenvalues:
 $$\lambda_1 =-d_b<0,~~\lambda_2 =-d_h<0.$$
Thus, we can conclude that $E_e$ is always locally asymptotically stable.   \\

Now we show its global stability as follows. Let $(B(t), H(t))$ be a solution of Model \eqref{BH}, then $(B(t), H(t))$ is bounded and positive for all $t>0$ by Theorem \ref{th1:pb}.
Define $\limsup_{t\to\infty}H(t)=H^{\infty}$.  By Lemma B in \ref{lemma-fluct}, there exists sequence $\{t_n\} \uparrow \infty$ such that $\lim_{n\to\infty}H(t_n)=H^{\infty}$,  and $\lim_{n\to\infty}H'(t_n)=0$, and same for $B(t)$. For any $\epsilon>0$, there exists $N$ such that for $n>N$, $H(t_n-\tau)\le H^{\infty}+\epsilon$ holds. Thus, according to Model \eqref{BH}, for $n>N$ we have
 \begin{eqnarray*}
 \begin{array}{ll}
H'(t_n)&= e^{- d_b \tau} \frac{rH(t_n-\tau) ^2}{K+H(t_n-\tau)^2+\alpha B(t_n-\tau)}-d_h H(t_n)\\
&\leq e^{- d_b \tau} \frac{rH(t_n-\tau) ^2}{K+H(t_n-\tau)^2}-d_h H(t_n)\\
&\leq e^{-d_b \tau}\frac{r(H^\infty+\epsilon)^2}{K+(H^\infty+\epsilon)^2}-d_h H(t_n).
\end{array}
\end{eqnarray*}
Let $n\to \infty$, we get
  \begin{eqnarray*}
0\leq e^{-d_b \tau}\frac{r(H^\infty+\epsilon)^2}{K+(H^\infty+\epsilon)^2}-d_h H^{\infty}.
\end{eqnarray*}

It follows by the arbitrariness of $\epsilon$ that
\begin{eqnarray*}
H^{\infty}\left(-d_h (H^\infty)^2+ e^{-d_b \tau} r H^\infty-d_h K \right)\ge 0.
\end{eqnarray*}
Then $d_h > \frac{re^{-d_b\tau}}{2\sqrt{K}}$, we know that for all $x, y\in \mathbb{R}$, $-d_h x^2+ e^{-d_b \tau} r x-d_h K<0$. Therefore, $H^{\infty}=0$, and hence from the positivity of solution we have $\lim_{t\to\infty}H(t)=0$. Furthermore, we obtain the following limiting equation
\begin{eqnarray*}
\frac{dB}{dt}=-d_b B(t),
\end{eqnarray*}
which implies that $\lim_{t\to\infty}B(t)=0$. Therefore, $E_{e}=(0,0)$  is globally asymptotically stable.\\
\end{proof}

\subsection*{Proof of Theorem \ref{th1:stability}}
\begin{proof}
Let $(B^*,H^*)$ be an equilibrium of Model \eqref{BH}. Then the linearization of the proposed model \eqref{BH} at the equilibrium $(B^*,H^*)$ is shown as follows:\\

In the case of the interior equilibrium $(B^*,H^*)=E_{i}$, from Equation \eqref{ht1}, we have
$$\frac{2rKH^*}{(K+(H^*)^2)^2}=\frac{rH^*}{(K+(H^*)^2)}\frac{2K}{(K+(H^*)^2)}=\frac{2Kd_he^{d_b\tau}}{(K+(H^*)^2)}.$$

The characteristic equation \eqref{th1:ch} evaluated at the positive interior equilibrium $(B^*,H^*)$ when $\alpha=0$ gives the following expression:

\begin{align*}
C(\lambda)
                     &= det \left(  \left[ \begin{array}{lcr}
               -d_b & \frac{2rKH^*}{(K+(H^*)^2)^2}\\
               0 &  -d_h
              \end{array} \right]
	    +  \left[ \begin{array}{lcr}
		  0 & - \frac{2rKH^*}{(K+(H^*)^2)^2}e^{-d_b\tau}\\
		    0& \frac{2rKH^*}{(K+(H^*)^2)^2}e^{-d_b\tau}
		  \end{array} \right]
		  \ast e^{-\lambda\tau} - \lambda \mathscr{I} \right)\\
		  &=det \left( \left[ \begin{array}{lcr}
               -d_b -\lambda & \frac{2rKH^*}{(K+(H^*)^2)^2}(1-e^{-(\lambda+d_b)\tau})\\ 0 & -d_h+\frac{2rKH^*}{(K+(H^*)^2)^2}e^{-(\lambda+d_b)\tau}-\lambda
              \end{array}\right]\right.\\
               &=( -d_b -\lambda)\left(-d_h+\frac{2rKH^*}{(K+(H^*)^2)^2}e^{-(\lambda+d_b)\tau}-\lambda\right)\\
               &=( -d_b -\lambda)\left(-d_h+\frac{2Kd_he^{d_b\tau}}{K+(H^*)^2}e^{-(\lambda+d_b)\tau}-\lambda\right)
               \end{align*}where $H^*=\frac{re^{-d_b\tau}\left(1\pm\sqrt{1-\left(\frac{2d_he^{d_b\tau}}{r}\right)^2K}\right)}{2d_h}.$ This implies that the stability of the interior equilibrium $(B^*,H^*)$ is determined by the eigenvalues of the following equation evaluated at $H^*$ since $\lambda=-d_b<0$
               $$-d_h+\frac{2Kd_h}{K+(H^*)^2}e^{-\lambda\tau}-\lambda=0\Leftrightarrow \lambda=-d_h+\frac{2Kd_h}{K+(H^*)^2}e^{-\lambda\tau}.$$

Let $A=-d_h$ and $B(H)=\frac{2d_h K}{K+H^2}$, then we have $B(H)>0>A$. At the mean time, we have

$$0<H^*_1<H^*_2 \mbox{ and } H^*_1H^*_2=K \Rightarrow  (H^*_1)^2<K \mbox{ and } (H^*_2)^2>K.$$
Therefore, we can obtain the following inequalities
$$A+B(H^*_1)=-d_h+\frac{2Kd_h}{K+(H_1^*)^2}>0 \mbox{ and }A+B(H^*_2)=-d_h+\frac{2Kd_h}{K+(H_2^*)^2}<0.$$
By applying Theorem 4.7 from Hal Smith \cite{smith2010introduction}, we can conclude that the interior equilibrium $E_1$ is always unstable and $E_2$ is always  locally asymptotically stable for any delay $\tau>0$.\\
\end{proof}
\subsection*{Proof of Theorem \ref{th4:stabilityctrical}}
\begin{proof}
According to Proposition \ref{p:eq1}, we know that Model \eqref{BH} has a unique interior $E=(B^{*}, H^{*})=\left( \frac{r\left(1 - e^{-d_b\tau}\right)}{2d_b},\sqrt{K}\right)$ when $d_h = \frac{re^{-d_b\tau}}{2\sqrt{K}}$. In order to study its stability, we define the following matrices:

\begin{eqnarray}\label{matrixUV}
U=
\left[ \begin{array}{cc}
-d_b & \frac{r}{2\sqrt{K}} \\
0 & -d_h
\end{array}
\right] \quad \mbox{and} \qquad V=\left[
\begin{array}{cc}
0 & -d_h \\
0 & d_h
\end{array} \right].
\end{eqnarray}
Let $L(\lambda)$ to be represented as follows
\begin{eqnarray}\label{zerochar}
L(\lambda)=\lambda+d_h-d_h e^{-\lambda\tau}
\end{eqnarray}
which has $\lambda=0$ as one of its eigenvalues. By applying Lemma A in \ref{Brauer Lemmas} to \eqref{zerochar}, except for the root $\lambda=0$, all roots of \eqref{zerochar} has negative real parts for all $0 \le \tau < \infty$. Since \eqref{zerochar} has a simple zero eigenvalue, we need to use the center manifold and normal form theory in \cite{faria1995normal} to obtain the local stability of $E=\left( \frac{r\left(1 -e^{-d_b\tau}\right)}{2d_b},\sqrt{K}\right)$.\

Let $\tilde{B}=B-\frac{r\left(1 - e^{-d_b\tau}\right)}{2d_b}$, $\tilde{H}=H-\sqrt{K}$, and still denote $\tilde{B}=B, \tilde{H}=H$, then the system \eqref{BH} becomes
\begin{eqnarray}\label{BH2Taylor}
\left\{\begin{array}{ll}
\frac{dB}{dt}&=-d_b B(t)+\frac{r}{2\sqrt{K}}H(t)-d_h H(t-\tau)-\frac{r}{2K}H^2(t)+d_h\sqrt{K} H^2(t-\tau)+O(3),\\
\frac{dH}{dt}&=-d_h H(t)+d_h H(t-\tau) -d_h\sqrt{K} H^2(t-\tau)+O(3).
\end{array}
\right.
\end{eqnarray}

Let $\Lambda=\{0\}$. From normal form theory in  \cite{faria1995normal}, there exists a one-dimension ordinary differential equation which has the same dynamical property as \eqref{BH2Taylor} near $0$. Rewriting \eqref{BH2Taylor} as
\begin{eqnarray}\label{BH2FDE}
\dot{z}(t)=l(z_t)+F(z_t)
\end{eqnarray}
where
$$z_t(\theta)=z(t+\theta)\in \mathcal{C}:=([-\tau,0],\mathbb{R}_+^2),$$
$\mathcal{C}$ is the phase space with the norm $|\phi|=\max_{-\tau\le \theta\le 0}|\phi(\theta)|$, and
\begin{eqnarray*}
l(\phi)=U\phi(0)+V\phi(-\tau),
\end{eqnarray*}
where $U$ and $V$ are given in \eqref{matrixUV}, and
\begin{eqnarray*}
F(\phi)=\left[\begin{array}{cc}
-\frac{r}{2K}\phi_2^2(0)+d_h\sqrt{K} \phi_2^2(-\tau)+O(3)\\
-d_h\sqrt{K} \phi_2^2(-\tau)+O(3)
\end{array}
\right].
\end{eqnarray*}
Take
$
\mu(\theta)=U\delta(\theta)-V\delta(\theta+\tau)$,
where
\begin{eqnarray*}
\delta(\theta)=\left\{\begin{array}{ll}
1,\quad \theta=0,\\
0,\quad \theta\neq 0.
\end{array}
\right.
\end{eqnarray*}
Then we have follows
\begin{eqnarray*}
l(\phi)=\int_{-\tau}^{0} d\mu(\theta)\phi(\theta).
\end{eqnarray*}
By the adjoint theory of FDE, the phase space $\mathcal{C}$ can be decomposed as $\mathcal{C}=P\oplus Q$, where $P=span\{\Phi(\theta)\}, \Phi(\theta)=(1,\beta)^T$, $\beta=\frac{d_b}{\frac{r}{2\sqrt{K}}-d_h}$. Taking the base $\Psi$ of adjoint space $P^*$ of $P$ satisfies $\langle\Psi,\Phi\rangle=1$, where $\langle\cdot,\cdot\rangle$ is a bilinear function defined in $\mathcal{C}^*\times \mathcal{C}$ by
\begin{eqnarray*}
\langle\psi,\phi\rangle=\psi(0)\phi(0)-\int_{-\tau}^{0} \int_0^s \psi(\theta-s)d\mu(s)\phi(\theta)d\theta.
\end{eqnarray*}
By a direct computation, we have
\begin{eqnarray*}
\Psi(s)=\left(0, (\beta (1+\tau d_h))^{-1}\right).
\end{eqnarray*}
Consider the following expand space
\begin{eqnarray*}
\mathcal{B}\mathcal{C}=\left\{\phi| \ \phi:[-\tau, 0]\to \mathcal{C}, \phi\  \mbox{is continuous in} \  [-\tau, 0) \ \mbox{and} \ \lim\limits_{\theta\to 0}\phi(\theta) \ \mbox{exists}\right\},
\end{eqnarray*}
then the abstract ODE in $\mathcal{B}\mathcal{C}$ associated with FDE \eqref{BH2FDE} can be written as the form
\begin{eqnarray}\label{AbstrODE}
\frac{d}{dt}u=Au+X_0F(u),
\end{eqnarray}
where
\begin{eqnarray*}
A\phi=\dot{\phi}+X_0(l(\phi)-\dot{\phi}(0)),\ \phi\in C^1([-\tau,0],\mathbb{R}_+^2),
\end{eqnarray*}
and
\begin{eqnarray*}
X_0=\left\{\begin{array}{ll}
I,\quad \theta=0,\\
0,\quad \theta\in [-\tau,0).
\end{array}
\right.
\end{eqnarray*}
The projection mapping $\pi: \mathcal{B}\mathcal{C}\to P$:
 \begin{eqnarray*}
\pi(\phi+X_0\alpha)=\Phi(\langle\Psi,\phi\rangle+\Psi(0)\alpha),
\end{eqnarray*}
leads to the decomposition $\mathcal{B}\mathcal{C}=P\oplus\mbox{Ker}{\pi}$. Decomposing $u$ in Equation \eqref{AbstrODE} in the form of  $u=\Phi x+y$ where $x\in\mathbb{R}$ and $y\in Q^1:=\mbox{Ker}{\pi}\cap D(A)=Q\cap C^1$. Then \eqref{AbstrODE} is equivalent to the system
\begin{eqnarray*}
\left\{\begin{array}{ll}
\dot{x}=\Psi(0)F(\Phi x+y),\\
\dot{y}=A_{Q^1}y+(I-\pi)X_0F(\Phi x+y).
\end{array}
\right.
\end{eqnarray*}
with
\begin{eqnarray*}
\Psi(0)F(\Phi x+y)=\frac{-d_h\sqrt{K}}{\beta(1+\tau d_h)}(\beta x+y_2(-\tau))^2+O(3).
\end{eqnarray*}
 Thus, the local invariable manifold of \eqref{BH2Taylor} at $0$ with the tangency with the space $P$ satisfies $y(\theta)=0$, the flow on this manifold is given by the following one-dimension ODE
 \begin{eqnarray}\label{flow}
\dot{x}(t)=\frac{-\beta d_h\sqrt{K}}{1+\tau d_h} x^2(t)+O(3).
\end{eqnarray}
This implies that the zero solution of \eqref{flow} is stable. Therefore, the interior equilibrium $E=(B^{*}, H^{*})$ is locally asymptotically stable for all $\tau>0$. The proof is complete.\\
\end{proof}

\subsection*{Proof of Theorem \ref{BaiBH2-00Stability}}
\begin{proof}
We linearize the equations of Model \ref{BH2}:
{\footnotesize
\begin{equation}
\begin{aligned}
D  \left(  \left[ \begin{array}{c}
                 \dot{B}(t)\\
                 \dot{H}(t)
                \end{array}
            \right] \right)
            \Bigg\vert_{(B^*,H^*)}
                       &= \left[ \begin{array}{lcr}
               -d_b & \frac{2rKH^*}{(K+(H^*)^2)^2}\\
               0 &  -d_h
               \end{array} \right]
	      \left[ \begin{array}{c}
	              B(t)\\
	              H(t)
	             \end{array} \right]
	    +  \left[ \begin{array}{lcr}
		  -e^{-d_b\tau} & 0 \\
		    e^{-d_b\tau}& 0
		  \end{array} \right]
		  \left[ \begin{array}{c}
		     B(t-\tau)\\
	             H(t-\tau)
			\end{array} \right].
			\label{JE}
	   \end{aligned}
	   \end{equation}
}

The matrix \eqref{JE} evaluated at the extinction equilibrium $E_e$  gives the characteristic equation

\begin{align*}
C(\lambda,\tau)
                     &= det \left(  \left[ \begin{array}{lcr}
               -d_b & \frac{2rKH^*}{(K+(H^*)^2)^2}\\
               0 &  -d_h
              \end{array} \right]
	    +  \left[ \begin{array}{lcr}
		 -e^{-d_b\tau} & 0 \\
		    e^{-d_b\tau}& 0
		  \end{array} \right]
		  \ast e^{-\lambda\tau} - \lambda \mathscr{I} \right)\\
		  &=det \left( \left[ \begin{array}{lcr}
               -d_b -e^{-d_b\tau}e^{-\lambda\tau}-\lambda & \frac{2rKH^*}{(K+(H^*)^2)^2}\\ e^{-d_b\tau}e^{-\lambda\tau} & -d_h-\lambda
              \end{array}\right]\right)\\
              &= ( -d_h -\lambda)(-d_b-e^{-d_b\tau}e^{-\lambda\tau}-\lambda)-\frac{2rKH^*}{(K+(H^*)^2)^2}e^{-(d_b+\lambda)\tau}
\end{align*}

At $(0,0)$, $C(\lambda,\tau)=( -d_h -\lambda)(-d_b-e^{-d_b\tau}e^{-\lambda\tau}-\lambda)$. Clearly, one characteristic root is $\lambda=-d_h<0$, others are the roots of the following equation
\begin{align}\label{Char-BH2}
\lambda+d_b+e^{-d_b\tau}e^{-\lambda\tau}=0.
\end{align}

When there is no delay, i.e., $\tau=0$, \eqref{Char-BH2} has only a negative characteristic root $\lambda==-d_b$, Model \ref{BH2} is asymptotically stable at $(0,0)$. Moreover, for every $\tau\ge 0$, \eqref{Char-BH2} has no nonnegative real root.

We assume $\lambda = iw$, $w >0$, is a root of \eqref{Char-BH2} for some $\tau>0$. Then, we have 
\begin{align}\label{cos-sin}
\cos(w\tau)=-d_b e^{d_b\tau}, \quad \sin(w\tau)=w e^{d_b\tau},
\end{align}
which gives 
\begin{align}\label{Imag-w}
w^2=e^{-2d_b \tau}-d_b^2.
\end{align}
It is clear that \eqref{Imag-w} has a positive real root
\begin{align}\label{Imag-root}
w=(e^{-2d_b \tau}-d_b^2)^{\frac{1}{2}}.
\end{align} 
if and only if $e^{-d_b \tau}>d_b$, i.e., $\tau<\tau^*:=\frac{1}{d_b}\ln\left(\frac{1}{d_b}\right)$, and $0<d_b<1$.

Notice that $\cos(w\tau)<0, \sin(w\tau)>0$, there is a unique $\theta, \frac{\pi}{2} <\theta <\pi$, such that $w\tau = \theta$  makes both equation of \eqref{cos-sin} hold.  Then, if $e^{-d_b \tau}>d_b$, we get a set of values of $\tau$ for which there are imaginary
roots:
\begin{align}\label{tau-k}
\tau_k=\frac{\theta+k\pi}{w},\quad k=0,1,2,\cdots,
\end{align} 
where $w$ is given by \eqref{Imag-root} and 
\begin{align}\label{theta}
\theta=\pi-\arctan\left(\frac{w}{d_b}\right).
\end{align} 

From \eqref{Char-BH2}, we have
\begin{align*}
(1-\tau e^{-(d_b+\lambda)\tau})\frac{d\lambda}{d\tau}=(\lambda+d_b)e^{-(d_b+\lambda)\tau},\quad \text{and}\quad \lambda+d_b=-e^{-(d_b+\lambda)\tau}.
\end{align*}
Thus,
\begin{align*}
\left(\frac{d\lambda}{d\tau}\right)^{-1}=-\frac{\tau}{\lambda+d_b}-\frac{1}{(d_b+\lambda)^2},
\end{align*}
 and
 \begin{align*}
S(\tau):&=\text{sign}\left\{\left(\frac{d(\text{Re}\lambda)}{d\tau}\right)\right\}_{\lambda=iw}
=\text{sign}\left\{\text{Re}\left(\frac{d\lambda}{d\tau}\right)^{-1}\right\}_{\lambda=iw}\\
&=\text{sign}\left\{-\text{Re}\frac{\tau}{\lambda+b}-\text{Re}\frac{1}{(b+\lambda)^2}\right\}_{\lambda=iw}\\
&=\text{sign}\{-\tau d_b^3 -d_b^2+w^2(\tau d_b-1)\}\\
&=\text{sign}\{-2d_b^2+(1-\tau d_b)e^{-2d_b \tau}\}\\
&=\text{sign}\{\phi(\tau)\}.
\end{align*}
Here,
\begin{align}\label{phi}
\phi(\tau)=-2d_b^2+(1-\tau d_b)e^{-2d_b \tau}.
\end{align}
Clearly,
\begin{align}\label{phi-D}
\phi'(\tau)=e^{-2d_b \tau}d_b(2\tau d_b-3).
\end{align}

In order to get the stability of the equilibrium $(0,0)$, we first claim:
 \begin{itemize}
   \item For all $\tau\ge\max\{0,\tau^*\}$, $(0,0)$ is asymptotically stable.
 \end{itemize}
In fact, we rewrite \eqref{Char-BH2} as the form
$\lambda=A+B e^{-\lambda\tau}$ with $A=-d_b, B=-e^{-d_b\tau}$. For any $\tau\ge\tau^*$, $B\ge -e^{-d_b\tau^*}=-d_b=A$. Thus, by applying Theorem 4.7 from Hal Smith, $(0,0)$ is local asymptotically stable.\\

Now, we consider the following two cases.

{\bf Case 1.}  $d_b\ge \frac{\sqrt{2}}{2}$.

This case is divided into two subcases: (i) $d_b\ge 1$ and (ii) $\frac{\sqrt{2}}{2}\le d_b< 1$.

(i)\quad $d_b\ge 1$. 

In this subcase, the above claim implies that $(0,0)$ is asymptotically stable for all $\tau\ge 0$. This also can be proved as follows. For all $\tau\ge 0$, $e^{-2d_b \tau}-d_b^2<0$ holds, and hence the equation \eqref{Imag-w} has no positive real root. This implies that the equilibrium $(0,0)$ has no stability switch as $\tau$ increases in $[0,\infty)$. Since $(0,0)$ is asymptotically stable at $\tau=0$, then it remains asymptotically stable for all $\tau\ge 0$.

(ii)\quad $\frac{\sqrt{2}}{2}\le d_b< 1$. 

Assume $\tau\in (0,\tau^*)$. Then $e^{-2d_b \tau}-d_b^2>0$ and \eqref{Imag-w} has unique positive real root \eqref{Imag-root}. In such case, $\tau d_b<1$ since $\tau d_b<\tau^* d_b=\ln\left(\frac{1}{d_b}\right)\leq \ln\sqrt{2}<1$, $\phi(\tau)<-2d_b^2+(1-\tau d_b)\le -\tau d_b<0$. It follows that $S(\tau)=-1$ for $\tau\in (0,\tau^*)$, i.e., possible stability switches from unstable to stable may occur. Since $(0,0)$ is asymptotically stable at $\tau=0$, then it remains asymptotically stable for all $\tau\in (0,\tau^*)$.

When $\tau\ge \tau^*$, our claim shows that $(0,0)$ is asymptotically stable. Therefore, $(0,0)$ is asymptotically stable for all $\tau\ge 0$.

{\bf Case 2.}  $0<d_b< \frac{\sqrt{2}}{2}$.

This case is also divided into two subcases: (i) $\frac{1}{e}\le d_b< \frac{\sqrt{2}}{2}$ and (ii) $d_b< \frac{1}{e}$.

(i)\quad $\frac{1}{e}\le d_b< \frac{\sqrt{2}}{2}$. 

For $\tau\ge \tau^*$, we know that $(0,0)$ is asymptotically stable.
Now, we assume $\tau\in (0,\tau^*)$, then \eqref{Imag-w} has unique positive real root \eqref{Imag-root}. In such case, $\phi(0)=1-2d_b^2>0$ and $\phi(\tau^*)=-2d_b^2+(1-\tau^* d_b)e^{-2d_b \tau^*}=-2d_b^2+(1-\tau^* d_b)d_b^2<0$, and, from \eqref{phi-D}, $\phi'(\tau)<0$ since $\tau d_b<\tau^* d_b=\ln\left(\frac{1}{d_b}\right)\leq 1$. Thus, there exists unique $\tau_{c}\in (0,\tau^*)$ such that 
\begin{description}
  \item (a)\quad For $\tau\in (0,\tau_c)$, $\phi(\tau)>0$ and hence $S(\tau)=1$. In this case, possible stability switches from stable to unstable may occur as $\tau$ increases in $(0,\tau_c)$.
  \item (b)\quad At $\tau=\tau_c$, $\phi(\tau)=0$ and hence $S(\tau)=0$;
  \item (c)\quad For $\tau\in (\tau_c,\tau^*)$, $\phi(\tau)<0$ and hence $S(\tau)=-1$. In this case, possible stability switches from unstable to stable may occur as $\tau$ increases in $(\tau_c,\tau^*)$.
\end{description}

Therefore, by the stability of $(0,0)$ at $\tau=0$ and $\tau\ge\tau^*$, we can conclude that $(0,0)$ is asymptotically stable for $\tau\in (0,\tau_0)$ or $\tau\ge \tau_1$, while unstable for $\tau\in (\tau_0, \tau_1)$, where $\tau_0, \tau_1$ are given by \eqref{tau-k}.

(ii) $d_b< \frac{1}{e}$. 

Since $\frac{1}{d_b}>e$, then $\frac{1}{d_b}<\tau^*=\frac{1}{d_b}\ln\left(\frac{1}{d_b}\right)$. Thus, we has the following two scenarios:
\begin{description}
  \item (1) $\tau\in\left(\frac{1}{d_b}, \tau^*\right)$. Since $\tau d_b\ge 1$, we have $\phi(\tau)<0$ and hence $S(\tau)=-1$, which implies that
possible stability switches from unstable to stable may occur as $\tau$ increases in $\left(\frac{1}{d_b}, \tau^*\right)$.
  \item (2) $\tau\in\left(0, \frac{1}{d_b}\right)$. In such case, $\phi(0)=1-2d_b^2>0$ and $\phi\left(\frac{1}{d_b}\right)=-2d_b^2<0$, and $\phi'(\tau)<0$. Thus, there exists unique $\tau_c\in \left(0, \frac{1}{d_b}\right)$       \begin{description}
  \item (a)\quad For $\tau\in (0,\tau_c)$, $\phi(\tau)>0$ and hence $S(\tau)=1$. .
  \item (b)\quad At $\tau=\tau_c$, $\phi(\tau)=0$ and hence $S(\tau)=0$;
  \item (c)\quad For $\tau\in \left(\tau_c,\frac{1}{d_b}\right)$, $\phi(\tau)<0$ and hence $S(\tau)=-1$. .
\end{description}
\end{description}

Therefore, by the stability of $(0,0)$ at $\tau=0$ and $\tau\ge\tau^*$, we can conclude that $(0,0)$ is asymptotically stable for $\tau\in (0,\tau_0)$ or $\tau\ge \tau_1$, while unstable for $\tau\in (\tau_0, \tau_1)$.\\
\end{proof}

\subsection*{Proof of Theorem \ref{BaiBH2-inEStability}}
\begin{proof}
Consider the positive equilibria of Model \ref{BH2}. $(B^*, H^*)$ is a positive equilibrium if and only if $B^*=d_h e^{d_b\tau}H^*$ and $H^*$ is a positive root of
\begin{align}\label{Hfrac-1}
\frac{r H}{K+H^2}=d_h(d_b e^{d_b\tau}+1),
\end{align}
or, equivalently,
\begin{align}\label{Hfrac-2}
d_h(d_b e^{d_b\tau}+1)H^2-rH+Kd_h(d_b e^{d_b\tau}+1)=0.
\end{align}
Clearly, if $r<2\sqrt{K}d_h(d_b e^{d_b\tau}+1)$, there is no positive equilibrium; if $r>2\sqrt{K}d_h(d_b e^{d_b\tau}+1)$, or, equivalently,
\begin{align}\label{Hfrac-3}
\tau<\frac{1}{d_b}\ln\left(\left(\frac{r}{2d_h\sqrt{K}}-1\right)/d_b\right),\quad \text{and}\quad r>2d_h\sqrt{K}(1+d_b),
\end{align}
Model \ref{BH2} has two positive equilibria $E_1=(B_1^*, H_1^*)$ and $E_2=(B_2^*, H_2^*)$~($H_1^*<H_2^*$), where
\begin{align*}
H^*_i=\frac{r\pm\sqrt{r^2-4 K \left(d_b d_h e^{d_b \tau }+d_h\right)^2}}{2 \left(d_b d_h e^{d_b \tau }+ d_h\right)},\, i=1,2.
\end{align*}
Let
\begin{align}\label{taustar}
\tau^*=\frac{1}{d_b}\ln\left(\left(\frac{r}{2d_h\sqrt{K}}-1\right)/d_b\right),\quad \text{subject to}\quad r>2d_h\sqrt{K}(1+d_b).
\end{align}

The characteristic equation at $(B_i^*, H_i^*)$ is
\begin{align}\label{CharIn-1}
C(\lambda,\tau)=\lambda^2+(d_b+d_h)\lambda+d_bd_h+(\lambda+d_h-\Phi(H^*)(\tau))e^{-d_b\tau}e^{-\lambda\tau}=0,
\end{align}
where
\begin{align}\label{PhiHtau}
\Phi(H^*)(\tau)=\frac{2rKH^*}{(K+(H^*)^2)^2}
\end{align}
and $H^*=H_1^*$ or $H_2^*$, depending on $\tau$.

Let
\begin{align}\label{PQ-1}
P(\lambda, \tau)=\lambda^2+P_1(\tau)\lambda+P_0(\tau),\quad Q(\lambda, \tau)=Q_1(\tau)\lambda+Q_0(\tau),
\end{align}
where
\begin{align}\label{PiQi}
&P_1(\tau)=d_b+d_h, \quad P_0(\tau)=d_bd_h,\\
&Q_1(\tau)=e^{-d_b\tau},\qquad Q_0(\tau)=(d_h-\Phi(H^*)(\tau))e^{-d_b\tau}.
\end{align}
Then the characteristic equation \eqref{CharIn-2} can be rewritten as follows
\begin{align}\label{CharIn-2}
C(\lambda, \tau)=P(\lambda, \tau)+Q(\lambda, \tau)e^{-\lambda\tau}=0.
\end{align}

First, we prove that $\lambda=0$ cannot be a root of \eqref{CharIn-2}, i.e., $P(0,\tau)+Q(0,\tau)\neq 0$, for any $\tau\in[0,\tau^*)$. In fact,
\begin{eqnarray}\label{CharIn-21}
\begin{array}{lll}
C(0, \tau)&=P(0, \tau)+Q(0, \tau)=P_0(\tau)+Q_0(\tau)\\
&=d_bd_h+(d_h-\Phi(H^*)(\tau))e^{-d_b\tau}\\
&=e^{-d_b\tau}(d_h(d_b e^{d_b\tau}+1)-\Phi(H^*)(\tau))\\
&=e^{-d_b\tau}\left(\frac{rH^*}{K+(H^*)^2}-\frac{2rKH^*}{(K+(H^*)^2)^2}\right)\\
&=e^{-d_b\tau}\Phi(H^*)(\tau)((H^*)^2-K).
\end{array}
\end{eqnarray}
Here, \eqref{Hfrac-1} is used in the third equation. Since $H_1^*H_2^*=K$ and $H_1^*<H_2^*$, we know that $(H_1^*)^2<K<(H_2^*)^2$. Thus,  $C(0,\tau)<0$ at $E_1=(B_1^*, H_1^*)$, and $C(0,\tau)>0$ at $E_2=(B_2^*, H_2^*)$. It follows that for all $\tau\in[0,\tau^*)$, $\lambda=0$ cannot be a root of \eqref{CharIn-2} at both $E_1$ and $E_2$.

Now, we consider the stability of $E_1$ and $E_2$ when $\tau=0$. At $\tau=0$, the characteristic equation \eqref{CharIn-2} becomes $P(0, \tau)+Q(0, \tau)=0$, i.e.,
 \begin{align}\label{CharIn-0tua}
\lambda^2+(P_1(0)+Q_1(0))\lambda+P_0(0)+Q_0(0)=0.
\end{align}
At $E_1=(B_1^*, H_1^*)$, since $P_0(\tau)+Q_0(\tau)<0$ for all $\tau\in[0,\tau^*)$, we have $P_0(0)+Q_0(0)<0$. At $E_2=(B_2^*, H_2^*)$, we have $P_0(0)+Q_0(0)>0$ since $P_0(\tau)+Q_0(\tau)>0$ for all $\tau\in[0,\tau^*)$. Thus, we can conclude that at $\tau=0$, $E_1$ is unstable and $E_2$ is locally asymptotically stable.

In order to determine the local stability of the interior equilibrium $E_i, i=1,2$ when $\tau\in(0,\tau^*)$, we proceed as follows Kuang's book Chapter 3 \cite{kuang1993delay}.

Let $\lambda= iw(\tau), w(\tau)>0$, be the root of \eqref{CharIn-2}, then we have
\begin{eqnarray}\label{PQ-iw}
\begin{array}{lll}
P(iw, \tau)=-w^2+iwP_1(\tau)+P_0(\tau),\quad &Q(iw, \tau)=iwQ_1(\tau)+Q_0(\tau),\\
P_{\text{R}}(iw, \tau)=P_0(\tau)-w^2,\quad &Q_{\text{R}}(iw, \tau)=Q_0(\tau),\\
P_{\text{I}}(iw, \tau)=wP_1(\tau),\quad &Q_{\text{I}}(iw, \tau)=wQ_1(\tau).
\end{array}
\end{eqnarray}
By Theorem 4.1 in Kuang's book \cite{kuang1993delay}, we look for the positive roots $w(\tau)>0$ of
 \begin{align}\label{Fwtau-1}
F(w,\tau)=|P(iw, \tau)|^2-|Q(iw, \tau)|^2=0,\quad \tau\in[0,\tau^*).
\end{align}
Since
\begin{align}\label{Fwtau-2}
F(w,\tau)&=w^2+w^2(-2P_0(\tau)+P_1^2(\tau)-Q_1^2(\tau))+P_0^2(\tau)-Q_0^2(\tau)\\
&=w^4+b(\tau)w^2+c(\tau),
\end{align}
where
\begin{align}\label{btau-ctau}
b(\tau)=-2P_0(\tau)+P_1^2(\tau)-Q_1^2(\tau),\quad c(\tau)=P_0^2(\tau)-Q_0^2(\tau),
\end{align}
equation \eqref{Fwtau-2} may have no positive root, one positive root $w_{+}(\tau)$ or $w_{-}(\tau)$, or two positive roots $w_{+}(\tau)$ and $w_{-}(\tau)$, depending on $b(\tau)$ and $c(\tau)$. $w_{\pm}(\tau)$  can be represent as follows:
\begin{align*}
w_{\pm}(\tau)=\left[\frac{1}{2}(-b(\tau)\pm \sqrt{b^2(\tau)-4c(\tau)})\right]^{\frac{1}{2}}.
\end{align*}

In order to determine the occurrence of stability switches, we need to determine the sign of $\frac{d\text{Re}\lambda}{d\tau}$ or $\text{Re}\left(\frac{d\lambda}{d\tau}\right)^{-1}$.
From Theorem 3.1 and its proof of Kuang\cite{kuang1993delay}, we have
\begin{align}\label{Sign}
\begin{array}{ll}
S(\tau):&=\text{sign}\left\{\left(\frac{d(\text{Re}\lambda)}{d\tau}\right)\right\}_{\lambda=iw}
=\text{sign}\left\{\text{Re}\left(\frac{d\lambda}{d\tau}\right)^{-1}\right\}_{\lambda=iw}\\
&=\text{sign}\{P_1^2(\tau)-2P_0(\tau)-Q_1^2(\tau)+2w^2\}\\
&=\text{sign}\{\pm \sqrt{b^2(\tau)-4c(\tau)}\}.
\end{array}
\end{align}

First, we consider the local stability of the interior equilibrium $E_1$ when $\tau\in[0,\tau^*)$. Since for all $\tau\in[0,\tau^*)$, $P_0(\tau)+Q_0(\tau)<0$, and $P_0(\tau)=d_bd_h>0$, we have $Q_0(\tau)=(d_h-\Phi(H^*)(\tau))e^{-d_b\tau}<0$. Thus, $P_0(\tau)-Q_0(\tau)>0$, $\tau\in[0,\tau^*)$. It follows that
\begin{align*}
c(\tau)=P_0^2(\tau)-Q_0^2(\tau)=[P_0(\tau)+Q_0(\tau)][P_0(\tau)-Q_0(\tau)]<0,\quad \tau\in[0,\tau^*).
\end{align*}
Therefore, \eqref{Fwtau-2} has unique positive root $w_{+}(\tau)$. From \eqref{Sign}, $S(\tau)=1$. Since $E_1$ is unstable at $\tau=0$,
no stability switch occur as $\tau$ increases in $[0,\tau^*)$, and $E_1$ is unstable for all $\tau\in[0,\tau^*)$.\\

Now, we consider the local stability of the interior equilibrium $E_2$ when $\tau\in[0,\tau^*)$. We have know that for all $\tau\in[0,\tau^*)$, $P_0(\tau)+Q_0(\tau)>0$ and $c(\tau)=[P_0(\tau)+Q_0(\tau)][P_0(\tau)-Q_0(\tau)]$. Thus, we consider the sign of $P_0(\tau)-Q_0(\tau)$. By a careful computation, we get
 \begin{align}\label{P0Q0}
P_0(\tau)-Q_0(\tau)=d_h e^{-d_b\tau}\varphi(\tau),
\end{align}
where
 \begin{align}\label{P0Q01}
\varphi(\tau)=(d_b e^{d_b\tau}+1)\left[2-\sqrt{1-\frac{4Kd_h^2 (d_b e^{d_b\tau}+1)^2}{r^2}}\right]-2.
\end{align}
It is clear that $\varphi(\tau)$ is strictly increasing with respect to $\tau\in[0,\tau^*), f(\tau^*)=2d_b e^{d_b\tau^*}$, and
\begin{align*}
\varphi(0)=(d_b +1)\left[2-\sqrt{1-\frac{4Kd_h^2 (d_b +1)^2}{r^2}}\right]-2.
\end{align*}

Now, we consider the following two cases.\\

{\bf Case 1.} $d_b\ge 1$. Then clearly $\varphi(0)>0$ and hence $\varphi(\tau)>0$ for all $\tau\in[0,\tau^*)$. Thus, $c(\tau)>0$. In addition, $b(\tau)=d_b^2+d_h^2-e^{-2d_b\tau}>0$. Therefore, for all $\tau\in[0,\tau^*)$, $F(w,\tau)\neq 0$. This implies that in such case no stability switch occur with $\tau$ increasing in $[0,\tau^*)$. Since $E_2$ is is locally asymptotically stable at $\tau=0$, we can conclude that $E_2$ is is locally asymptotically stable for all $\tau\in[0,\tau^*)$ if $d_b\ge 1$.\\

{\bf Case 2.} $0<d_b<1$. Then, $(1+d_b)^2-4d_b^2>0$. When $r>2d_h\sqrt{K}(1+d_b)$, we have
\begin{align}\label{varphi}
\begin{array}{ll}
\varphi(0)>0&\Leftrightarrow \sqrt{1-\frac{4Kd_h^2 (d_b +1)^2}{r^2}}<\frac{2d_b}{1+d_b}\\
&\Leftrightarrow r<\frac{2d_h\sqrt{K}(1+d_b)^2}{\sqrt{(1+d_b)^2-4d_b^2}}.
\end{array}
\end{align}
Thus, when $2d_h\sqrt{K}(1+d_b)<r<\frac{2d_h\sqrt{K}(1+d_b)^2}{\sqrt{(1+d_b)^2-4d_b^2}}$, $\varphi(0)>0$ and hence for all $\tau\in[0,\tau^*)$, $c(\tau)>0$.

This case is also divided into two subcases: (i) $d_b^2+d_h^2\ge 1$, (ii) $d_b^2+d_h^2< 1$.\\

{\bf (i)} $d_b^2+d_h^2\ge 1$. It follows  $b(\tau)=d_b^2+d_h^2-e^{-2d_b\tau}>0$. Thus, in such case,  $F(w,\tau)\neq 0$ holds for all $\tau\in[0,\tau^*)$. This implies that in such case no stability switch occur with $\tau$ increasing in $[0,\tau^*)$. Since $E_2$ is locally asymptotically stable at $\tau=0$, we can conclude that $E_2$ is is locally asymptotically stable for all $\tau\in[0,\tau^*)$ if  $2d_h\sqrt{K}(1+d_b)<r\le\frac{2d_h\sqrt{K}(1+d_b)^2}{\sqrt{(1+d_b)^2-4d_b^2}}$.

If $r>\frac{2d_h\sqrt{K}(1+d_b)^2}{\sqrt{(1+d_b)^2-4d_b^2}}$, then $\varphi(0)<0$ and there exists unique $\tau_c\in (0,\tau^*)$ such that $c(\tau)<0$ for $\tau\in(0,\tau_c)$ and $c(\tau)>0$ for $\tau\in(\tau_c,\tau^*)$. Thus, if $\tau\in(\tau_c,\tau^*)$, then $F(w,\tau)= 0$ has no positive root; if $\tau\in(0,\tau_c)$, then $F(w,\tau)= 0$ has a unique positive root $w_{+}$ and $S(\tau)=1$. Since $E_2$ is stable at $\tau = 0$, from Theorem 3.1 of Kuang \cite{kuang1993delay}, we know that the stability of $E_2$ switches just once in $[0,\tau_c)$ from stable to unstable.

{\bf (ii)} $d_b^2+d_h^2< 1$. We divide this subcase into the following two scenarios:\\

{\bf (a)}\quad $\frac{d_b^2}{\left(\frac{(1+d_b)^2}{\sqrt{(1+d_b)^2-4d_b^2}}-1\right)^2}<d_b^2+d_h^2< 1$, which implies that $$2d_h\sqrt{K}\left(1+\sqrt{\frac{d_b^2}{d_b^2+d_h^2}}\right)<2d_h\sqrt{K}\frac{(1+d_b)^2}{\sqrt{(1+d_b)^2-4d_b^2}}.$$
 This scenario is again divided into following three cases:\\

(1)\quad  $2d_h\sqrt{K}\left(1+\sqrt{\frac{d_b^2}{d_b^2+d_h^2}}\right)<r\le 2d_h\sqrt{K}\frac{(1+d_b)^2}{\sqrt{(1+d_b)^2-4d_b^2}}$.

 Since $2d_h\sqrt{K}\left(1+\sqrt{\frac{d_b^2}{d_b^2+d_h^2}}\right)>2d_h\sqrt{K}(1+d_b)$, from \eqref{varphi}, $\varphi(0)\ge0$ and hence for all $\tau\in[0,\tau^*)$, $c(\tau)\ge0$.

 When $\frac{1}{2d_b}\ln\left(\frac{1}{d_b^2+d_h^2}\right)\le\tau<\tau^*$, $b(\tau)\ge0$. Thus, for all $\tau\in[0,\tau^*)$, $F(w,\tau)\neq 0$.

 When $0<\tau< \frac{1}{2d_b}\ln \left(\frac{1}{d_b^2+d_h^2}\right)$, we have $b(\tau)< 0$. Thus, we consider $b^2(\tau)-4c(\tau)$. By a simple computation, we have
 \begin{align*}
\begin{array}{ll}
b^2(\tau)-4c(\tau)&=(P_1^2(\tau)-Q_1^2(\tau))(P_1^2(\tau)-Q_1^2(\tau)-4P_0(\tau))+4Q_0^2(\tau)\\
&=\phi(\tau)+4Q_0^2(\tau).
\end{array}
\end{align*}
Here, $\phi(\tau)=\left((d_b+d_h)^2-e^{-2d_b\tau}\right)\left((d_b+d_h)^2-e^{-2d_b\tau}-4d_b d_h\right)$. It is clear that
\begin{align*}
\begin{array}{ll}
\phi(\tau)\geq 0 \Leftrightarrow \tau\le \frac{1}{2d_b}\ln\left(\frac{1}{(d_b+d_h)^2}\right)\quad \text{or} \quad \tau\ge \frac{1}{2d_b}\ln\left(\frac{1}{(d_b-d_h)^2}\right).
\end{array}
\end{align*}
Thus, if $\frac{1}{2d_b}\ln \left(\frac{1}{(d_b+d_h)^2}\right)<\tau< \frac{1}{2d_b}\ln \left(\frac{1}{d_b^2+d_h^2}\right)$, then $\phi(\tau)<0$. If $0<\tau\le \frac{1}{2d_b}\ln \left(\frac{1}{(d_b+d_h)^2}\right)$, then $\phi(\tau)\ge0$ and hence
$b^2(\tau)-4c(\tau)>0$, which implies that $F(w,\tau)=0$ have two positive roots $0<w_{-}<w_{+}$. Thus, from Theorem 3.1 of Kuang\cite{kuang1993delay}, the stability of $E_2$ can change a finite number of times at most as $\tau$ is increased $\tau\in[0, \tau^*)$, and eventually it becomes unstable.\\

(2)\quad $2d_h\sqrt{K}(1+d_b)<r\le 2d_h\sqrt{K}\left(1+\sqrt{\frac{d_b^2}{d_b^2+d_h^2}}\right)$.

From \eqref{varphi}, $\varphi(0)>0$ and hence for all $\tau\in[0,\tau^*)$, $c(\tau)>0$. In this case,
\begin{align*}
\tau^*=\frac{1}{d_b}\ln\left(\left(\frac{r}{2d_h\sqrt{K}}-1\right)/d_b\right)\leq \frac{1}{2d_b}\ln \left(\frac{1}{d_b^2+d_h^2}\right).
\end{align*}
It follows $b(\tau)=d_b^2+d_h^2-e^{-2d_b\tau}<0$ for all $\tau\in(0, \tau^*)$. Note that
\begin{align*}
\tau^*=\frac{1}{d_b}\ln\left(\left(\frac{r}{2d_h\sqrt{K}}-1\right)/d_b\right)\leq \frac{1}{2d_b}\ln \left(\frac{1}{(d_b+d_h)^2}\right)\Leftrightarrow r\le 2d_h\sqrt{K}\left(\frac{d_b}{d_b+d_h}+1\right).
\end{align*}
Thus,
\begin{itemize}
  \item when $2d_h\sqrt{K}(1+d_b)<r\le 2d_h\sqrt{K}\left(\frac{d_b}{d_b+d_h}+1\right)$, $\phi(\tau)\ge 0$ and hence $b^2(\tau)-4c(\tau)>0$ for all $\tau\in[0, \tau^*)$. It yields that $F(w,\tau)=0$ have two positive roots $0<w_{-}<w_{+}$. Thus, from Theorem 3.1 of Kuang \cite{kuang1993delay}, the stability of $E_2$ can change a finite number of times at most as $\tau$ is increased in $[0, \tau^*)$, and eventually it becomes unstable.
  \item when $2d_h\sqrt{K}\left(\frac{d_b}{d_b+d_h}+1\right)<r\le 2d_h\sqrt{K}\left(1+\sqrt{\frac{d_b^2}{d_b^2+d_h^2}}\right)$, we have $\tau^*>\frac{1}{2d_b}\ln\left(\frac{1}{(d_b+d_h)^2}\right)$. If $ \frac{1}{2d_b}\ln\left(\frac{1}{(d_b+d_h)^2}\right)<\tau< \tau^*$, then $\phi(\tau)< 0$. If $0<\tau\le \frac{1}{2d_b}\ln\left(\frac{1}{(d_b+d_h)^2}\right)$, $\phi(\tau)\ge 0$ and hence $b^2(\tau)-4c(\tau)>0$. Thus, in such case, from Theorem 3.1 of Kuang \cite{kuang1993delay} the stability of $E_2$ can change a finite number of times at most as $\tau$ is increased in $[0, \tau^*)$.
\end{itemize}

(3)\quad $r>\frac{2d_h\sqrt{K}(1+d_b)^2}{\sqrt{(1+d_b)^2-4d_b^2}}$.

 In this case, $\varphi(0)<0$ and there exists unique $\tau_c\in (0,\tau^*)$ such that $c(\tau)<0$ for $\tau\in(0,\tau_c)$ and $c(\tau)>0$ for $\tau\in(\tau_c,\tau^*)$. Thus, if $\tau\in(0,\tau_c)$, then $F(w,\tau)= 0$ has a unique positive root $w_{+}$ and $S(\tau)=1$. Since $E_2$ is stable at $\tau = 0$, from Theorem 3.1 of Kuang \cite{kuang1993delay}, we know that the stability of $E_2$ switches once in $[0,\tau_c)$ from stable to unstable.

{\bf (b)}\quad $d_b^2+d_h^2\le \frac{d_b^2}{\left(\frac{(1+d_b)^2}{\sqrt{(1+d_b)^2-4d_b^2}}-1\right)^2}$, which implies that $$2d_h\sqrt{K}\left(1+\sqrt{\frac{d_b^2}{d_b^2+d_h^2}}\right)\ge 2d_h\sqrt{K}\frac{(1+d_b)^2}{\sqrt{(1+d_b)^2-4d_b^2}}.$$
This scenario is again divided into following two cases:\\

(1)\quad $2d_h\sqrt{K}(1+d_b)<r\le \frac{2d_h\sqrt{K}(1+d_b)^2}{\sqrt{(1+d_b)^2-4d_b^2}}$.

From \eqref{varphi}, $\varphi(0)>0$ and hence for all $\tau\in[0,\tau^*)$, $c(\tau)>0$. In this case, we also have
$\tau^*\leq \frac{1}{2d_b}\ln \left(\frac{1}{d_b^2+d_h^2}\right)$.
It follows $b(\tau)=d_b^2+d_h^2-e^{-2d_b\tau}<0$ for all $\tau\in(0, \tau^*)$.

If $2d_h\sqrt{K}\left(\frac{d_b}{d_b+d_h}+1\right)<\frac{2d_h\sqrt{K}(1+d_b)^2}{\sqrt{(1+d_b)^2-4d_b^2}}$, similar to the arguments above, we get that no matter $2d_h\sqrt{K}(1+d_b)<r\le 2d_h\sqrt{K}\left(\frac{d_b}{d_b+d_h}+1\right)$ or $2d_h\sqrt{K}\left(\frac{d_b}{d_b+d_h}+1\right)<r\le 2d_h\sqrt{K}\frac{(1+d_b)^2}{\sqrt{(1+d_b)^2-4d_b^2}}$ the stability of $E_2$ can change a finite number of times at most as $\tau$ is increased in $[0, \tau^*)$.

If $\frac{2d_h\sqrt{K}(1+d_b)^2}{\sqrt{(1+d_b)^2-4d_b^2}}<2d_h\sqrt{K}\left(\frac{d_b}{d_b+d_h}+1\right)$, then when $2d_h\sqrt{K}(1+d_b)<r\le \frac{2d_h\sqrt{K}(1+d_b)^2}{\sqrt{(1+d_b)^2-4d_b^2}}$, $\phi(\tau)\ge 0$ for all $[0,\tau^*)$ and hence $b^2(\tau)-4c(\tau)>0$ for all $\tau\in[0, \tau^*)$. It yields that $F(w,\tau)=0$ have two positive roots $0<w_{-}<w_{+}$. Thus, from Theorem 3.1 of Kuang \cite{kuang1993delay}, the stability of $E_2$ can change a finite number of times at most as $\tau$ is increased in $[0, \tau^*)$, and eventually it becomes unstable.

 (2)\quad $r>\frac{2d_h\sqrt{K}(1+d_b)^2}{\sqrt{(1+d_b)^2-4d_b^2}}$.

  Similar to the arguments above, $\varphi(0)<0$ and there exists unique $\tau_c\in (0,\tau^*)$ such that the stability of $E_2$ switches once in $[0,\tau_c)$ from stable to unstable.

The proof is completed.

\end{proof}

\section*{Acknowledgements}
The work of Y.K. is also partially supported by NSF-DMS (1716802); NSF-IOS/DMS (1558127), and The James S. McDonnell Foundation 21st Century Science Initiative in Studying Complex Systems Scholar Award (UHC Scholar Award 220020472).\\

\begin{appendix}

\section{Some Important Lemmas}\label{Brauer Lemmas}

\newtheorem*{thmA}{Lemma A}
\newtheorem*{thmB}{Lemma B}

Consider the characteristic equation of the form
\begin{eqnarray}\label{char}
p(z)+e^{-z\tau}q(z)=0
\end{eqnarray}
where $p$ and $q$ are polynomials with real coefficients and $\tau>0$ is the delay. The following result was given by Brauer \cite{brauer1987absolute}.

\begin{thmA}\label{Brauer}
Suppose that $p(z)$ and $q(z)$ are analytic in some open set containing $z\geq 0$, and satisfying the following conditions:\\
(i)\quad $p(z)\neq 0$, $Rez\geq 0$,\\
(ii)\quad $\overline{p(-iy)}=p(iy),  \overline{q(-iy)}=q(iy), 0\leq y<\infty$,\\
(iii)\quad $p(0)+q(0)=0$,\\
(iv)\quad $|q(iy)|<|p(iy)|$ for $0<y<\infty$,\\
(v)\quad $\lim_{|z|\to\infty, Rez\geq 0} |q(z)/p(z)|=0$.\\
Then except for the roots $z=0$, all roots of \eqref{char} are in $Rez<0$ for all $0\leq \tau<\infty$.
\end{thmA}

We need the fluctuation lemma due to Hirsh, Hanisch, and Gabriel\cite{hirsch1985differential}.

\begin{thmB}[Fluctuation Lemma]\label{lemma-fluct}
Let $f: \mathbb{R}_+\to \mathbb{R}$ be a differentiable function. If $\liminf_{t\to\infty} f(t) < \limsup_{t\to\infty} f(t)$, then there are sequences $\{t_m\} \uparrow \infty$ and $\{s_m\} \uparrow \infty$ such that
 \begin{eqnarray*}
\begin{array}{ll}
 f(t_m) \to \limsup_{t\to\infty}f(t),\quad  f'(t_m) \to 0\quad \mbox{as}\quad m \to\infty, \\
 f(s_m) \to \liminf_{t\to\infty}f(t),\quad  f'(s_m) \to 0\quad \mbox{as}\quad m \to\infty.
 \end{array}
\end{eqnarray*}
\end{thmB}

\subsection*{Proof of Proposition \ref{p:eq1}}
\begin{proof}

{Let $m=d_b r$, $n=d_h^2e^{d_b\tau}(e^{d_b\tau}-1)$, and $c=4d_b^2d_h^2Ke^{2d_b\tau}$, then:
$$H_1^*=\frac{e^{-d_b\tau}}{2d_bd_h}(m-\alpha n-\sqrt{(m-\alpha n)^2-c}),$$ and $$H_2^*=\frac{e^{-d_b\tau}}{2d_bd_h}(m-\alpha n+\sqrt{(m-\alpha n)^2-c}).$$
So, function $g=m-\alpha n-\sqrt{(m-\alpha n)^2-c}$ and function $f =m-\alpha n+\sqrt{(m-\alpha n)^2-c}$ will decide two equilibrium functions are increasing or decreasing functions of $\alpha$. After that,
$$g'(\alpha)= n(-1+\frac{m-n\alpha }{\sqrt{(m-n\alpha)^2-c}}),$$ and
$$f'(\alpha)= n(-1+\frac{n\alpha - m}{\sqrt{(m-n\alpha)^2-c}}).$$
Since $H_1^*>0$, $m>n\alpha$ and $m-n\alpha>\sqrt{(m-\alpha n)^2-c}$, therefore $g'(\alpha)>0$ and $f'(\alpha)<0$, i.e. $H_1^*$ is monotonically increasing, and $H_2^*$ is monotonically decreasing. }\\

\red{Next, we consider parameter $r$. Then $$g'(r)= d_b(1-\frac{d_br-n\alpha }{\sqrt{(d_br-n\alpha)^2-c}})<0,$$ and
$$f'(r)= d_b(1+\frac{d_br-n\alpha }{\sqrt{(d_br-n\alpha)^2-c}})>0.$$
Therefore, $H_1^*$ is monotonically decreasing by $r$, and $H_2^*$ is monotonically increasing by $r$. }\\

\red{Afterwards, if $K$ increases, only $c$ will increase. Then $\sqrt{(m-n\alpha)^2-c}$ will decrease. Therefore, $H_1^*$ is monotonically increasing by $K$, and $H_2^*$ is monotonically decreasing by $K$.}\\

\red{When we consider $d_b$ and $d_h$, we can simplified the model to Model (\ref{BH3}), i.e. $\alpha=0$. Then we let $H_1^*$ be the function $p=\frac{e^{-d_b \tau } \left(r-\sqrt{r^2-4 d_h^2 K e^{2 d_b \tau }}\right)}{2 d_h}$, and $H_2^*$ be the function $q=\frac{e^{-d_b \tau } \left(r+\sqrt{r^2-4 d_h^2 K e^{2 d_b \tau }}\right)}{2 d_h}$. So, $$p'(d_h)=\frac{r e^{-d_b \tau } \left(\frac{r}{\sqrt{r^2-4 d_h^2 K e^{2 d_b \tau }}}-1\right)}{2 d_h^2}>0,$$ and $$q'(d_h)=\frac{r e^{-d_b \tau } \left(-\frac{r}{\sqrt{r^2-4 d_h^2 K e^{2 d_b \tau }}}-1\right)}{2 d_h^2} <0.$$ Therefore, $H_1^*$ is monotonically increasing by $d_h$, and $H_2^*$ is monotonically decreasing by $d_h$. }\\

\red{Finally, let us see $d_b$. Then $$p'(d_b)=\frac{r   e^{-d_b \tau } \left(\frac{r}{\sqrt{r^2-4 d_h^2 K e^{2 d_b \tau }}}-1\right)\tau}{2 d_h}>0,$$ and $$q'(d_b)=\frac{r  e^{-d_b \tau } \left(-\frac{r}{\sqrt{r^2-4 d_h^2 K e^{2 d_b \tau }}}-1\right)\tau}{2 d_h}<0.$$ Therefore, $H_1^*$ is monotonically increasing by $d_b$, and $H_2^*$ is monotonically decreasing by $d_b$.} 
\end{proof}
\end{appendix}


\bibliographystyle{unsrt}

\end{document}